\documentclass[%
reprint,
amsmath,amssymb,
aps,
pra,
]{revtex4-2}
\usepackage[dvipsnames]{xcolor}

\usepackage{graphicx}
\usepackage{dcolumn}
\usepackage{bm}
\usepackage{hyperref}

\usepackage{soul}

\newcommand{\fl}[1]{\textcolor{green}{#1}}
\newcommand{\versiontwo}[1]{\textcolor{magenta}{v2: #1}}

\renewcommand{\st}[1]{}
\renewcommand{\versiontwo}[1]{} 
\renewcommand{\fl}[1]{} 

\newcommand\bk{{\bf k}}
\newcommand\bx{{\bf x}}
\newcommand\egaldef{\stackrel{\mbox{\upshape\tiny def}}{=}}
\newcommand\egallaw{\stackrel{d}{=}}
\newcommand\EE{\mathcal{E}}
\newcommand\E{{\mathbb E}}
\newcommand\R{{\mathbb R}}
\newcommand\Z{{\mathbb Z}}
\newcommand\LL{{\mathcal L}}
\newcommand\I{{\mathbb I}}
\newcommand\Tr{{\rm Tr}}

\newcommand\x{{\bf x}}

\newcommand\C{\mathcal{C}}

\def\DD{\displaystyle} 

\newcommand\1{\leavevmode\hbox{\rm 
\small1\kern-0.35em\normalsize1}}
\newcommand\ind[1]{\1_{\{#1\}}}        

\newcommand\ie{\textit{i.e.}~}
\newcommand\eg{\textit{e.g.}~}

\begin{document}
	
\preprint{APS/123-QED}
	
\title{Learning and extrapolating scale-invariant processes}

\author{Anaclara Alvez$^{1,2}$}
\author{Cyril Furtlehner$^{1,3}$}
\author{François Landes$^{1,2}$}
\affiliation{$^1$Laboratoire Interdisciplinaire des Sciences du Numérique (LISN)}
\affiliation{$^2$Université Paris-Saclay}
\affiliation{$^3$INRIA-Saclay}
	
\date{\today}
	
\begin{abstract}
Machine Learning (ML) has deeply changed some fields recently, like Language and Vision. 
In the case of complex systems, spectacular breakthroughs happened too (\textit{e.g.}~for protein folding) and more are expected to come.
Our question is: how and to which extent can one regress scale-free processes, \ie processes displaying power law behavior, like earthquakes or avalanches?
The events one is interested in predicting are the large ones, \textit{i.e.}~events that are typically rare in the training set, so we are basically in the extrapolation regime. 
While some recent works also tackle scale-free systems by proposing generative models closely aligned with the renormalization group framework,
here instead we explore the problem of prediction in the extrapolation regime. We consider two paradigmatic 
problems that are statistically self-similar.
The first one is a $2$-dimensional  fractional 
Gaussian field obeying linear dynamics, self-similar by construction  and amenable to exact analysis. 
The second one is the Abelian sandpile model, 
exhibiting self-organized criticality. 
        
The emerging paradigm of geometric deep learning 
shows that including known symmetries into the model's architecture is key to success (as translation invariance for images). 
Here one may hope to extrapolate only by leveraging scale invariance, 
which is however a peculiar symmetry, as it involves possibly non-trivial coarse-graining operations and anomalous scaling.
We perform experiments on various existing architectures like U-net, Riesz network 
(scale invariant by construction), or our own proposals: a wavelet-decomposition based Graph Neural Network (with discrete scale symmetry), a Fourier embedding layer and a
Fourier-Mellin Neural Operator.
Based on these experiments and a complete characterization of the linear case, we identify the main issues relative to spectral biases and coarse-grained representations,
and discuss how to alleviate them with the relevant inductive biases.		
\end{abstract}

\maketitle

	
\section{\label{sec:intro}Introduction}
    
In the last decade machine learning (ML) and more specifically deep neural networks (DNN) have thoroughly renewed the research perspectives in many fields like natural language processing and computer vision. Despite indisputable successes, the introduction of ML approaches to describe or predict complex physical systems remains a challenge~\cite{carleo2019machine,meng2025physics}, driven by the specificity of physics data. Unlike natural image or language datasets, physics data are governed by symmetries, conservation laws, causal relations, and complex systems frequently involves rare, highly non-linear events that are critical to their dynamics.
These characteristics impose specific statistical constraints that traditional, general purpose ML models are not inherently designed to handle, requiring the introduction of innovative approaches to ensure both accuracy and interpretability.
By complex physics we mean systems displaying qualitatively similar phenomena over a broad range of scales, where all or many scales are relevant and thus cannot be decoupled. Such systems can be found across the natural and social sciences \cite{newman2005power}, in phenomena ranging from seismic activity~\cite{gutenberg1956earthquake}, to neural avalanches~\cite{beggs2003neuronal}, or financial market fluctuations~\cite{gabaix2003theory}.
Setting aside multifractality—a hallmark of complexity—we observe that even monofractal systems, which by definition exhibit self-similarity under scale transformations, already pose significant challenges to standard machine learning models.
The purpose of this work is to discuss two important  aspects of this problem.
The first aspect concerns \textit{spectral bias}~\cite{rahaman2019spectral}, that most ML models suffer from, and which becomes a critical issue in this context. Generically most ML models learn only the low frequencies, as these dominate the loss, making them unable or very slow to learn also the high frequency features. This problem will appear as a crucial point in the examples we present. 
The second aspect deals with the inability of standard ML models to extrapolate. 
While conventional machine learning models excel at interpolation within the training distribution, they struggle to extrapolate beyond it or into its tails.
For complex systems the rare events are usually the extreme ones, that play a crucial role in the physical process: they cannot be neglected.

Formally we can define the ability of an ML model to \textit{extrapolate} as follows. 
Given some physical process in the form of a function $f_{GT}(\varphi)=\psi$ (deterministic or not) connecting a pair  $(\varphi,\psi)$ of input-output quantities (not specified here, but think for instance of a field configuration before and after an avalanche), and a corresponding real observable  $s = g(\varphi,\psi)\in\R$ (\eg the size of the avalanche),  we want to be able to train a neural network $f_\theta$ on a subset ${\mathcal S}_{\rm train}\subset \{(\varphi,\psi), \ind{g(\varphi,\psi) < s^{\rm th}} \}$ corresponding to observation below some threshold $s^{th}\in \R$  and still be able to generalize well to any pair $(\varphi,\psi)$, in particular contained in the subset $\ind{g(\varphi,\psi) \ge s^{\rm th}}$ which are typically those which are too rare to be observed during training. 
Intuitively this seems possible if the true process $f_{GT}$ obeys a form of scale-symmetry, such that large-scale events ($s>s^{th}$) are similar and share the same properties as the small scales ones, up to some well adapted rescaling. The question of how to inform the ML model with the
appropriate scale symmetry is the subject of the present work. 

The practical importance of symmetries in Neural Networks has been recognized since the beginning, with the early introduction by Y.~Le~Cun of convolutional networks\cite{lecun1989backpropagation,li2021survey}, that ensure (approximate) translation symmetry. 
Another more recent example is Graph Neural Networks \cite{scarselli2008graph}, that are built so as to ensure node-permutation equivariance, since node labeling should not impact modeling or predictions. 
For general groups, the idea of building a neural network that is exactly equivariant by construction, has then been formalized in a series of papers~\cite{bronstein2021geometric}. 
Let us recall the definition of equivariance. An operator (\eg a Neural Network) $f : X \xrightarrow{} Y$ 
is equivariant with respect to a group $G$ if and only if: 
    \begin{align}\label{def:deter_equi}
        \forall g \in G : \hspace{1cm} \rho_Y(g) \circ f = f \circ \rho_X(g)        
    \end{align}
    where $\rho_X,\rho_Y$ are the representations of the group $G$ in the vector spaces $X,Y$. Intuitively, equivariance means that if the input is transformed according to $g$,
    the output also transforms according to the same transformation $g$.
Invariance is a special case of equivariance where one chooses a trivial representation $\forall g\in G, \rho_Y(g)=\mathbb{I}$, thus restricting the expressivity of the operator $f$.

Depending on the complexity of the symmetry group, the number of generators can be very large and possibly infinite. Since standard implementations are based on the regular representation, \ie one latent variable per generator, the size of the equivariant layer can be potentially infinite, hence a tradeoff has to be found between the computational or memory burdens and the faithfulness of the representations (projection of the input onto the basis corresponding to the generators). 
For the rotation group, useful in many scientific applications, 
there is a large body of work, from the early theoretical papers \cite{thomas_tensor_2018} up to the diffusion of two dedicated torch packages e3nn~\cite{e3nn_software} and Nvidia's \textit{cuEquivariance} (closed source).
In this steerable network approach~\cite{thomas_tensor_2018}, the point is to have representations built in the appropriate basis, \ie that corresponding to the irreducible representation of the group of interest, and perform products in the proper space, so as to maintain equivariance.

In the context of Computer Vision, the notions of  scale invariance or equivariance have recently been introduced~\cite{worrall2019deep, sosnovik2019scale}. In appendix \ref{app:scale_equivariance} we recall these definitions and show to which extent they apply to statistically self-similar processes.
It is worth noticing that many works focus on the problem of correctly classifying images regardless of the size of the object of interest in the picture, \ie deal  with an invariant task. 
In this sense, they deal with a markedly different problem from the one we want to address. 
Spatial transformers~\cite{jaderberg2015spatial} is a general purpose method to perform arbitrary affine transformation of arbitrary input feature map to facilitate classification for instance, but is not as such able to generalize to unseen scales. Some other methods are based on multi-scale channels where either the input~\cite{kanazawa2014locally} or the (steerable) filters are replicated at different scales~\cite{sosnovik2019scale}. 
Some methods use continuous filters to generate multiscale representation of the input~\cite{lin2017feature,adelson1984pyramid} like Gaussian derivative filters~\cite{lindeberg2022scale} or Riesz ones~\cite{barisin2024riesz}, the latter yielding scale invariant networks by construction. 
Fourier input is also considered in~\cite{rahman2024truly} in order to correct for antialiasing when input feature maps are downscaled with continuous filters.

Among these methods only a few of them may readily generalize to unseen scales, notably the Riesz networks~\cite{barisin2024riesz} can be successfully used to segment cracks of all sizes on tomographic images of concrete, even when trained solely on a subset of crack sizes. A key element to this success is of course that cracks are self-similar objects: the shapes of  small cracks is distributed similarly to that of larger cracks.

Independently from this line of work, for complex systems, connections between the renormalization group (RG) flow and neural networks have emerged as a compelling alternative approach to conventional supervised learning models.
While such methods are typically limited in scope, notable exceptions exist -- such as proposals to construct generative models directly from RG equations~\cite{marchand2023multiscale,guth2022wavelet}.
Beside image processing, the development of scale invariant/equivariant  architectures in the context of analyzing and forecasting critical systems with non-trivial scaling behavior appears to us as a virgin ground.

Our context is distinct from the usual computer vision setup, where the output, \eg a classification label, is invariant w.r.t. to the symmetry group. In our context we instead seek an equivariance in a statistical sense
where Eq.~(\ref{def:deter_equi}) is now replaced by 
\begin{align}
    \forall g \in G : \hspace{1cm} \rho_Y(g) \circ f \stackrel{d}{=} f \circ \rho_X(g)   \label{def:stat_equi}   
\end{align}
meaning that a change of the input by a group element does change the output distribution.
This stems from the requirement that for a self-similar field, scaling operations—such as zooming in or out—necessitate generating missing components of the data
and accounting for anomalous scaling behavior. Specifically, when the input is rescaled, either its fine (small-scale) and coarse (large-scale) components, as well as those of the output, must be synthesized to preserve the invariance of their joint distribution. 
The simplest example is zooming in on a Brownian motion: to 'look closer,' one must sample small scale Brownian paths, as the process is inherently stochastic.
For the type of data we study (e.g., fractional Gaussian fields), scaling introduces non-trivial, exponent-dependent behavior (e.g., tied to the Hurst exponent) that must  explicitly be accounted for. Our goal is to design a neural network that inherently incorporates these scaling laws.

We investigate these questions under the following functional regression framework where 
we have a pair of input-output fields $(\varphi,\psi)$ 
both defined on a finite square lattice $\Omega\subset {\mathbb Z}^2$ of 
size $\vert\Omega\vert = L^2$. The fields are related by 
\begin{align}\label{def:functional_regression}
\psi(\x) = F[\varphi](\x) +\epsilon(\x)
\end{align}
where $F$ is an unknown operator, $\epsilon$ is a noise field also defined on the square lattice,
and all $\varphi$, $\psi$ and $\epsilon$ belong to some family of scale-free fields with
long range spatial correlations and self-similarity characterized by critical exponents. $F$
and the various fields are assumed to be such that equation~(\ref{def:functional_regression}) is equivariant
w.r.t. scale symmetry, equation~(\ref{def:stat_equi}). Typically, noise in this context arises from edge effects—where scaling either artificially introduces or
loses information at the edges of the domain.
Given a certain number of observations $\{(\varphi^{(s)},\psi^{(s)}),s=1,\ldots M\}$, the goal is then to predict $\psi$ from $\varphi$.
Note that this regression problem has recently been framed using the neural operator approach in the  literature~\cite{anandkumar2020neural,li2020fourier,lu2021learning,kovachki2023neural}.

Concretely we consider examples with different levels of difficulty, out of which lessons can be learned.   
The first self-similar toy-problem of functional regression we consider  involves Fractional Gaussian fields
(FGF)~\cite{lodhia2016fractional} subject to a linear functional map $F$, where the extrapolation task is defined by training on data with truncated spectrum and testing on data with full spectrum. 
In the cases of primary interest—specifically, when masking either the high-frequency or low-frequency ends of the spectrum—this corresponds, in direct space,
to super-resolution or extrapolation to larger scales, respectively. In practice, this means testing on examples at scales not encountered during training.

Our second example concerns non-linear regression in the context of self-organized criticality: the sandpile model~\cite{dhar1990self, dhar1999abelian} *
is an avalanche process paradigmatic of scale free-processes, where the size distribution of events $S$ follows a $1/S$ scaling. The extrapolation is
then defined as the ability to predict large avalanches when training solely on small and intermediate ones.

Equipped with these two quite different test cases, we investigate how to inform the network with the scale symmetry,
in order to first fit the training set in its full (broad) spectrum, and possibly extrapolate to an extra piece of the spectrum, under the assumption of a self-similar task.
Inspired by the mathematical framework of steerable networks  that some of us worked with~\cite{pezzicoli2024rotation, pezzicoli2024statistical}, we are looking to work in
the appropriate bases, such that operations are by construction scale-invariant or equivariant.


\section{\label{sec:setting} Datasets and Tasks:}	
In this section we present the two physical systems to be studied -- the Fractional Gaussian field (FGF) and the Abelian Sandpile Model (ASM) -- and specify the tasks we want to solve.
The FGF serves as a relatively basic, well-understood example of a scale invariant system, to which we add simple dynamics. The ASM is a toy model for self-organized criticality
and avalanche phenomena, more complex due to the fact that, despite having avalanche processes with long-range correlations (large susceptibility), the spatial correlations decay rapidly.

\subsection{Fractional Gaussian field}\label{sec:FGF}
We first consider a simple multi-dimensional Fractional Gaussian field (FGF). Formally it is defined in the continuum in arbitrary dimension as~\cite{lodhia2016fractional}
\begin{align*}
\varphi = \bigl(-\Delta)^{-\beta/4} W
\end{align*}
where $\Delta$ is the Laplace operator, $W$ a white noise on $\R^d$ and $\beta > 0$ the fractional exponent. It is fully characterized by the covariance structure of its Fourier modes given by 
\begin{align*}
\E[\widetilde\varphi(\bk)\widetilde\varphi(\bk')] \propto S(\bk)\delta(\bk-\bk'),
\end{align*}
with what is usually called power spectrum:
\begin{align}
S(\bk) \propto |\bk|^{-\beta}.\label{eq:power_spectrum}
\end{align}
Formally its Boltzmann distribution reads 
\begin{align}\label{def:FGF}
P(\varphi) = \frac{1}{Z} e^{-\frac{1}{2}\int d^2\bk \mathcal{E}(\bk)}
\end{align}
with the energy given by
\begin{align}
   \mathcal{E}(\bk) \egaldef k^\beta \vert\widetilde\varphi(\bk)\vert^2 \label{def:Energy}
\end{align}
With this energy, the Boltzmann distribution is invariant under rescaling 
Note that the standard Gaussian free field corresponds to $\beta=2$. The FGF is also commonly parametrized by the Hurst exponent $H=\frac{\beta-d}{2} \in[0,1]$
to quantify the roughness and self-similarity of the field. The real space covariance structure in this range of parameters is then given by 
\begin{align*}
  \E[\varphi(\x)\varphi(\x')] \propto \vert\x-\x'\vert^{2H}
\end{align*}
and is typically long-range, tunable by $H$.
According to this scaling, we see that the FGF displays self-similarity, in the sense that
\begin{align}
\varphi(s\bx) &\stackrel{d}{=} s^\frac{\beta-d}{2} \varphi(\bx) ,		\label{eq:scaling_FGF1} \\[0.2cm]
\widetilde\varphi(\bk/s)&\stackrel{d}{=} s^\frac{\beta+d}{2} \widetilde\varphi(\bk).\label{eq:eq8scaling_FGF2}
\end{align}
These equations can be seen as the renormalization equation for the FGF, which involves no other
exponent than the anomalous dimension of the field.

To be concrete, in the following we will discretize our FGF $\varphi: \Omega\subset \mathbb Z^2\to \mathbb{C}$ on an $L\times L$ lattice, which amounts to generate a  collection of  
random Fourier coefficients $\widetilde\varphi(\bk)$: 
\begin{align}\label{eq:FGF}
\varphi(\mathbf x) = \frac{1}{|\Omega|}\sum_{\mathbf k} e^{i2\pi \mathbf k\cdot \mathbf x} \widetilde\varphi(\mathbf k),
\end{align}
with $\bk = (k_x,k_y)$ parameterized by $L^2$ pairs of in integers as $\bk_{i,j} = (-\frac{1}{2}+\frac{i}{L},-\frac{1}{2}+\frac{j}{L})$
with $i,j=0,\ldots L-1$. 
Each coefficient's real and imaginary part is chosen to follow a normal distribution:
\begin{align}
   Re[\widetilde\varphi(\mathbf k)] \egallaw Im[\widetilde\varphi(\mathbf k)] \egallaw\mathcal N(0, |\mathbf k|^{-\beta}),
\end{align}
except for the choice $\widetilde\varphi(\mathbf 0)=0$, so that field average is zero by construction:
$\sum_\bx \varphi(\bx) = 0 $. 

Using this field, we now consider two different simple linear processes preserving the FGF distribution, hence displaying a form of equivariance.
The first one corresponds to a phase-mixing dynamics and the second one to a local spectral flow associated to  scale transform and rotations: 
\begin{align}
\frac{\partial \widetilde{\varphi}(\bk, t)}{\partial t} &= -i\nu(\bk) \widetilde{\varphi}(\bk, t)\qquad (\text{Phase Mixing}) \label{eq:phase_mixing}\\[0.2cm]
\frac{\partial \widetilde{\varphi}(\bk, t)}{\partial t} &= i\Bigl(\eta T_{\rm scale}(\beta)+\omega T_{\rm rot}\Bigr)\widetilde{\varphi}(\bk, t)
         \qquad (\text{Spectral flow}) \label{eq:eq12spectral_flow}
\end{align}
where $\nu(\bk)$ is an arbitrary phase function, while $T_{\rm scale}$ and $T_{\rm rot}$ are two (commuting) self-adjoint operators 
\begin{align}
    T_{\rm scale}(\beta) &\egaldef -i\Bigl(\bk\cdot\nabla_\bk+1+\frac{\beta}{2}\Bigr) \label{def:Tscale_eq13} \\[0.2cm]
    T_{\rm rot} &\egaldef i\bigl(k_x\partial_{k_y}-k_y\partial_{k_x}\bigr) \label{def:Trot}
\end{align}
corresponding respectively to scale and rotation transformations with speed controlled respectively by parameters $\eta,\omega\in\R$. These two processes represent the only local
linear and stationary transformations that we can think of that preserve the FGF distribution; we consider them separately because they do not commute and want
to see how to treat them specifically. The corresponding operators (in particular by definition of $T_{\rm scale}(\beta)$) are indeed self-adjoint with respect to the inner product
\begin{align*}
    \langle \varphi_1,\varphi_2\rangle_\beta  = \int d^2\bk k^\beta \varphi_1(\bk)\varphi_2^*(\bk)
\end{align*}    
induced by the norm defining the energy~(\ref{def:Energy}).
Using the fact that $e^{a \bk\cdot \nabla_\bk}\varphi(\bk)=\varphi(e^{a}\bk)$, the 
solutions to these processes are found in closed form,
\begin{align}
\widetilde{\varphi}(\bk, t) &= \widetilde{\varphi}(\bk, 0) e^{-i\nu(\bk) t}, \quad (\text{Phase Mixing}) \label{eq:eq15_phaseMixing}\\
\widetilde{\varphi}(\bk, t) &= \widetilde{\varphi}(\mathcal{R}_{\omega t}\bk s(t) , 0)s(t)^{\beta/2+1},~ (\text{Spectral flow})\label{eq:eq16_spectralflow}
\end{align}
with $\mathcal{R}_\alpha\bk = \bigl(k_x\cos(\alpha)+k_y\sin(\alpha),-k_x\sin(\alpha)+k_y\cos(\alpha)\bigr)$ 
and $s(t) = e^{\eta t}$. Note that this process  is similar in spirit  to the linear model of Kolmogorov cascade introduced recently~\cite{apolinario2023linear}.
This means the evolution is simply a change of phase at every $\bk$ with respect to the initial configuration $\widetilde\varphi(\bk, 0)$ in the first case, while in the second there
is a combined radial and orthoradial current $J(\bk,t) = k^{\beta}\vert\phi(\bk,t)\vert^2(\eta\bk-\omega\bk^{\perp})$ of energy flow spiraling toward large [resp. small]
scales when $\eta>0$ [resp. $\eta<0$]. The inference task then consists in regressing $\psi(\bk)=\varphi(\bk,1)$ from $\varphi(\bk,0)$. 
The phase mixing  illustrates the case of a purely local transformation (in $\bk$ space) while for the spectral flow we have an example of a non-local transformation (at least in $\bk$ space).
Both conserve the energy~(\ref{def:Energy}).  
In addition, in both cases the scaling properties~(\ref{eq:scaling_FGF1}), (\ref{eq:eq8scaling_FGF2})  are preserved by the dynamics. 
In the case of (\ref{eq:eq15_phaseMixing}) because a phase shift does not affect the power spectrum (which in turn controls the scaling property).
In the case of (\ref{eq:eq16_spectralflow}), trivially for the rotation part, and for pure scaling we can check consistency: setting $\mathcal{R}_\alpha=I$ in (\ref{eq:eq16_spectralflow})
then injecting  (\ref{eq:eq8scaling_FGF2}), one obtains $\widetilde{\varphi}(\bk,t)=\widetilde \varphi(s\bk,0)s^{\beta/2+1} \egallaw \varphi(\bk,0)$, meaning the spectral flow preserves the distribution.
As previously noted, the energy is conserved, except possibly at the boundaries of the momentum space, which means that some energy has to be injected into the system at small scales
to compensate for the energy leaving the system at large scales for instance when $\eta>0$. 
	
\begin{figure}[t]
\includegraphics[width=\columnwidth]{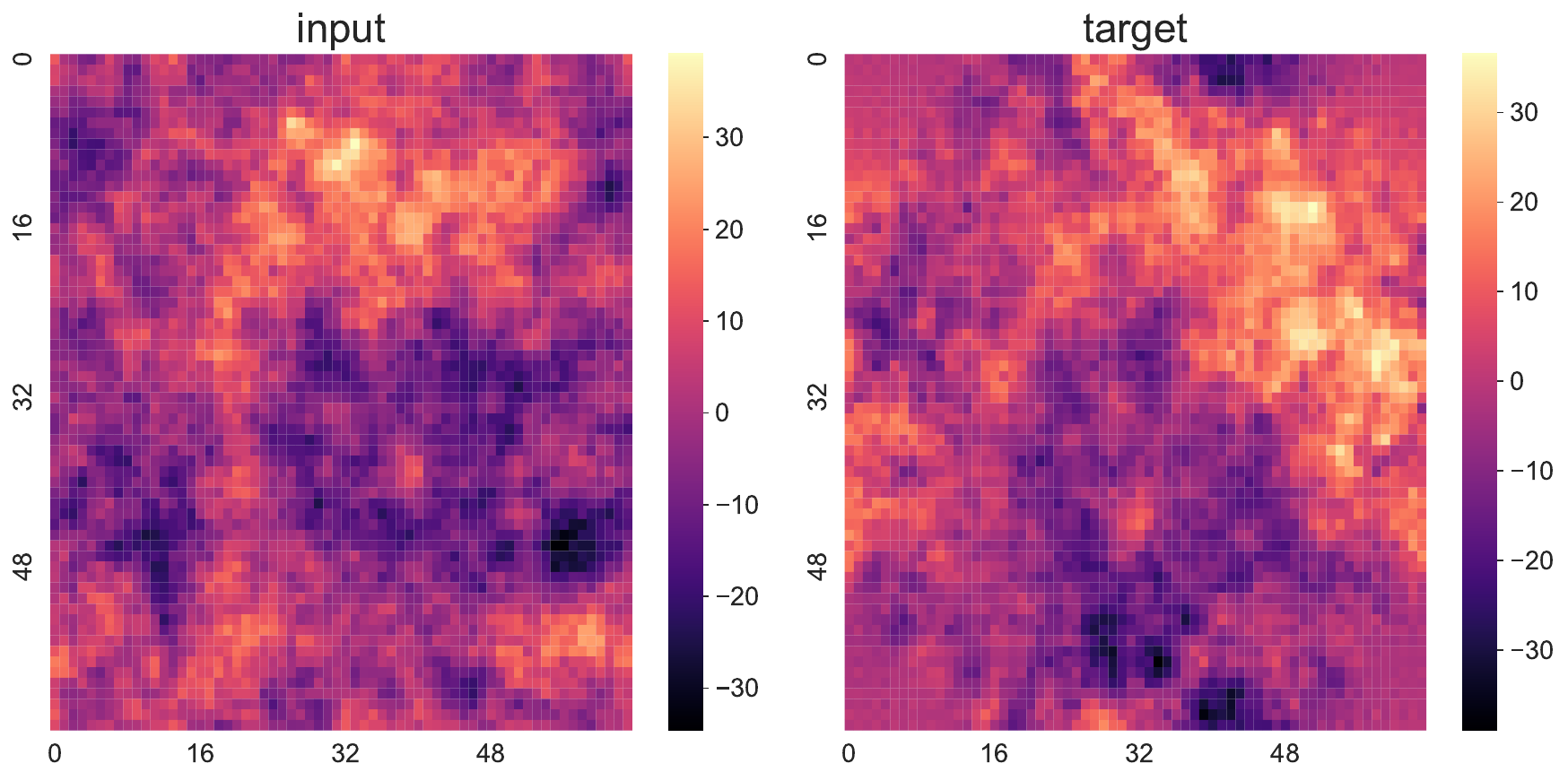}
\caption{\label{fig:rescaling_task} Sample input and output with lattice size $L^2 = 64\times 64$ for the FGF spectral flow inference  task. The transformation applied corresponds
  to a rescaling of factor $s=0.9$ and a rotation of $\pi/4$.}
\end{figure}

In Fig.~\ref{fig:rescaling_task} we show a pair of input and output samples for the FGF spectral flow process. Note that the corners of the image are distorted due to the rotation
which will induce some limitation in the performance of models corresponding to idealistic conditions.
    
Our goal is to evaluate the model’s ability to extrapolate across scales, ensuring that the test set includes scales not present in the training data. 
To achieve this, we set the Fourier coefficients within a band $k_1< k < k_2$  to zero for all samples in the training set. At test time, the model is evaluated on samples containing
the full range of frequencies. This thereby includes both  super-resolution experiment and large scale extrapolation when hiding respectively high and low frequencies.
In order to measure the level of extrapolation we define an extrapolation factor for both the super-resolution and large scale extrapolation as follows.
If $[1/k_{\rm max},1/k_{\rm min}]$ is the range of considered scales in the test set and if $1/k_{\rm occ}$ designate the scale above [resp. below] which the spectrum is occulted
in the train set of the large scale extrapolation [resp. super-resolution] experiment 
we define, for a scaling factor $s<1$ of the transformation the extrapolation factor as
\begin{align}\label{def:f_extrapol}
    f_{\rm extrapol} \egaldef
    \begin{cases}
    \DD \frac{s k_{\rm occ}}{k_{\rm min}} \qquad\text{(large scale extrapolation)}\\[0.3cm]
    \DD \frac{s k_{\rm max}}{k_{\rm occ}} \qquad\text{(super-resolution)}
    \end{cases}
\end{align}
which corresponds to the ratio  between the largest scale $1/k_{\rm min}$ that can be predicted in the test and largest one $1/sk_{\rm occ}$ 
in the train [resp. the smallest in the train $1/k_{\rm occ}$ and the smallest one $1/k_{\rm occ}$ in the train] in the large scale extrapolation [resp. super-resolution] experiment.

\subsection{Abelian Sandpile Model}
The Abelian sandpile model (ASM) is a paradigmatic example of \textit{self-organized criticality}, a concept introduced by Bak, Tang, and Wisenfeld in 1987 \cite{bak1987self}
to group together physical systems that, without needing to fine-tune any control parameter, exhibit critical-point like behavior. 
In practice, this often relates to the dissipation parameter being vanishing. These systems spontaneously evolve from any configuration towards a set of
states that display scale invariance and power law distributions of physical quantities \cite{turcotte1999self, watkins201625}.
There exists a certain number of variations of the ASM, here we focus on one of them \cite{dhar1990self}.
	
The ASM is defined on a square lattice with $N$ sites. To each site $i$ we associate a variable $z_i$ representing the height of the sandpile at each site,
and a threshold value ${z_c}_i$. The dynamics consists in two steps:
\begin{enumerate}
\item Select a site at random with uniform probability, and add a grain of sand to that site
\begin{align}
z_i \mapsto z_i + 1
\end{align}
\item If at any site $z_i > {z_c}_i$, that site becomes unstable. \textit{Topple} any unstable sites
\begin{align}
z_j \mapsto z_j + \Delta_{ij} \quad\textrm{for all } j = 1,\dots, N\label{eq:toppling}
\end{align}
with $\Delta_{ii} = -4$ in the diagonal, $\Delta_{ij} = 1$ if $i$ and $j$ are nearest neighbours, and $\Delta_{ij} = 0$ everywhere else.

Repeat until all sites are stable. The sequence of topplings that occur until stability is reached is called an \textit{avalanche}.
\end{enumerate}
In practice we will take the critical height to be ${z_c}_i=4$.
Given this value of  
the Laplacian, during the avalanche the energy is conserved in the bulk of the lattice and is dissipated at the boundary. This dissipation ensures that avalanches remain finite.

After a possible transient regime, the dynamics reaches a set of typical configurations (recurrent under deterministic drive or uniform random drive) and the ASM displays
critical behavior, in the form of power law distributions. 
There are mainly  two independent critical exponents that control these power laws. The first one corresponds to the \textit{distribution of avalanche sizes} $P(s)$,
where $s$ is defined as the number of sites that topple during an avalanche (its area). The power law is written as \cite{pruessner2012self}
\begin{align}
	P(s)\sim s^{-\tau}.
\end{align}
The second one, often called dynamical exponent, is related to the \textit{duration of the avalanches}, defined as the number of consecutive updates that need to be done to reach stability.
It has the scaling law:
\begin{align}
	t\sim r^z,
\end{align}
where $r$ is the linear size (radius) of the avalanche. There are  other critical exponents, such as those corresponding to the avalanche duration distribution
and the total number of topplings distribution, 
for which there are scaling relations in some limit cases~\cite{christensen1993sandpile, priezzhev1996formation, dhar1999abelian}.
	
An approximate spatial renormalization group scheme for the ASM was proposed in the 90's by Vespignani et.~al.~\cite{vespignani1995renormalization},
in which the stochastic process that defines the dynamics is renormalized, rather than the field itself. This scheme yields values of the critical exponents that are in overall agreement 
with numerical simulations.
	
In rough terms, the idea of this renormalization scheme is to define coarse-grained variables. Blocks at a given scale are said to be stable if, 
when energy is transferred to it, no energy is transferred to its neighbors (topplings inside the block at smaller scales may still occur). Similarly,
a block is said to be critical if it will topple when energy is transferred to it, 
and will transfer energy to its neighbours at the same scale. The dynamics can be characterized by the density $\rho$ of critical cells,
and the probabilities $\mathbf{p} = (p_1, p_2, p_3, p_4)$, where $p_i$ is the probability that energy will be transferred to $i$ neighbours when a cell topples.
Imposing certain rules on how to go from one scale to the next one, RG equations can be obtained for $(\rho, \mathbf{p})$.
However, the approximations used in this procedure are not fully controlled \cite{dhar1999abelian}
and some discrepancies were found between the predicted scaling behaviour and what is observed in simulations.

Predicting the next avalanche in these systems is a very challenging task, although not impossible, even in the SOC setting where temporal and spatial correlations are very weak. For instance, it is shown on real experimental sandpile avalanche data in~\cite{ramos2009avalanche} that large scale events have
both temporal and structural precursor signatures. And for the abelian sandpile model, the temporal precursor signature has actually been shown to be effective in predicting large scale events~\cite{PhysRevE.80.026124}. Our study investigates to which extent ML techniques can identify structural precursor patterns. 
On one hand the prediction task is simplified by giving the position of the seed, but on the other hand the nature of the task itself is more complex as it consists in regressing the whole shape of the avalanche, not only its size.

\begin{figure}[t]
\includegraphics[width=\columnwidth]{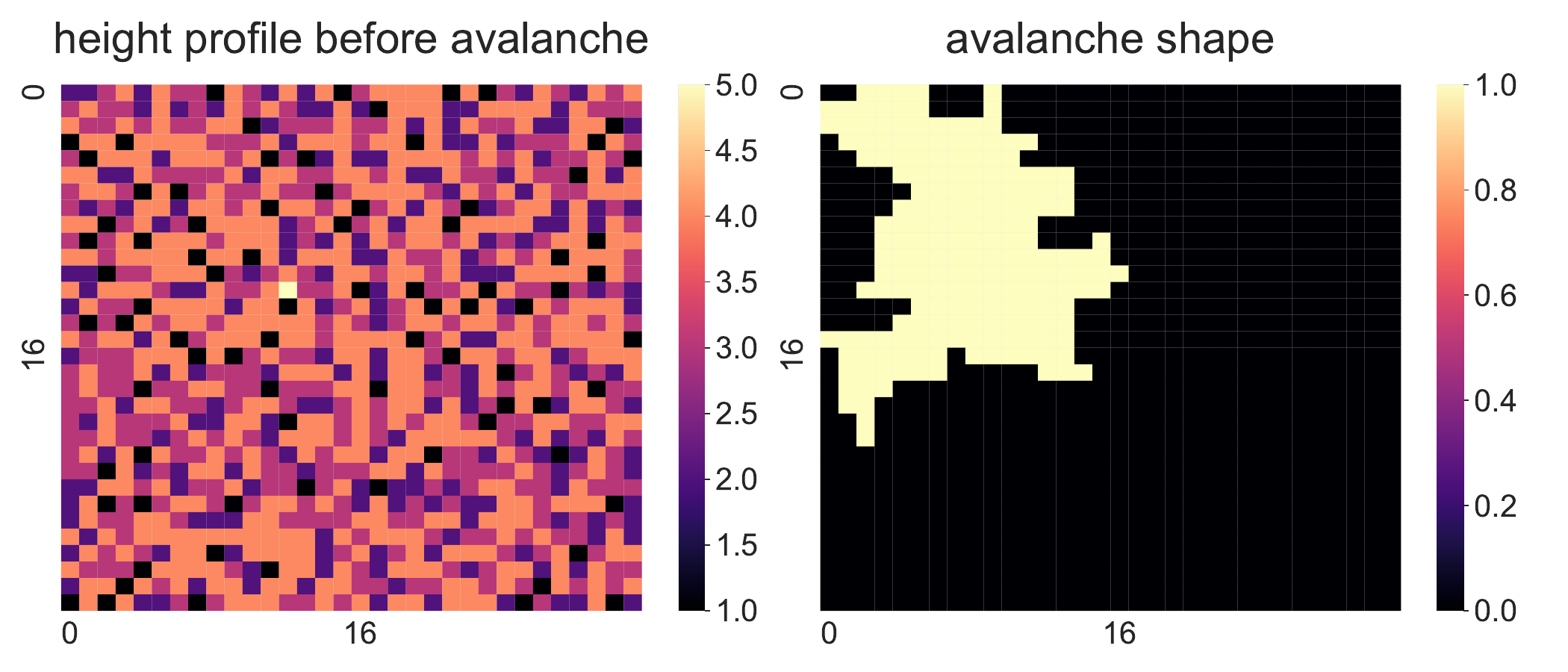}
\caption{\label{fig:avalanche_task} Sample input and label with lattice size $32\times 32$ for the regression task in the ASM. In the input, the unstable site corresponding
to the epicenter of the avalanche has a height of $5$. The output is a binary map, the ones marking the sites that toppled during the avalanche.}
\end{figure}

More precisely, the task we want to solve is the following: starting from the unstable configuration right before an avalanche occurs, we want to predict which sites will topple (the shape of the avalanche). In Fig.~\ref{fig:avalanche_task} we show a sample input-label pair. The input field $\varphi({\mathbf x})$, shown on the left, is an $L\times L$ lattice of discrete variables corresponding to the height of the sandpile at each site $\mathbf x$. There is one single unstable site at height $\varphi(\mathbf{x_0}) = 5$, marking the starting point of the avalanche. The target field $\psi(\bx)$ is a binary map, where the sites with value $\psi(\bx)=1$ are the ones that toppled during the avalanche that results from the initial configuration $\varphi(\mathbf x)$. Our model's output $\widehat \psi(\mathbf x)$ can be interpreted as the probability of site $\bx$ to topple during the avalanche. We can threshold this map to make it binary and obtain a predicted avalanche shape.
    
Although the process given by eq.~\eqref{eq:toppling} to go from the input to the output is deterministic and relatively simple, the rapidly decaying correlations in the input map makes it impossible to guess from the naked eye, and difficult to learn for a neural network. 
	
Extrapolation to large scale avalanches is of particular interest, as it provides a realistic baseline for applications to real systems in which large events are rare. 
An example is earthquake prediction, for which large-scale events are scarce \cite{gutenberg1956earthquake}. 
In order to test extrapolation in the ASM, 
we train on samples with a maximum avalanche size $S$, leaving out the largest avalanches (or vice versa for the smallest). We then test on events of all sizes.
Note that while in the case of the FGF the input and target fields share the same self-similarity properties (same exponent $\beta$ for input and target fields),
here instead the scaling properties differ between input and output, which are different types of fields, with different critical exponents.
This greatly increases the difficulty of the extrapolation problem.
 

\section{\label{sec:methods}Machine Learning Models}
Here we describe the neural network architectures we designed to solve the proposed tasks, along with known models that serve as baselines. 
Considering our problem setting~(\ref{def:functional_regression}), models can be described in the framework of neural operators~(NO)~\cite{anandkumar2020neural}:
\begin{align}\label{def:NO}
   \widehat \psi[\varphi](\x)  = \sigma\Bigl(\sum_{\x' \in \Omega} F_\theta(\x,\x') \varphi(\x')\Bigr)
\end{align}
where $\sigma$ is an activation function and $F_\theta$ is a linear operator, to be learned.
In our case there is no bias term, or it can be absorbed in $F$.
As we shall see, except from the U-Net, all models considered in the following can be cast in this framework or its iterated version (\ie many 
layers of NO~(\ref{def:NO})).
In order to realize extrapolation across scales, we explore different ways of inserting scale invariance in this framework. The Fourier embedding network is a model designed specifically for the FGF problem, that operates in Fourier space, and shares weights across all frequencies. Then we propose the Fourier-Mellin neural operator (FMNO), as a way to obtain scale invariance via convolutions in scales. The Riesz network \cite{barisin2024riesz} makes use of scale-equivariant operations too. Finally, we propose a GNN architecture that operates on a wavelet representation of the data, somewhat inspired by the idea of \cite{marchand2023multiscale}. Additionally, we use a U-net without built-in scale invariance as a baseline for all tasks.
	
\subsection{U-net}
	
The U-net \cite{ronneberger2015u} is a widely used architecture designed originally for segmentation tasks. It follows an encoder–decoder structure: the encoder operates by performing consecutive convolutions and coarse-grainings on the input, then the decoder upsamples the representation obtained to the original input size. The information at each encoding layer is recorded and then used by the decoder at the corresponding level, thus preserving the fine-grain information and mixing it with higher-level features. For a more detailed explanation, see Appendix~\ref{app:unet}.
	
The hierarchical scale structure of this network makes it a reasonable choice to treat the problems we are interested in. However, as the weights learned at different scales are independent from each other, it does not have any type of scale-invariance, so it is not expected to extrapolate to scales unseen during training. In this work we use it as a baseline to compare with the architectures that do have some type of implementation of scale invariance.

\subsection{Riesz network}
	
One way of constructing a scale-invariant neural network is to impose the constraint that all operations within it must commute with rescalings. This approach is used in \cite{barisin2024riesz} to build the Riesz network, an architecture designed for segmenting pictures of multiscale cracks in concrete. This architecture is based on the Riesz transform, an operator that commutes with continuous scaling transformations.
	
Given a continuous map $\varphi:\mathbb R^2 \to \mathbb R$, the Riesz transform $\mathcal R_j:L(\mathbb R^2)\to L(\mathbb R^2)$ is defined as
\begin{align}
	\mathcal R_j[\varphi](\mathbf x) = C\lim_{\varepsilon\to 0}\int_{\mathbb R^2\setminus \mathcal B_\varepsilon}d\mathbf x' \frac{x'_j\varphi(\mathbf x-\mathbf x')}{|\mathbf x'|^3},
\end{align}
where $j=1,2$ indicates the 2 spatial directions and $ \mathcal B_\varepsilon$ is the ball of radius $\varepsilon$ centered at the origin. In Fourier space the Riesz transform takes the simpler form
\begin{align}
	\tilde{\mathcal R_j}[\varphi](\mathbf k) = -i\frac{k_j}{|\mathbf k|}\widetilde\varphi(\mathbf k).
\end{align}
This operator commutes with naive rescaling transformations, \ie 
\begin{align}
	\mathcal{R}_i\left[\varphi\left( \frac{\cdot}{s} \right)\right] = [\mathcal{R}_i \varphi]\left(\frac{\cdot}{s}\right).
\end{align}
For proof of this property, refer to \cite{barisin2024riesz}. One can define higher order Riesz transforms by successively applying $\mathcal R_j$. In particular we have three second-order Riesz transforms given by
\begin{align}
	\mathcal R^{(m, n)}[\varphi](\mathbf x) = \mathcal R_1^m\circ\mathcal R_2^n[\varphi](\mathbf x),\quad m+n=2;\, m, n\in\mathbb N.
\end{align}

The basic component of the Riesz network is the Riesz layer. We consider for simplicity linear combinations of the two first-order and the three second-order Riesz transforms
\begin{align}
	J[\varphi] = C_0 + \sum_{k = 1}^{2} C_k \mathcal R_k[\varphi] + \sum_{m + n = 2} C_{mn} \mathcal R^{(m, n)}[\varphi].
\end{align}
The coefficients $(C_0, C_k, C_{mn})$ of the linear combination will be the learnable parameters of the network. Higher order Riesz transforms could in principle be included,
but they increase computational cost and are not needed for the scope of this work.
	
The Riesz layer number $\ell$ will take an input with a number $C(\ell)$ of channels $\varphi^{(\ell)} = (\varphi_1^{(\ell)}, \dots, \varphi_{C(\ell)}^{(\ell)})$,
and output a number of $C(\ell+1)$ of channels via
\begin{align}
	\varphi_c^{(\ell+1)} = \sum_{c'=1}^{C(\ell)} J_\ell^{(c, c')}[\varphi_{c'}^{(\ell)}],\label{eq:singleLayerRiesz}
\end{align}
meaning that a given channel $c$ at layer $\ell+1$ is connected to the channel $c'$ at layer $\ell$ via the application and subsequent linear combination of the five Riesz
filters considered, and the result is summed over all input channels. $J_\ell^{(c, c')}$ is defined as
\begin{align}
	J_\ell^{(c,c')}[\varphi] = C_0 + \sum_{k = 1}^{2} C_k^{(\ell, c, c')} \mathcal R_k[\varphi] \nonumber\\ + \sum_{m + n = 2} C_{mn}^{(\ell, c, c')} \mathcal R^{(m, n)}[\varphi],
\end{align}
allowing for different learnable parameters for each connexion between an input channel $c'$ and an output channel $c$, and for each layer $\ell$.
Layer $\ell$ thus has $6C(\ell)C(\ell+1)$ real parameters to be learned. 
	
The Riesz network is then built by successive applications of Riesz layers alternated with non-linearities. Note that the non-linearities should also preserve
the scale-equivariance property -- this is the case for rectified linear units activations (ReLU) ~\cite{barisin2024riesz}, which we will use in this work.
Any function that is piecewise linear in $]-\infty,0[$ and in $]0,\infty[$ fulfills this criterion, which means in practice that we can use only ReLU and leaky ReLU.
In practice, the Riesz layer is implemented in Fourier space, so the network is very similar to a Fourier neural operator~\cite{li2020fourier} in the sense that, for instance,
for a 2-layer network we have
\begin{align}
       \widehat \psi[\varphi](\x)  = \sigma\Bigl(\mathcal{F}^{-1} \tilde{J}_2  \mathcal{F}
       \sigma\Bigl(\mathcal{F}^{-1} \tilde{J}_1  \mathcal{F}(\varphi(\bx))\Bigr)\Bigr),
\end{align}
where $\varphi^{(\ell+1)} = \tilde{J}_\ell  \varphi^{(\ell)}$ is the operation done in \eqref{eq:singleLayerRiesz}, and $\mathcal{F}, \mathcal{F}^{-1}$ are the Fourier transform and
its inverse, applied channel-wise. The number of layers and channels is chosen depending on the task, see appendix \ref{app:experiment_details} for details.

\subsection{Fourier embedding network}

In the FGF phase mixing task, the main difficulty lies in the spectral bias. The high frequency components make it hard for standard architectures to perform the regression. However,
the time evolution we want to learn is local in Fourier space, \ie each Fourier coefficient $\varphi(\bk)$ is multiplied by a factor depending only on its own mode $\mathbf k$,
not on the other $\varphi(\bk'), \bk'\neq \bk$. Taking this into consideration we propose a model that operates in Fourier space, with the form 
\begin{align}
	F_\theta[\widetilde \varphi] = g_\theta \odot \widetilde\varphi
	\label{eq:learned_fn}
\end{align}
where $g_\theta(\mathbf k)$, a function of the frequencies of the lattice, is learned and multiplied element wise with $\widetilde\varphi$. 
The ground truth imposed by the dynamics corresponds to
\begin{align}
	g(\mathbf k) = e^{-i\nu(\bf k)},
\end{align}
with the concrete choice  
\begin{align}\label{eq:nuk}
   \nu(\bk) = \nu \vert\bk\vert^2    
\end{align}
made for the experiments, $\nu$ being a real parameter.
In practice we use an MLP for $g_\theta(\bk)$, but rather than applying it directly to the frequency modes $|\bk|$, inspired by~\cite{ma2025robustifying},
we first embed these modes themselves in Fourier. The embedding is $f(\mathbf k) = (f_1(\mathbf k), \dots, f_P(\mathbf k))$ with
\begin{align}
	f_p(\mathbf k) = e^{i2\pi |\mathbf k| p / P},
\end{align}
where we often set $P=100$.
This embedding is designed to facilitate the training of the MLP. It depends only on the modulus of $\mathbf k$, not its direction, so as to impose rotational symmetry. 
The network then reads:
\begin{align}
	F_\theta(\widetilde \varphi(\bk)) = \mathrm{MLP}_\theta(f(|\mathbf k|))\odot \widetilde{\varphi}(\bk),
\end{align}
where the MLP's output is a single scalar (complex) value.
Note that the same MLP is applied at every value of $\bk$, meaning that the weights are shared across all modes.
This network cannot perform well on the rescaling nor rotation tasks because it is local in $\bk$ space by construction
and specialized to the phase shifts task. Hence we do not perform experiments in these directions.

\subsection{Fourier-Mellin network}
In addition to the issue of spectral bias, the spectral flow of the FGF presents an additional problem,
since the process is now non-local in Fourier space. As explained in more detail in section \ref{sec:theory},
the appropriate basis for this task is the Fourier-Mellin (FM) basis, in which the kernel that describes the rescaling and rotating processes is diagonal.

We propose a model that operates in the Fourier-Mellin basis:
\begin{align}
    F_\theta\bigl[\varphi_{(\lambda, \mu)}^\star\bigr] =  \theta_{(\lambda,\mu)}\odot \varphi_{(\lambda,\mu)}^\star,
\end{align}
where $\varphi_{(\lambda, \mu)}^\star$ denotes the Fourier-Mellin transform of the field $\varphi^\star=\mathcal{F}[\widetilde\varphi^\star]$, where 
\[
  \widetilde\varphi^\star (u,\alpha) \egaldef \widetilde\varphi\Bigl[\bk = \bigl(e^u\cos(\alpha),e^u\sin(\alpha)\bigr)\Bigr] 
\]
is the Fourier component as function of $u=\log\vert\bk\vert$ and $\alpha=\text{angle}(\bk/|\bk|)$.
Then $\lambda$ and $\mu$  denote respectively the Fourier-Mellin radial and angular momentum that are conjugate to $u$ and $\alpha$. 
Each Fourier-Mellin mode being multiplied by a weight, this is a linear model in FM space.
The number of independent parameters is thus the number of components used when projecting on the FM basis.

Thus, the challenge is not on the ML side, but lies in the implementation of the Fourier-Mellin transform for discrete data. As in \cite{derrode2001robust},
we do it by first performing a change of basis from cartesian $(\bk)$ to logpolar coordinates $(u,\alpha)=(\log |\mathbf k|, \text{angle}(\bk/|\bk|))$,
followed by the application of a Fourier transform. 
Our input field $\varphi(\bx)$ is homogeneous and the origin does not play any special role, apart from being the center of the spectral flow task we consider.
On the contrary, the origin of $\widetilde\varphi(\bk)$ is not arbitrary, as $\bk\sim 0$ corresponds to the low frequencies. The logpolar setting is thus well motivated.  
This coordinate change implies interpolating between the input's cartesian grid and a logpolar grid. As this interpolation is not invertible, some numerical error is
introduced when inverting the transformation.

To minimize the reconstruction error of the coordinate change, we introduce the Dirichlet kernel, which gives the interpolation between the two grids:    
\begin{align}
       \widetilde\varphi_{\mathbf k^*} = \sum_{\mathbf k\in \Omega}D(\mathbf k^* - \mathbf k)\widetilde\varphi_{\mathbf k},
\end{align}
with $\{\mathbf k^*\in \Omega^*\}$ denoting the points of the logpolar grid, $\{\mathbf k\in \Omega\}$ the points of the cartesian grid, and
\begin{align}
  D(\mathbf q) = D_\mathrm{1d}(q_x)D_\mathrm{1d}(q_y),
\end{align}
with the 1d Dirichlet kernel $D_\mathrm{1d}$ defined as
\begin{align}
   D_\mathrm{1d}(q) = \begin{cases}
   \frac{1}{L} \frac{\sin (\pi L q)}{\sin(\pi q)} & q \ne 0 \\
    1  & q = 0
  \end{cases}
\end{align}
with $L$ the linear size of the Cartesian grid. Note that the field being interpolated must be periodic, which is not the case for the radial component in logpolar coordinates,
due to the power spectrum. Therefore, we multiply by the inverse spectrum in order to have a statistically periodic gaussian field, and take the Fourier-Mellin
transform of the field with constant variance. We restore the power spectrum in the decoder step, after going back to cartesian coordinates. The details of this
method are explained further in the Appendix \ref{app:model_details}. 

The Fourier-Mellin transform is then numerically implemented by computing
\begin{align}
     \varphi_{(\lambda,\mu)}^\star = \frac{1}{M}\sum_{\bk,\bk^\star\in \Omega,\Omega^\star} 
     D(\bk^\star - \bk)|\bk|^{\beta/2}\varphi_{\bk}e^{i(2\pi \log |\bk^\star|\lambda + \alpha \mu)},\label{eq:FM_coefs}
\end{align}
where $M$ is the number of Fourier-Mellin modes chosen and $\alpha=\text{angle}(\bk^\star)$. 
In particular, we can choose the resolution of the logpolar grid, which can be different from the original grid.
The inverse transformation is numerically done by means of the pseudo-inverse $D^\dagger$ of the Dirichlet kernel, using its SVD, so we have
\begin{align}
    \widetilde\varphi_{\bk} 
     =\frac{1}{|\bk|^{\beta/2}}  &\sum_{\bk^\star\in \Omega^\star} D^\dagger(\bk^\star - \bk) \notag\\[0.2cm]
     &\times\Bigl(\frac{1}{M}\sum_{(\lambda,\mu)}\varphi_{(\lambda,\mu)}^\star e^{-i(2\pi \log |\bk^\star|\lambda + \alpha \mu)}\Bigr).
\end{align}
The pseudo-inversion of the kernel needs to be done with high precision to avoid encoding-decoding errors. 
We find empirically that  to have a good reconstruction of the field, the logpolar grid needs to have around 3 times more points in each direction than the cartesian grid
(resulting in about 9 times more points).

\subsection{Wavelet-GNN}
We propose an original architecture consisting of a Graph Neural Network (GNN) operating on a graph-structured representation of the input in wavelet space. This takes inspiration from the Wavelet Conditional Renormalization Group approach developed in \cite{marchand2023multiscale}, where the wavelet representation is used to build a hierarchical generative model that is applied to physical systems, such as the $\varphi^4$ model and gravitational lensing data.
	
The wavelet transform is a decomposition of a signal into a basis of functions called \textit{wavelets}. There are different families of wavelet bases. 
In this work, we considered the use of the Daubechies wavelets,
which are functions that are both localized in direct and Fourier space~\cite{mallat1999wavelet}, making them well-suited to represent data at different scales.
In practice though, on our finite and relatively small grid, to avoid excessive boundary effects, we resort to ``db1'', which is equivalent to the Haar wavelet,
which is localized in space, but not in the frequency domain.
We use the library made available in~\cite{selesnick2005dual}, which allows for backpropagation through the wavelet transform in PyTorch.
	
A $2$-dimensional discrete wavelet transform applied to an input $\varphi_0= \varphi$ of size $L\times L$ gives a set of wavelet coefficients $\bar \varphi_1$ of size $L/2\times L/2$, with 3 channels each roughly representing the horizontal, vertical, and diagonal\footnote{What we call diagonal is actually what comes from combining horizontal and vertical basis elements.} details, \ie the fast degrees of freedom. The transform also yields a coarse-grained map $\varphi_1$ of the input, of size $L/2 \times L/2$, that contains the slow degrees of freedom. This totals to $(L/2)^2(3+1) = L^2$ coefficients.
	
\begin{figure}[ht]
    \includegraphics[width=\columnwidth]{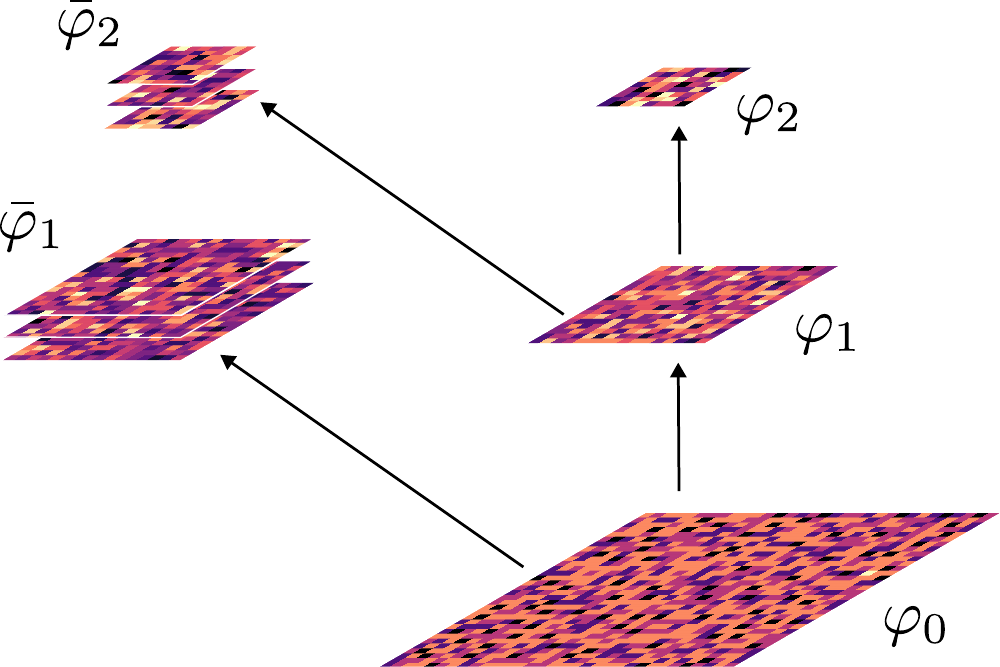}
\caption{\label{fig:wavelet_decomposition} Wavelet decomposition of the input. $\varphi_0$ can be perfectly recovered from $\varphi_2, \bar\varphi_2, \bar\varphi_1$. 
At each level, the number of ``pixels'' is divided by $4$.}
\end{figure}

The wavelet transform is invertible, so the original map $\varphi_0$ can be perfectly recovered from $(\bar\varphi_1, \varphi_1)$. 
Applying the wavelet transform again to $\varphi_1$, we can again separate the fast degrees of freedom at the next scale, obtaining $(\bar\varphi_2, \varphi_2)$.
Iterating this procedure, we build an explicit scale representation of the input, as illustrated in Fig.~\ref{fig:wavelet_decomposition}. 
Strictly speaking, reconstruction only needs the set of coefficients $(\varphi_{\log_2(L)}, \bar\varphi_{\log_2(L)},\bar\varphi_{\log_2(L)-1}, \ldots,\bar\varphi_{1})$,
where we assume that $L$ is a power of $2$. In practice, it is convenient to also use the coarse-grained maps
$(\varphi_{\log_2(L)},\varphi_{\log_2(L)-1}, \ldots,\varphi_{1})$ as input features.
	
For our purposes, given an input sample of size $L\times L $, with $L = 2^J$ a power of 2, we arrange the wavelet coefficients and coarse-grained maps into a graph
with a tree structure. There are  $2^{J-1}\times 2^{J-1}$ leave nodes in the tree (level $\ell=1$), and to each such node $i$ is assigned a 4-component feature vector,
$(\bar\varphi_1(i), \varphi_1(i))$. Subsequent levels of the tree are built in the same way, each level $\ell$ containing $2^{J-\ell}\times 2^{J-\ell}$
nodes and corresponding to a certain scale of the wavelet representation, with the root node (level $\ell=J$) containing the 4 features $(\bar\varphi_{J}, \varphi_{J})$. 
  
\begin{figure}[h]
\includegraphics[width=.8\columnwidth]{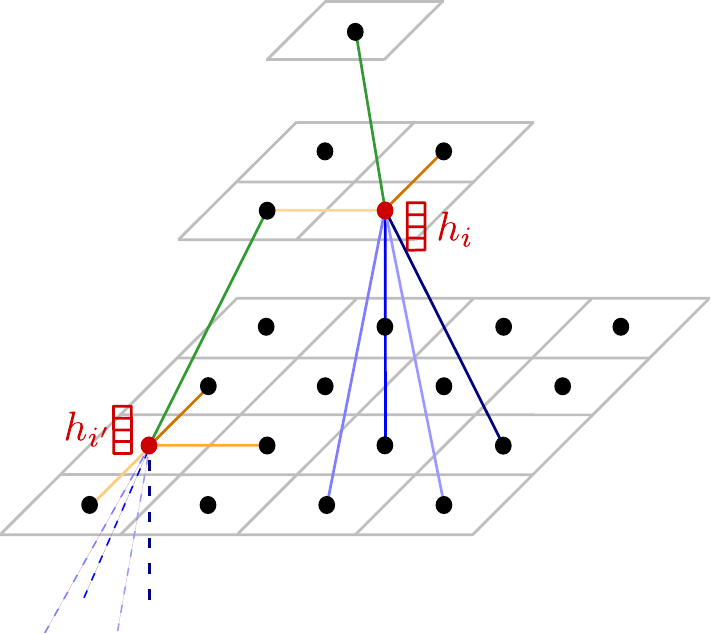}
\caption{\label{fig:wavelet_graph} Graph representation of a  $4\times 4$ input map, where we only show the edges outgoing from two nodes (highlighted in red).
       Each level in the tree corresponds to a scale of the wavelet decomposition (3 levels here). The 9 edge types are colored differently,
       in green for the child to parent relation, in shades of blue for the parent-children relations, and shades of orange/yellow for the neighbors relations.
       The node features $h_i$ contain the three wavelet coefficients and the value of the coarse-grained field at each position at the given scale.
       In dashed lines we show the edges that would continue if the tree continued downward.}
\end{figure}

Directed edges are drawn (see Fig.~\ref{fig:wavelet_graph} for a sk
etch) between nearest neighbors over the current $2^\ell\times 2^\ell$ lattice. Directed edges are also drawn between levels,
in such a way that nodes connected together correspond to the same spatial region. In the bulk of the tree, a node at level $\ell$ will have four children nodes at level $\ell - 1$,
and one parent node at level $\ell + 1$. We assign a tag to each (directed) edge, indicating the type of connection it represents.
There are 9 different tags: one for the parent node, 4 possible nearest neighbours and 4 possible children.
Having these 9 possible independent relations enables the model to distinguish spatial orientation and scale direction. 

This representation of the data allows to use GNN architectures in which the basic operations are given by a relational graph convolutional operator.
We pick one of the simplest instantiations of such a convolution~\cite{schlichtkrull2018modeling}:
\begin{align}
\mathbf h_i' = \sum_{r\in\mathcal R}\frac{1}{|\mathcal N_r(i)|}\sum_{j \in \mathcal N_r(i)} \Theta_r\cdot \mathbf h_j,\label{eq:GNN_conv}
\end{align}
where $i$ is now the node index (not directly related to spatial location), $\mathbf h_i$ is the feature vector at node $i$, $r$ is the tag index,
and $\mathcal N_r(i)$ is the set of $r$-type neighbors of $i$. 
The learnable weight matrices $\Theta_r$ are distinct for each tag type but shared across all similar edges. 
Note that our approach using this very simple layer can be readily extended to more expressive architectures, \eg using an MLP to build the message from node $j$
to node $i$, as long as the message somehow takes into account the nature of the edge.

One convolutional GNN layer consists in the nodes update defined by~(\ref{eq:GNN_conv}), followed by a non-linearity applied node-wise and feature-wise.
This design allows for a simple empirical implementation of scale invariance: by sharing the weights $\Theta_r$ across the levels of the tree, the information transfer
is done identically at each scale. 
In practice, we allow for different weights at the top $\Theta_r^{J-1}$ and at the bottom level $\Theta_r^1$, to account for the boundary effects related to these 2 special scales.
    
To obtain an output map with the same size as the input, we choose to have 4 channels in the nodes of the last representation (after the last convolution layer),
so as to be able to interpret them as wavelet coefficients and coarse-grained features. We then perform an inverse wavelet transform, using the implementation in ~\cite{selesnick2005dual}.
When performing the inverse transform, we discard the coarse-grained map $\varphi$ of all nodes except from the root, using the coefficients $\bar\varphi$ of all nodes, and inverse
iteratively from root to leaves.
We can count the parameters in a layer: the bulk nodes of the tree share $9$ matrices $\Theta_r$ of shape $c_{out}\times c_{in}$, where $c_{in},c_{out}$ are
the number of channels in input and output, respectively. The weight matrices at the top and bottom of the tree are independent from the bulk one: $\Theta_r^{L-1}$ has only 4 tags,
children, and $\Theta_r^{1}$ has 5 tags (neighbors, parent). The total is thus $18 c_{out} c_{in}$ parameters.

\subsection{Loss function}
For the FGF, we perform a pixel-wise regression of a complex-valued map and thus use by default the mean-squared error (MSE) loss function:
\begin{align}\label{def:MSE}
\mathcal{L}(\psi,\widehat \psi) \egaldef \frac{1}{N L^2}\sum_{s=1}^N \Vert \widehat\psi^{(s)}-\psi^{(s)}\Vert^2 
\end{align}
which can be indifferently written in the direct pixel or in its Fourier representation 
\begin{align}
\Vert \widehat\psi^{(s)}-\psi^{(s)}\Vert^2 &= \sum_{\x\in\Omega} \vert \widehat\psi_\x^{(s)}-\psi_\x^{(s)}\vert^2  \\[0.2cm]
&= \sum_{\bk\in\Omega} \vert \widehat{\widetilde\psi}_\bk^{(s)}-\widetilde\psi_\bk^{(s)}\vert^2.
\end{align}
In practice, for the phase mixing task we operate in Fourier space to save a gradient retro-propagation through the fast Fourier transform (FFT). 

In the phase mixing task, spectral bias is strong, but our Fourier embedding network is able to tackle it.
For the spectral flow task also, the gradient descent over this loss suffers from a strong spectral bias induced by the power law behavior of the spectrum. 
For the Fourier-Mellin network, in order to avoid that, we can redefine the loss by changing the norm: instead of the $L^2$ norm, we use that induced by the energy
function~(\ref{def:Energy}). In discrete form, the loss then corresponds to:
\begin{align}
\mathcal{L}(\psi,\widehat \psi) &= \frac{1}{NL^2}\sum_{\bk\in\Omega} k^\beta\bigl\vert \widehat{\widetilde\psi}_\bk^{(s)}-\widetilde\psi_\bk^{(s)}\bigr\vert^2 \\[0.2cm]
&\simeq \frac{1}{NL^2}\sum_{{(\lambda,\mu)}\in\Omega^\star} \bigl\vert \widehat{\psi}_{(\lambda,\mu)}^{\star(s)}-\psi_{(\lambda,\mu)}^{\star(s)}\bigr\vert^2 
\end{align}
where the last equation involves the Fourier-Mellin coefficients, see equation~(\ref{eq:FM_coefs}). In the continuous limit this last equation should be an equality,
but in practice, due to finite size effects and discretization problems discussed in the appendix \ref{app:scale_equivariance}, the relation is only approximate. Still it is more efficient to
use the Fourier-Mellin representation of the loss, again to save the retro-propagation through several FFTs and an interpolation layer. Similarly, the loss for the U-Net and Riesz
network in the spectral flow task is taken in direct space.

In the case of the ASM, the task can be seen as a pixel-wise binary classification, so we use the Binary Cross Entropy (BCE) Loss. The network outputs a map giving the ``probability''
of each site to belong to the avalanche about to unfold. Here, BCE reads:
\begin{align}
    \mathcal{L}(\psi, \widehat{\psi}) 
    = - \frac{1}{NL^2} \sum_{\bx, s} 
    &\Bigl[
        w_{\bx} \psi_{\bx}^{(s)} \log \sigma(\widehat{\psi}_{\bx}^{(s)})\nonumber \\[0.2cm]
        &+ \left(1 - \psi_{\bx}^{(s)}\right)
        \log\big(1 - \sigma(\widehat{\psi}_{\bx}^{(s)})\big)
    \Bigr],    \label{eq:BCE}
\end{align}
where we average over all pixels $\bx$ in the lattice and over samples $s$, $\sigma$ is the sigmoid function (applied to force output into the range $[0, 1]$), and  $w_\bx$ is a
reweighting factor that boosts the term corresponding to positive pixels. Indeed, the ASM dataset is highly imbalanced: since most avalanches are small,
the binary target maps contain mostly zeros, so the reweighting factor is needed to prevent the network from learning the trivial solution of predicting zeros everywhere.
In practice,  we use a constant factor $w_\bx=w$ given by the ratio between the total number of positive and negative pixels in the target fields $\psi_\bx$ from the training dataset
\begin{align}\label{def:w}
    w = \frac{\sum_{s, \bx} (1-\psi_\bx^{(s)})}{\sum_{s,\bx} \psi_\bx^{(s)}}.
\end{align}
On our finite dataset, it evaluates to $w \simeq 12.25$.

	
\section{\label{sec:results}Results}

\subsection{FGF phase mixing task}
	
We present the results obtained with the U-Net, the Riesz network, and the Fourier embedding network on the FGF task. The dataset consists of 50k samples with lattice size $L=128$,
with the target being the time evolution of the field $\varphi(\mathbf k, t=1)$, for a phase function~(\ref{eq:nuk}) with parameter $\nu = 50$.

We use a 4-level U-Net with a total of 485,673 parameters. For the Riesz network, we use 4 layers, with a total of 30,057 parameters. For the Fourier embedding network, we use an embedding of
$P=100$ features, and a 2-layer MLP with both hidden layers of size 64, giving a total of 10,689 parameters.

\begin{figure}[t]
\includegraphics[width=\columnwidth]{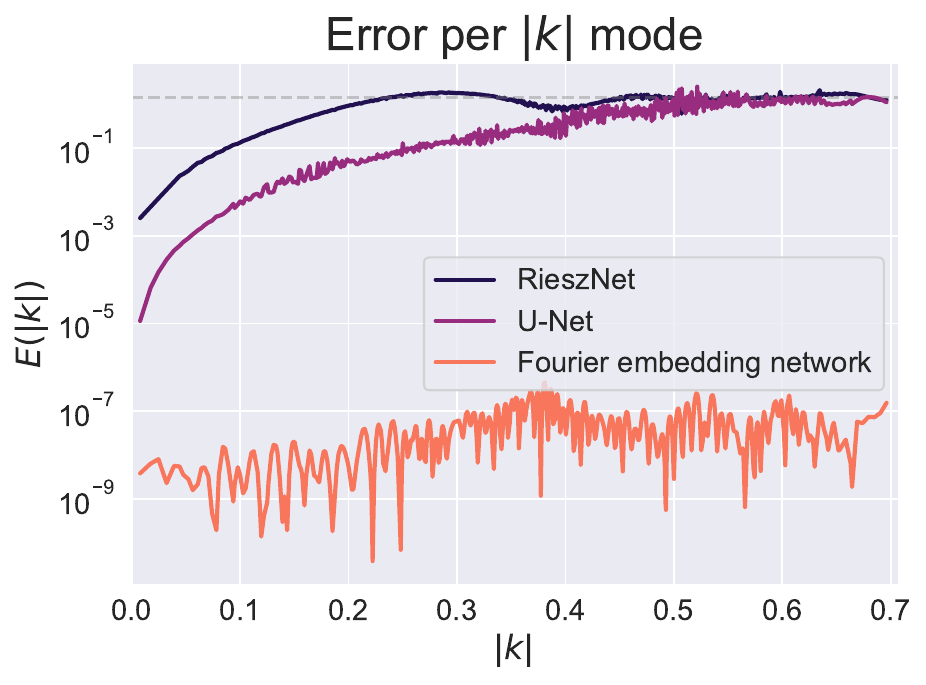}
\caption{\label{fig:fgf_error_riesz_unet} 
Normalized Mean Squared Error $E(|\mathbf k|)$ (equation~(\ref{eq:normalizedPS})) evaluated on the test set 
for the Riesz network, the U-net, and the Fourier embedding network.
The last one perfectly recovers the signal.
The dashed gray line corresponds to the random predictor.}
\end{figure}

We evaluate the performance of each model by computing the average mean squared error
as a function of $|\mathbf k|$, normalized by the power spectrum
\begin{align}
E(|\bk|) = \left(  \frac1N \sum_s|\psi^s_\bk-\widehat\psi^s_\bk|^2 \right)\vert\bk\vert^\beta,\label{eq:normalizedPS}
\end{align}
as shown in Fig.~\ref{fig:fgf_error_riesz_unet}.
Both U-Net and Riesz network only learn the low frequencies, which illustrates the problem of spectral bias. In contrast, using an equivalent (respectively the same) loss function,
the Fourier embedding network manages to appropriately deal with spectral bias. 

\begin{figure}[h]
\includegraphics[width=\columnwidth]{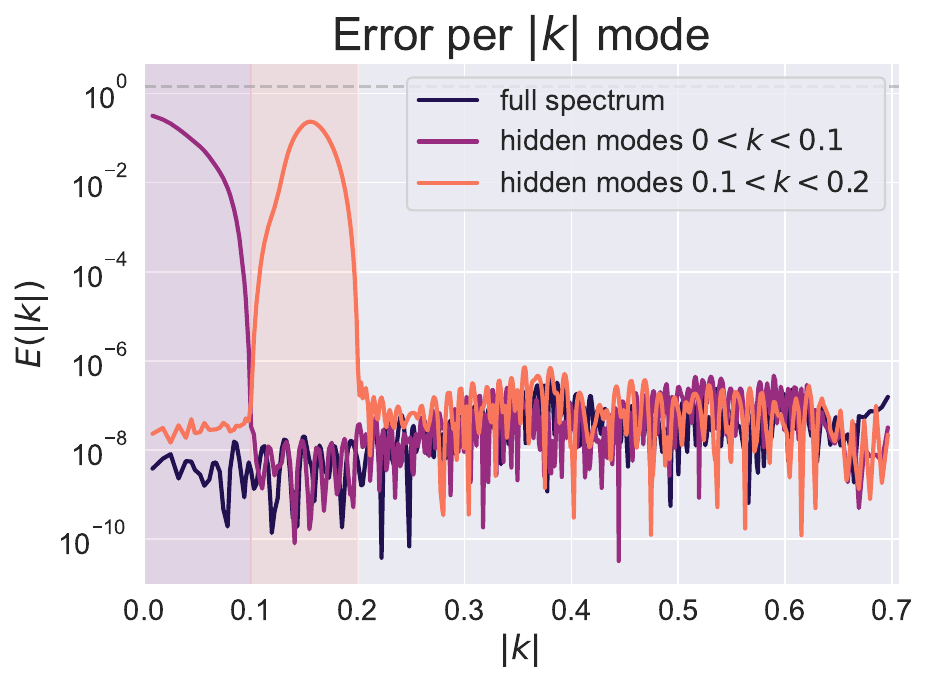}
\caption{\label{fig:fgf_error_fourier} 
Normalized Mean Squared Error $E(|\mathbf k|)$ (equation~(\ref{eq:normalizedPS})) evaluated on the test set, for the 
Fourier embedding network in the extrapolation setup.  
We compare the baseline training where all modes are present in the training set (\textit{full spectrum}) to two extrapolation cases where ranges of frequencies are hidden.
Hidden frequency ranges are highlighted with shaded areas.
The dashed gray line corresponds to the random predictor.}
\end{figure}

For the Fourier embedding network, which is the only network that manages to perform well in the standard task, we additionally perform two extrapolation experiments,
by hiding a range of frequencies in the training set. 
In the first one, we hide (set to $0$) the lowest part of the spectrum $0<|\mathbf k|<0.1$, which corresponds to extrapolation to large scales. 
In the second one, we hide the range $0.1<|\mathbf k|<0.2$, corresponding to interpolation to a set of middle scales. The results are shown in Fig.~\ref{fig:fgf_error_fourier}.

We observe here that the network learns for the three cases all the frequencies present in the training set. However, in both extrapolation experiments the test error
is high for the hidden range of frequencies, which indicates that the model is struggling to extrapolate across scales.
It still performs significantly better than a random predictor, while both the U-Net and the Riesz Network are unable (by design for the U-Net) to extrapolate.
    	
\begin{figure}[h]
\includegraphics[width=.8\columnwidth]{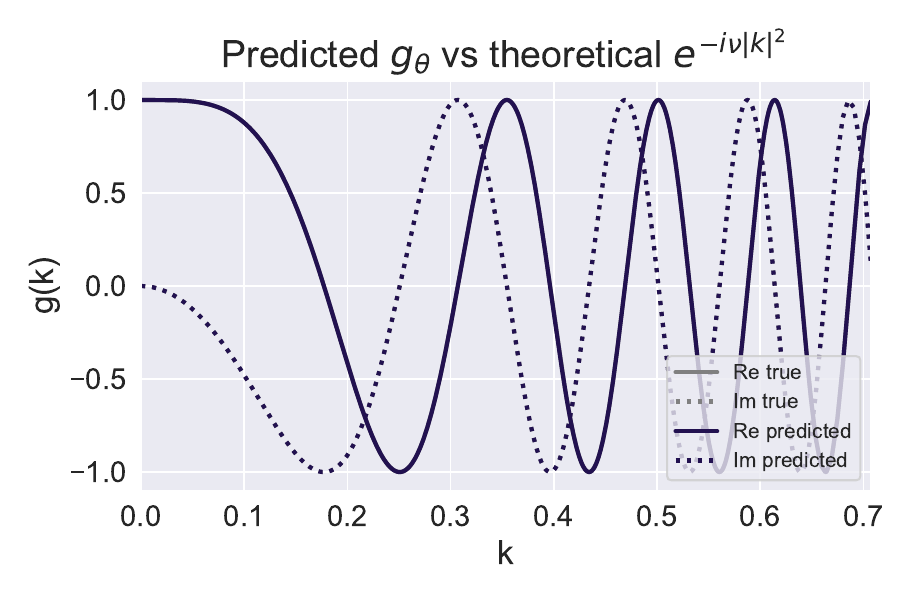}
\includegraphics[width=.8\columnwidth]{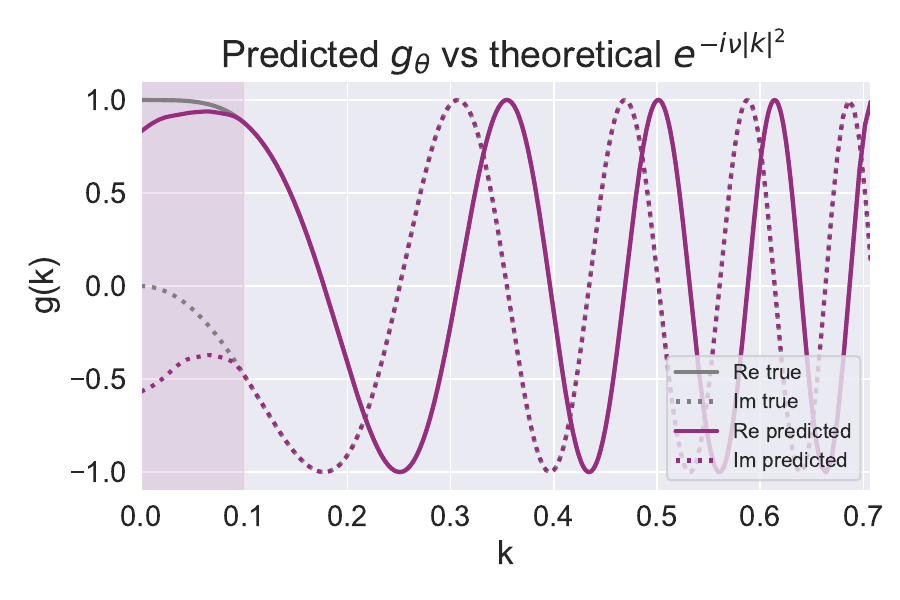}
\includegraphics[width=.8\columnwidth]{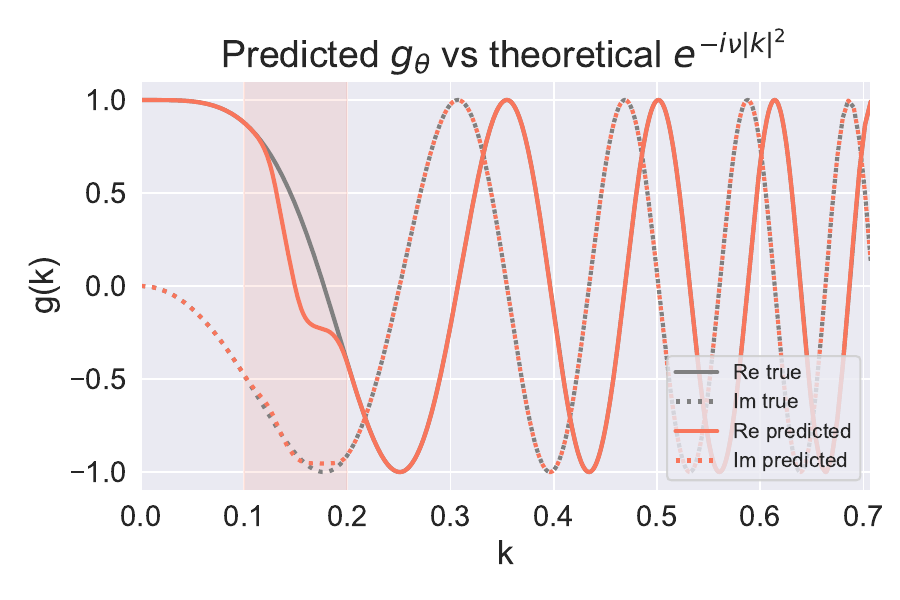}
\caption{\label{fig:fgf_learned_fn} 
Function $g_\theta(|\bk|)$ (equation~\ref{eq:learned_fn}) learned by the Fourier embedding network compared to the ground truth $g(\bk)=e^{-i\nu|k|^2}$.
In the first panel (standard experiment), the true and predicted values overlap so perfectly that they cannot be distinguished. 
In the other panels (extrapolation experiments), the mismatch are contained within the range of frequencies that was hidden during training.}
\end{figure}

For these three experiments, we can compare the learned function $g_\theta(\mathbf k)$, as given by eq.~\ref{eq:learned_fn} to the ground truth $g(\bk)=e^{-i\nu |\mathbf k|^2}$.
In Fig.~\ref{fig:fgf_learned_fn} we see that the output of the model perfectly matches the target in the baseline experiment without extrapolation, but deviates from it in
the hidden range for the extrapolation experiments.

In the FGF task, the energy distribution over scales is broad (see Eq.~\ref{eq:power_spectrum}), which makes the learning difficult, as evidenced by the failure of U-Net and
Riesz networks to succeed. Designing an architecture that explicitly operates in a Fourier representation, we induce the proper architectural bias, which proves to be key to correct fitting. 

In summary, we have introduced a testbed for extrapolation to unseen scales together with a baseline model, the Fourier Embedding Network (FEN), which shows the difficulty of
this out-of-domain generalization task. 
We believe that the key to solving this issue is to encode the proper scaling property inherent to the task into the architecture.

\subsection{FGF spectral flow regression task}\label{sec:spectral_flow_exp}

For this task we use a dataset of 250 samples with lattice size $L=64$. 
To perform scale transforms and rotations, data are prepared using the  analytic continuation  in $\bk$ space provided by the
Dirichlet kernel (equation~\ref{eq:Dirichlet_interpolation} in Appendix~\ref{app:model_details}).
The rescaling is done with a factor of $s=0.9$ and we experiment on two cases: a dataset with no rotation, and one with a rotation of angle $\alpha=\frac{\pi}{4}$. We test the U-Net
and the Riesz network, as well as the FM network designed for this task. For the U-Net and Riesz network we use the same architecture as in the phase mixing task,
while for the FM network we use $M=256$ Fourier-Mellin modes in both the radial and angular components to build the latent representation of the data, which gives $M^2 = 256^2$ weights to be learned.

\begin{figure}[t]
\includegraphics[width=.8\columnwidth]{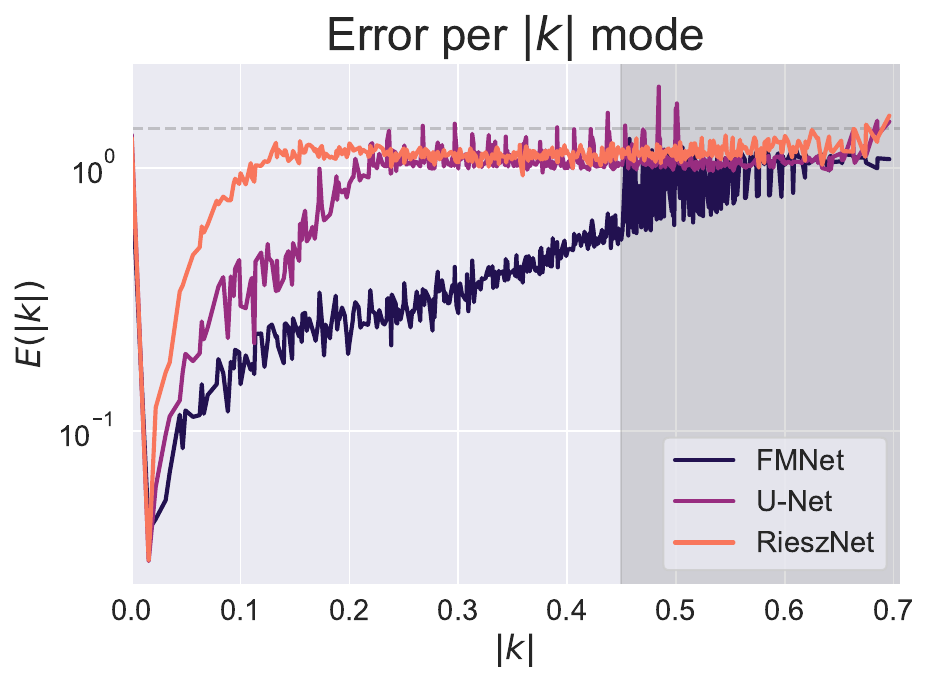}
\includegraphics[width=.8\columnwidth]{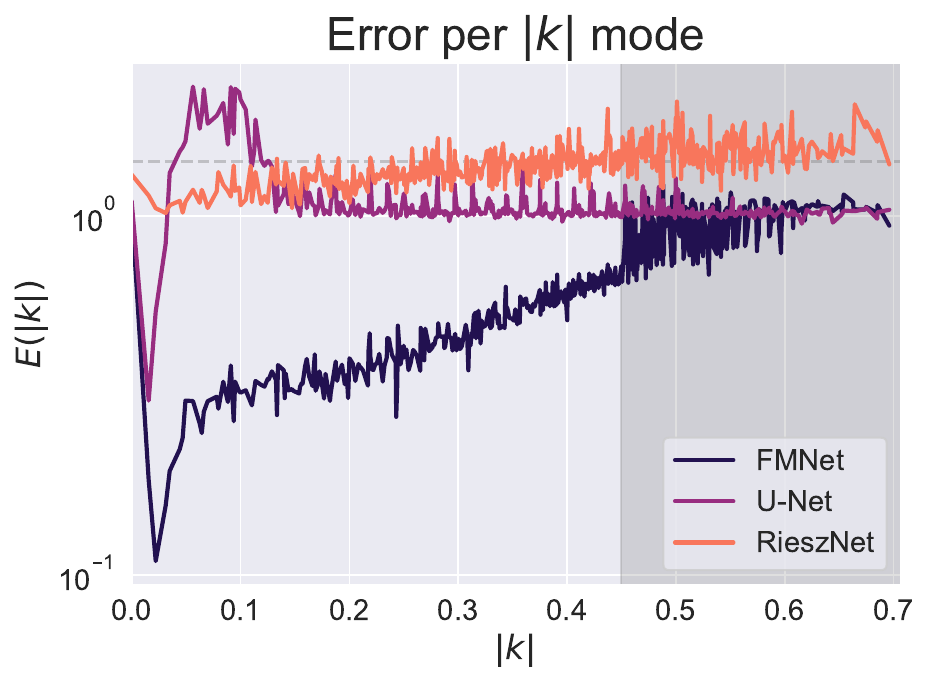}
\caption{\label{fig:model_error_comparison} 
Normalized Mean Squared Error $E(|\mathbf k|)$ (equation~(\ref{eq:normalizedPS})) evaluated on the test set, for the 
Riesz network, U-net and FM network on the spectral flow FGF task, with a rescaling factor of $0.9$. 
The first panel corresponds to scaling without rotation, and the second one to scaling combined with a rotation of $\pi/4$.
The dashed gray line corresponds to the random predictor.
The white line at $E(|\bk|)=1=10^0$ corresponds to a constant predictor $\widehat \psi=0$: error above this line can be considered a failure.}
\end{figure}

In Fig.~\ref{fig:model_error_comparison} we compare the three models on the baseline task without extrapolation, where all frequencies are present in the training data. 
As before, we measure the relative error in Fourier space, as a function of $|\mathbf k|$. 
In the case of pure scaling (no rotation), the U-net and the Riesz network manage to learn on the low end of the spectrum, but perform significantly worse than the FM network,
which despite also showing some spectral bias, is able to learn the intermediate frequencies up to $|\mathbf k|= 0.45$, which is the scale at which noise is inserted in
the data in the spectral flow process. 
In the case of scaling combined with rotation, we observe that both the U-net and Riesz network fail completely, while the FM network maintains its performance.
This failures are likely due to the process being non local both in real and Fourier space, in particular the linear layers of the Riesz network or convolution
layers of the U-net act locally in $\bk$ space. 

\begin{figure}[t]
\includegraphics[width=\columnwidth]{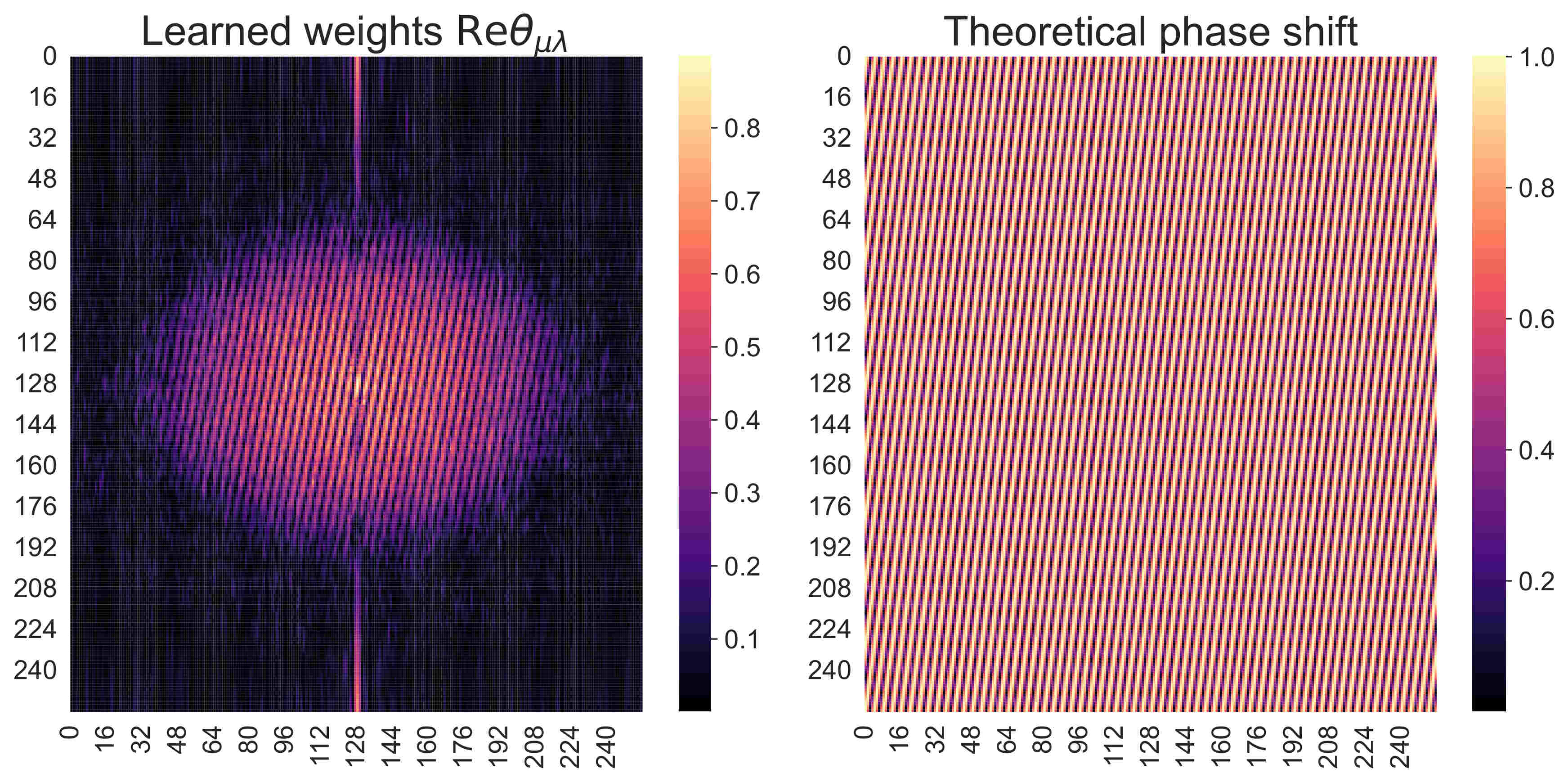}
\caption{\label{fig:learned_weights} Learned weights of the FM network on an FGF transformed dataset with scaling factor $s=0.9$ and rotation angle $\alpha=\pi/4$ (left),
compared to the theoretical phase shift in FM space corresponding to this transformation (right). 
The training data consist of samples of size $L^2=64\times 64$, and we use $M^2 = 256\times 256$ Fourier-Mellin features.}
\end{figure}

Additionally, in Fig.~\ref{fig:learned_weights} we plot the weights learned by the FM network. Note that the process corresponds to a translation in logpolar coordinates,
the rotation being a translation in the angular coordinate, and the rescaling a translation in the radial coordinate. Since the Fourier-Mellin transform is a Fourier transform
applied in logpolar coordinates, this means that the transformation can be seen as a phase shift in Fourier-Mellin space. 
We observe that indeed, the weights learned match the theoretical phase shift that defines the process, at least around the center, where the FM modes are stronger.
The difference between the two is due to residual spectral bias in Fourier Mellin caused by distortion of densities when performing the cartesian to logpolar interpolation. In particular
the angular dependency of the field looks smoother at large scale than at small ones in the log-polar representation.
The fact is that the loss of the learned model is lower than the loss that can be obtained from the theoretical phase shift. 
The theoretical phase shift is actually not optimal because the scale transform and rotation done on the input field to generate the output data is done 
as already said by interpolation on the square lattice, not on the log-polar lattice. As a result the phase-shift in FM corresponding to these transformation is only
approximate and the loss that we obtain is actually the lowest one that can be achieved with a purely local model in FM. 

\begin{figure}[t]
\includegraphics[width=.8\columnwidth]{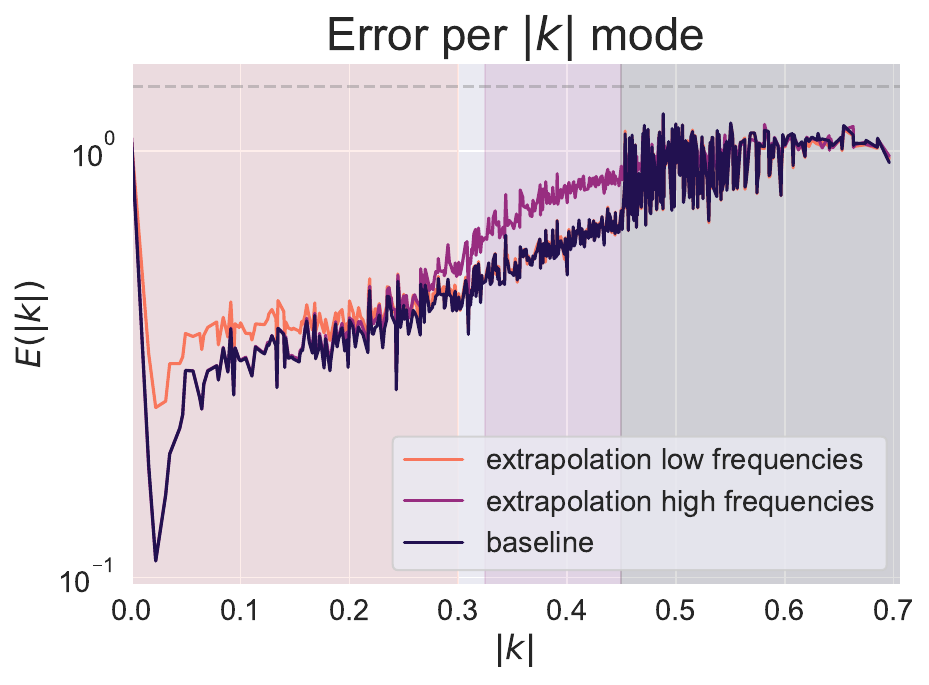}
\caption{\label{fig:extrapolation_error_comparison} 
Normalized Mean Squared Error $E(|\mathbf k|)$ (equation~(\ref{eq:normalizedPS})) evaluated on the test set, for the 
FM network on the spectral flow FGF task, with a rescaling factor of $0.9$, and a rotation angle of $\pi/4$. We compare the baseline training where all
modes are present in the training data with two extrapolation settings, where we hide the high frequencies (from $k_{\rm occ}=0.325$) and the low frequencies (up to $k_{\rm occ}=0.3$)
respectively with corresponding extrapolation factors up to $f_{\rm extrapol}=17.3$ and $f_{\rm extrapol}=1.29$. 
The dashed gray line corresponds to the random predictor.
The white line at $E(|\bk|)=1=10^0$ corresponds to a constant predictor $\widehat \psi=0$.}
\end{figure}

We consider two extrapolation settings: extrapolation to high frequencies \ie super-resolution, where we hide the high end of the spectrum
in the training data; and extrapolation to low frequencies, where we hide the low end. 
For the first case, frequencies above $|\mathbf k|=0.45$ correspond to the noise inserted of the rescaling, so the cut should be made below that range. 
In figure Fig.~\ref{fig:extrapolation_error_comparison} we compare the baseline experiment and the two extrapolation cases. 
The FM Network is able to extrapolate both to high and low frequencies: although the error is slightly higher than the baseline in the hidden range,
it is still significantly below the order of constant or random predictors. 
The model is especially successful at extrapolating to low frequencies, where even though the hidden range is larger, the curve is very close to the baseline.

\begin{figure}[t]
\includegraphics[width=.8\columnwidth]{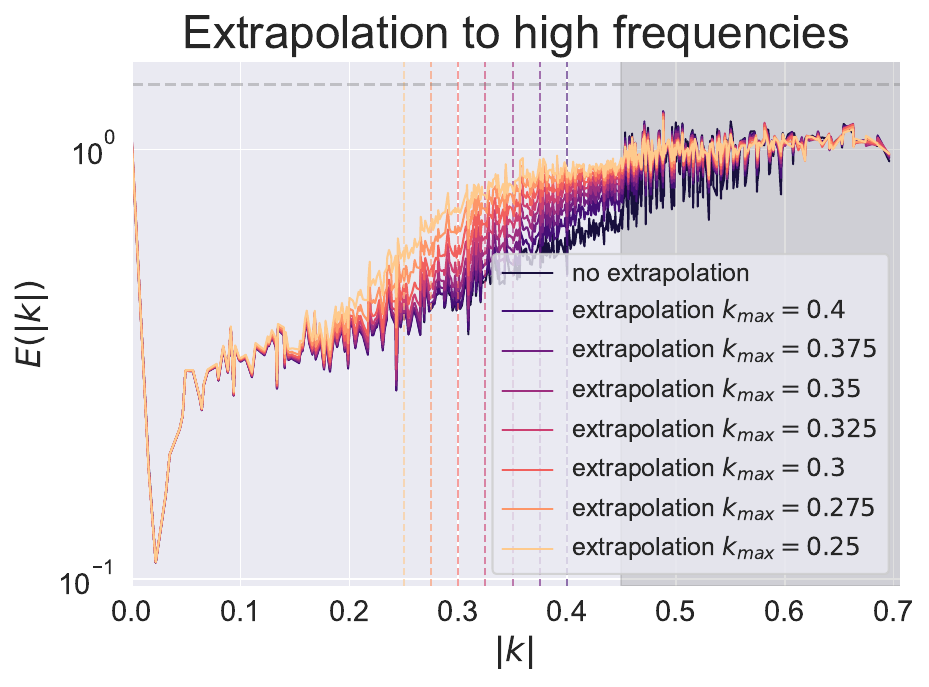}
\includegraphics[width=.8\columnwidth]{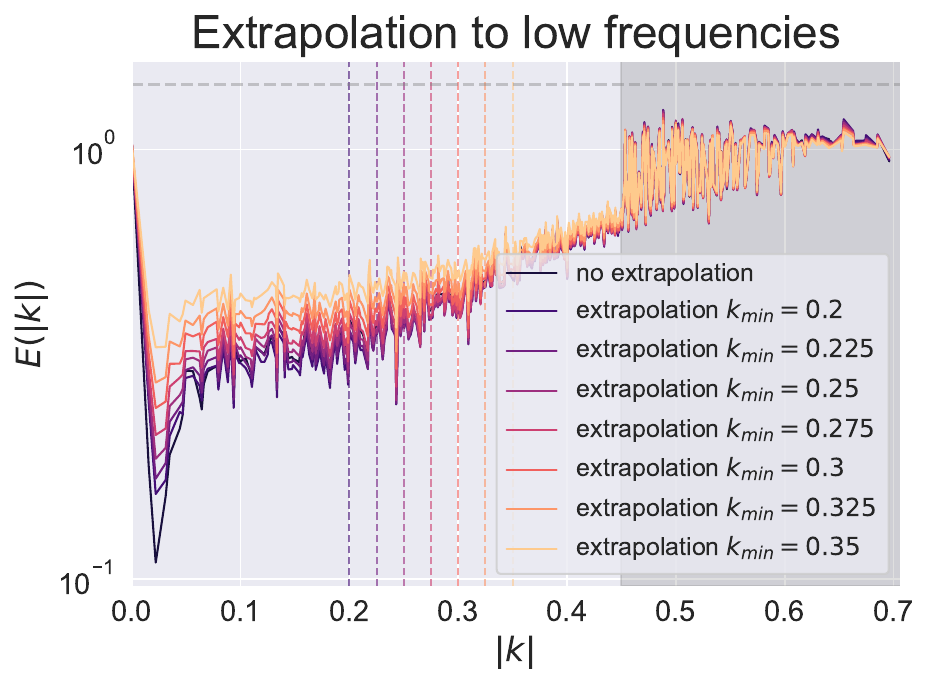}
\caption{\label{fig:extrapolation_low_high_freq} 
    Normalized Mean Squared Error $E(|\mathbf k|)$ (equation~(\ref{eq:normalizedPS})) evaluated on the test set, for the 
    FM network on the spectral flow FGF task, with a rescaling factor of $s=0.9$, and a rotation angle of $\pi/4$. 
    We compare the error curves for the two extrapolation settings (extrapolation to high (top) and low (bottom) frequencies respectively) as we increase the hidden frequency range.
    The range of extrapolation are therefore between $f_{\rm extrapol}\in[11.5,23]$ and $f_{\rm extrapol}\in [1.29,2.35]$ respectively for large scale extrapolation and the super-resolution settings.}
\end{figure}

We compare the error curves as a function of the width of the hidden range, for both extrapolation settings, in Fig.~\ref{fig:extrapolation_low_high_freq}. 
Again, the performance is remarkably good, especially for extrapolation to low frequencies, even when a large portion of the spectrum is hidden during training. 
We remark that the number of points in the input corresponding to a given value of $|\mathbf k|$ increases with $|\mathbf k|$, with only a few input points
accounting for the lower frequencies. This explains why we are able to hide a larger range of low frequencies.

We have shown that the FM model manages to alleviate the problem of spectral bias that affects the U-net and Riesz network at the spectral flow task,
dealing well with the non-locality induced by the rotation, which causes both the other models to fail completely. 
Additionally and most importantly, this model performs remarkably well at extrapolation when we hide a range of frequencies from the training data, especially
for large scale extrapolation with extrapolation factors of the order of $f_{\rm extrapol}> 10$ are easily obtained, while super-resolution also work 
but less neatly with extrapolation factors of the order of $f_{\rm extrapol} <10$.
This is expected as the model is designed to learn the FM phases which correspond to fully delocalized features in $\bk$ space, hence appropriate for extrapolation.
The residual spectral bias in FM make it however more difficult for super-resolution than for for large scale extrapolation, also simply because the density of modes
hence of information is much larger at small than at large scale.

One message from these empirical observations is that finding a representation where the operator is nearly diagonal and avoiding a large spectral bias are two requirements
that need to be considered to obtain good results. 
As will be more clearly stated in Section~\ref{sec:theory},
the first requirement reduces drastically the complexity of the regression (from $L^4$ to $L^2$) while the second one ensures that all scales are learned
at the same time, yielding better extrapolation. 
The  locality helps to reduce the spectral bias by reducing the size of the feature matrix, but it is not necessarily directly achievable by symmetry arguments, since translation 
and roto-symmetry do not commute. 
So for more complex processes than those considered in II.A \ie processes mixing (\ref{eq:phase_mixing}) and (\ref{eq:eq12spectral_flow}), the way out could be to alternate
representations diagonal to the translation  symmetry and diagonal to the roto-scale symmetry.

\subsection{Abelian Sandpile Model}

For the ASM we consider a dataset consisting of $10^6$ samples with lattice size 32. The data is generated by running the dynamics and registering one event every hundred
avalanches to provide some decorrelation. Note that the number of training samples is much smaller than the size of the configuration space of the ASM,
which scales exponentially with the lattice size \cite{dhar1999abelian}. We present the results obtained with the U-Net, the Riesz network, and the Wavelet GNN.
	
\begin{figure}[t]
\includegraphics[width=.6\columnwidth]{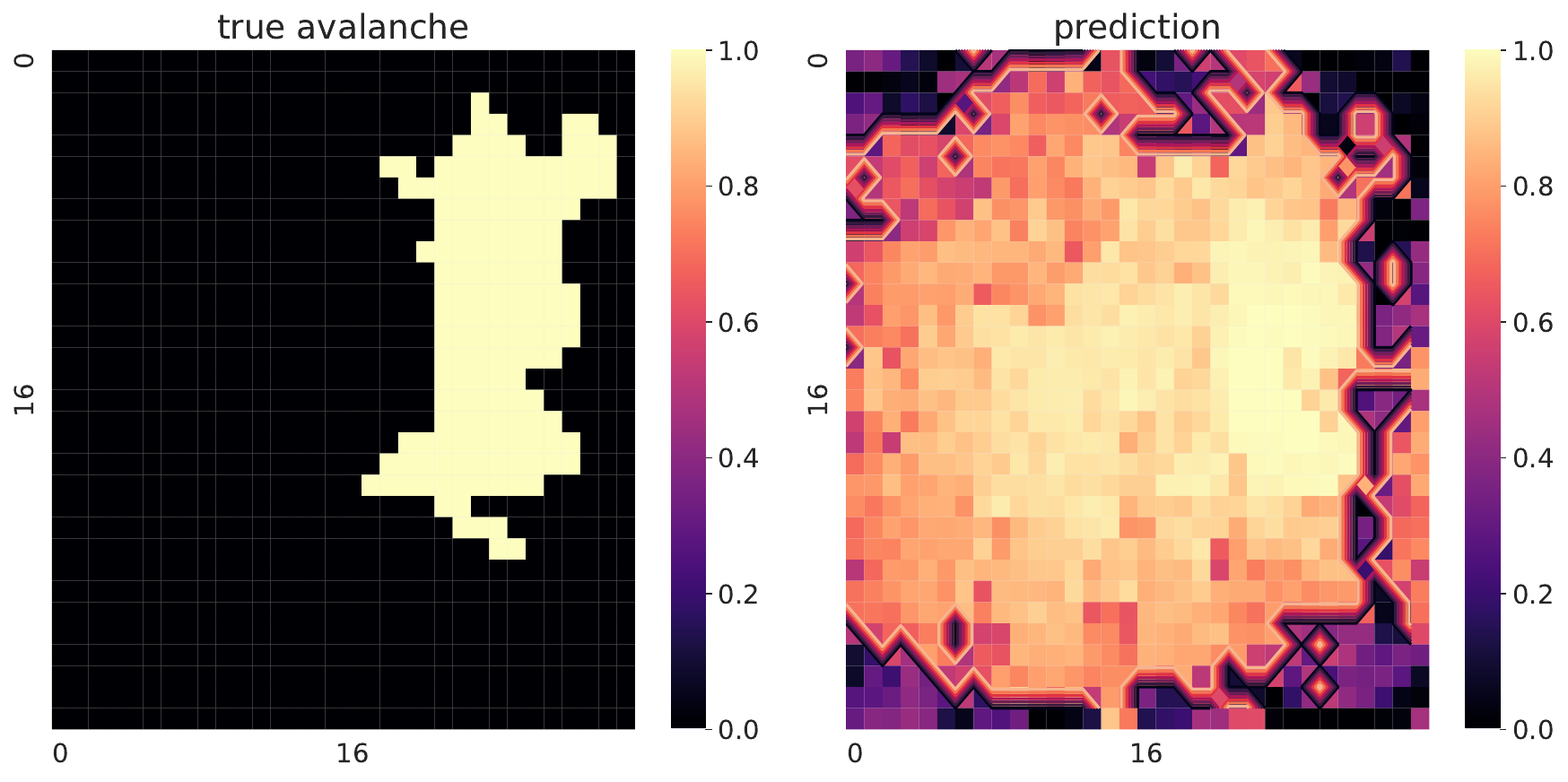}
\includegraphics[width=.6\columnwidth]{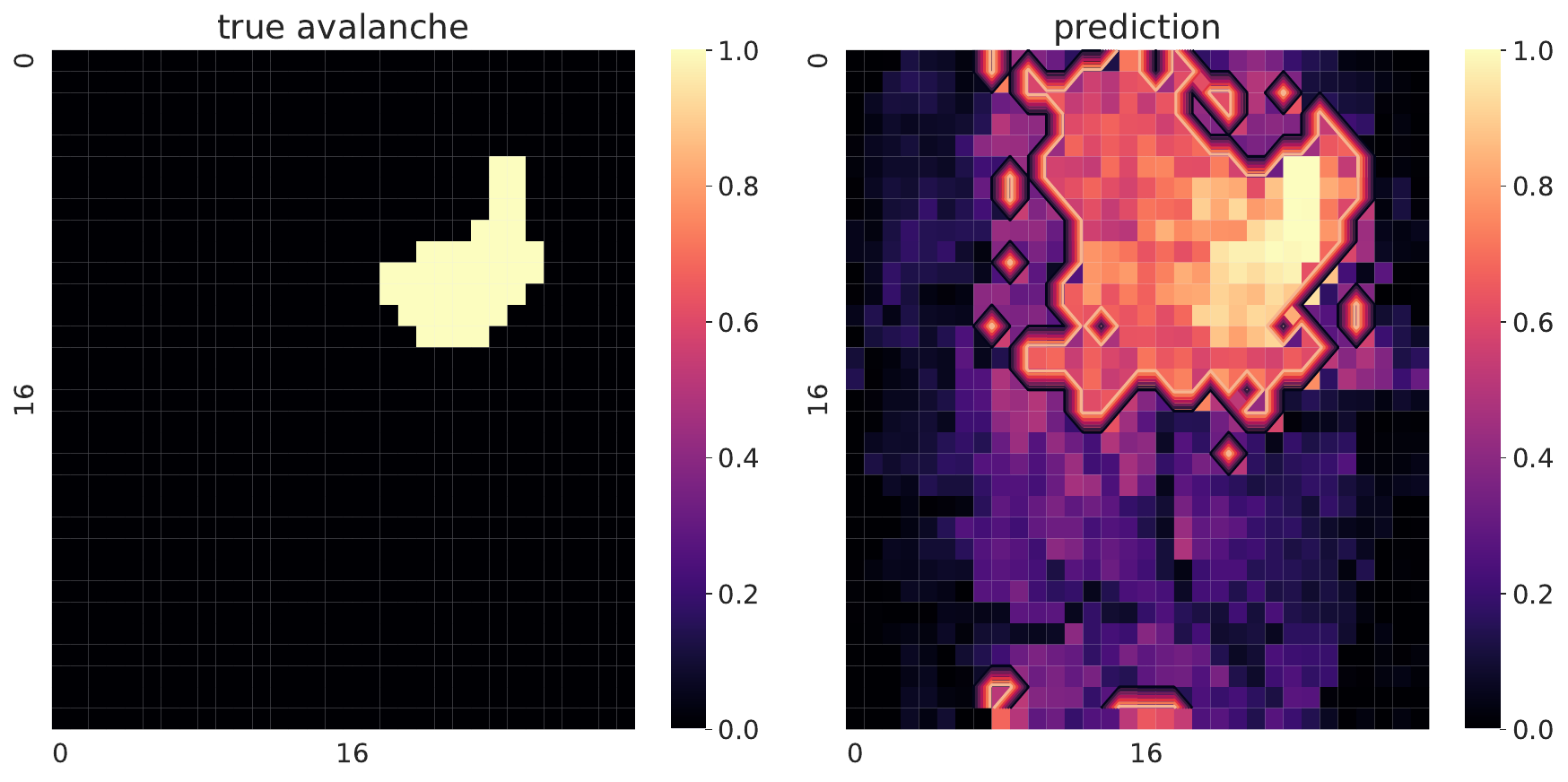}
\includegraphics[width=.6\columnwidth]{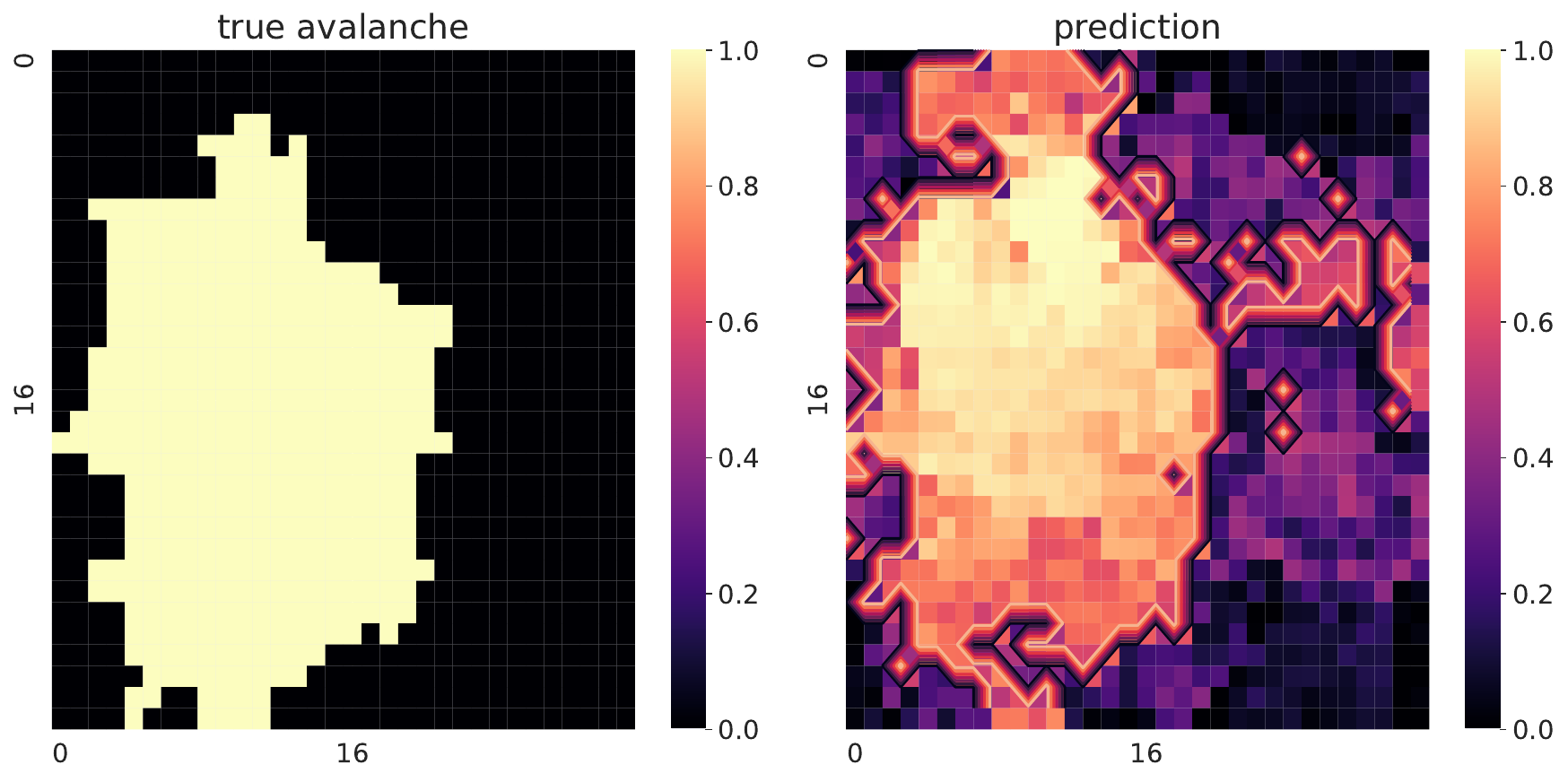}
\caption{\label{fig:asm_sample_outputs} Sample outputs of the Riesz network for the ASM task, on lattice size $32\times 32$. The left panel corresponds to the target outputs and
the right panel to the predicted ones.}
\end{figure}

In Fig.~\ref{fig:asm_sample_outputs} we show sample outputs obtained from the Riesz network compared to their respective ground truth. 
The output map is interpreted as representing the probability for each site to topple during the avalanche. 
As expected, the model is able to capture the starting point of the avalanche as well as some of its  shape.
There is a visible bias towards larger avalanches, which is due to the boosting of the positive term in the loss function.

\begin{figure}[t]
\includegraphics[width=.8\columnwidth]{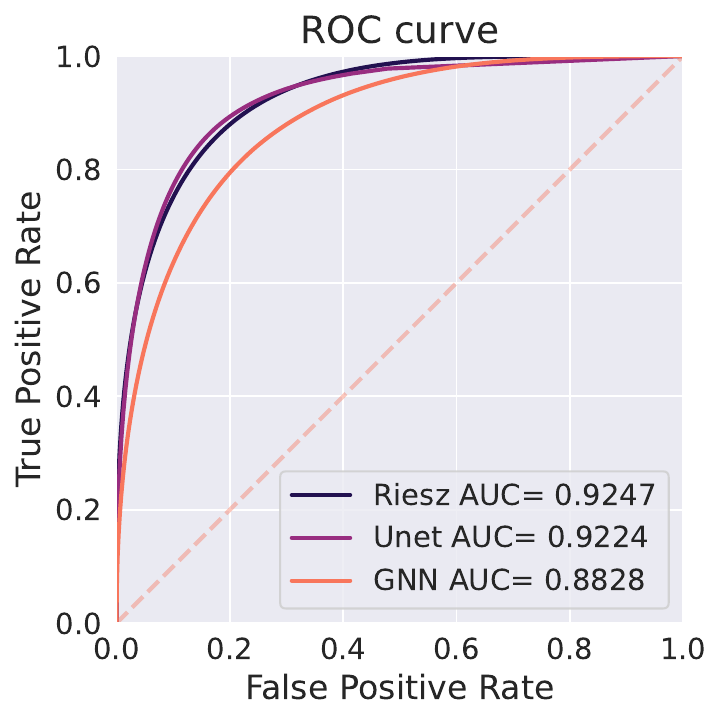}
\caption{\label{fig:asm_roc_curve} ROC curve comparison for the U-Net, Riesz network and Wavelet GNN.}
\end{figure}

To obtain a crisp prediction, we determine an optimal threshold by means of the receiver operating characteristic (ROC)
curve, which is the curve obtained by varying the threshold of the binary classifier and plotting the true positive rate (TPR) against the false positive rate (FPR).
The area under the curve (AUC) gives a (threshold-independent) metric of the quality of the classifier, with a perfect classifier having an AUC of $1$, and a random one, $0.5$.
In Fig.~\ref{fig:asm_roc_curve} we show the ROC curves obtained for each of the three models. 
The Riesz network gives the best performance out of the three models, but only by a small margin compared to the U-Net.
The Wavelet GNN has the lowest AUC of the three, but the performance is still comparable to that of the other two.

\begin{figure}[t]
    \includegraphics[width=\columnwidth]{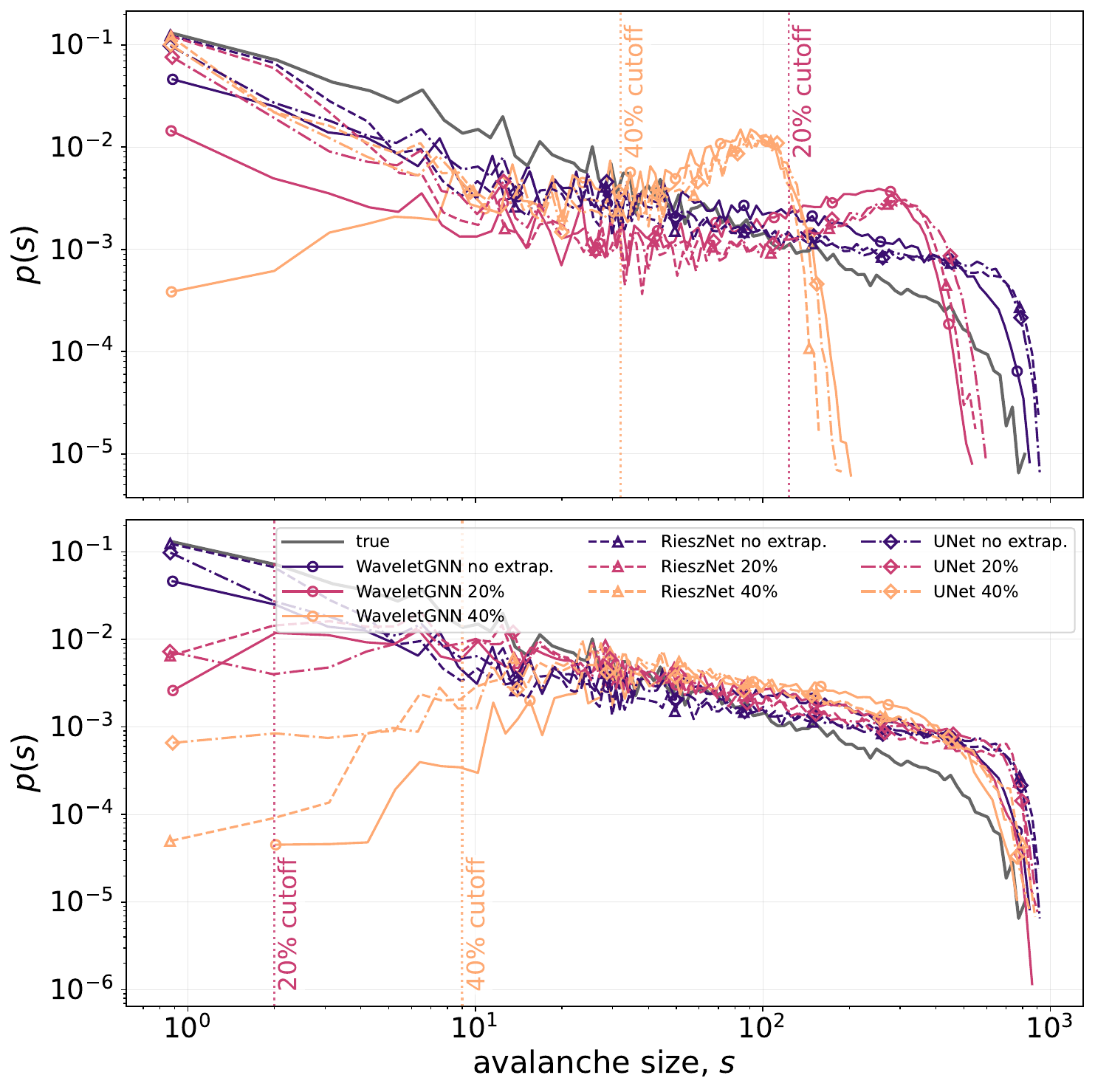}
\caption{\label{fig:asm_size_distribution} 
Size distribution of the predictions from the U-Net, Riesz network and Wavelet GNN, compared to the empirical distribution in the training data.
The top panel shows the results of large-scale extrapolation: the pink curves show the distribution of the three models trained with the largest $20\%$ hidden,
the yellow ones show the results of training with the largest $40\%$ hidden, and the purple ones show the baseline experiments. The bottom panel shows analogous
results for the small-scale extrapolation experiments.}
\end{figure}

The optimal threshold is obtained by maximizing the balanced accuracy, defined in terms of the TPR and the true negative rate (TNR) as $A=\frac12(\mathrm{TPR} + \mathrm{TNR})$.
The avalanche prediction obtained by applying this threshold is marked as a contour on the samples in Fig.~\ref{fig:asm_sample_outputs},
and is used to compute the avalanche size shown in Figs.~\ref{fig:asm_size_distribution}.

We also run extrapolation experiments,  where we hide ranges of avalanche sizes in the train set. Large [resp. small]-scale extrapolation corresponds to hiding largest [resp. smallest]
avalanches quantiles.
Each time we re-compute the factor $w$~(\ref{def:w}) so as to  boost the active sites by the appropriate factor. 
In Fig.~\ref{fig:asm_size_distribution} we compare the predicted avalanche size distributions obtained for all models and extrapolation modes, each time using the optimal threshold
with respect to the ROC curve. We display the empirical (true) distribution obtained from the original test set samples.

In the baseline case without extrapolation, all three models are biased towards predicting larger avalanches than the true distribution, confirming the qualitative observation of the output samples. 
Qualitatively, for small avalanches, the  Wavelet GNN better matches the slope of the true distribution, although its stronger excess of larger events shifts the distribution down. 
We report balanced accuracy and another metric (Jaccard's index) as function of size in Appendix~\ref{app:experiment_details}.

In the extrapolation cases, all models under-perform similarly: they fail to predict event sizes far from the training range,
shifting the distribution towards smaller events in the large-scale extrapolation case, and towards larger ones for small-scale extrapolation. 
Note however that the bias towards large events still allows models trained on sizes up to around $30$ to predict avalanches up to size $200$.

To conclude, we note that none of the models, even the Riesz network and the Wavelet GNN, which implement different forms of scale invariance, are able to properly extrapolate. 
This preliminary investigation thus leaves open the question as to how one can appropriately inform a model about this kind of scale invariance. 
Directions for future work include the design of more subtle architectural constraints, \textit{e.g.}~by informing the network about anomalous scaling behavior of the system
through its input and output map correlation length exponents, similarly to what we did with the Fourier-Mellin transform. This could be possibly combined with a more careful
design of the loss function, \textit{e.g.}~ by reweighting samples with avalanche size $s$ by a factor $\sim P(s)^{-1}$. Other strategies  more in spirit of the spatial
RG flow~\cite{NNRGflow} are also considered for future work.

\section{Theoretical analysis}\label{sec:theory}
The main point of the previous section is to empirically investigate, on a functional regression task in linear and non-linear settings, a few ways to inform a neural network with scale invariance, 
and how it can help extrapolate. 
The present section develops theoretical points, focusing on the linear regression task, to clarify the following points: 
(i) how known anomalous scaling properties of input-output relationships are constraining  the linear neural operator (Sections~\ref{sec:Scale_linear} and \ref{sec:Proj_fspace}); 
(ii) obtaining the learning dynamical equations of this functional regression setting; how does the spectral bias influences the dynamics and can we identify a proper scaling emerging from these equations in the asymptotic regimes, in terms of size of input-output, number of samples and number of features (Sections \ref{sec:linearized_dyn} and \ref{sec:Asymp_dynamics}); (iii) what is the expected behavior of the model in the extrapolation regime in the ideal case where where the operator is diagonal in Fourier-Mellin representation (Section~\ref{sec:Extrapolation_asymp})

\subsection{Scale Equivariant Linear models}\label{sec:Scale_linear}
We denote by ${\rm FGF}(\beta)$ the fractional field distribution corresponding to exponent $\beta$.
Let us first state the property that should obey an equivariant functional map w.r.t. scale symmetry and rotation, leaving aside translation for the moment.
We assume this functional $F$ to take a $2$-d field $\varphi\sim FGF(\beta)$ as input and to output another
$2$-field $\psi\sim FGF(\gamma)$ with exponents $\beta$ and $\gamma$, possibly distinct. Using the Fourier representation this corresponds to have
(from now on we drop the tilda in $\widetilde\varphi$ for denoting Fourier components of the  field) 
\begin{align}\label{eq:map}
\psi[\varphi](\bk) = \int d^2\bk' F(\bk,\bk')\varphi(\bk')
\end{align}
with $\bk = (k_x,k_y)$ a $2$-d vector which we also represent in polar coordinates as $(k=|\bk|,\alpha)$.  
Assuming rotation symmetry we have
\begin{align*}
F(\bk,\bk') = F(k,k',\alpha-\alpha')
\end{align*}
The equivariant property corresponds to having both $\varphi$ and $\psi$ satisfying 
the scale symmetry~(\ref{eq:eq8scaling_FGF2}) but with possibly distinct exponents.
In the Fourier representation this reads
\begin{align}
\varphi_s(\bk) = \frac{1}{s^{1+\beta/2}}\varphi\Bigl(\frac{\bk}{s}\Bigr) &\stackrel{d}{=} \varphi(\bk) \label{def:varphiseq52}\\[0.2cm]  
\psi_s(\bk) = \frac{1}{s^{1+\gamma/2}} \psi\Bigl(\frac{\bk}{s}\Bigr) &\stackrel{d}{=} \psi(\bk) \label{def:ys}
\end{align}
where these equality hold both in distribution, meaning that the rescaled fields $\varphi_s\sim FGF(\beta)$ and $\psi_s\sim FGF(\gamma)$ follows the same distributions as the
original ones as can be checked from the definition~(\ref{def:FGF},\ref{def:Energy}).
Now imposing equivariance on $F$ means that  $\varphi_s$ and $\psi_s$ should also be related by~(\ref{eq:map}). This then leads to (see Appendix~\ref{app:scale_equivariance_linear}) impose 
the following form for $F$:
\begin{align}\label{eq:scale_form}
F(\bk,\bk') = \frac{1}{k^{\gamma/2}}
\frac{f\Bigl(\frac{k'}{k},\alpha-\alpha'\Bigr)}{kk'}{k'}^{\beta/2}.
\end{align}
A convention is chosen here such that the kernel $f\bigl(\frac{k'}{k},\alpha-\alpha'\bigr)/(kk')$ is a map from the Hilbert space  $\mathcal{H}_{L_2}$ of square-integrable functions onto
itself, with respect to the standard $L_2$ norm.
In this form we have $F_\theta = D_{\rm F}^{-\gamma/2} D_{\rm FM} D_{\rm F}^{\beta/2}$ where $D_{\rm F}$ is a diagonal operator in the Fourier representation with $D_{\rm F}(k) = k$,
while $D_{\rm FM}$ is strictly scale invariant and diagonal in the ``Fourier-Mellin'' representation. This representation uses the following orthonormal basis w.r.t. to the standard $L_2$ inner product
\begin{align}\label{def:Fourier-Mellin}
\phi_{(\lambda,\mu)}(k,\alpha) \egaldef \frac{1}{2\sqrt{\pi\log(L)}}\frac{e^{i(2\pi\lambda\log(k)+\mu\alpha)}}{k}
\end{align}
indexed by $\lambda\in\R$ and $\mu\in\Z$ being respectively the eigenvalue of the $2$-d scale and rotation operator~(\ref{def:Tscale_eq13},\ref{def:Trot}), with $\beta=0$
as far as the scale operator is concerned (see Appendix~\ref{app:Fourier_Mellin}). Here the normalization corresponds to assume an IR cutoff for $k$ equal to $1/L$. 
Then the relation between $f$ and $D_{\rm FM}$ basically corresponds to a Fourier transform in scale.

To conclude this section let us remark that the form of $F$ insuring equivariance can be relaxed by allowing the functional to be equivariant
up to some arbitrary phase shift \ie gauge transformation. This does still preserve the scaling properties of the output. In this respect we extend our family of operators to  
\begin{align*}
F(\bk,\bk')  = \frac{1}{k^{\gamma/2}}\frac{\bigl\vert f\bigl(\frac{k'}{k},\alpha-\alpha'\bigr)\bigr\vert}{k k'}{k'}^{\beta/2}\ e^{i\bigl(\omega_{\rm out}(\bk) -\omega_{\rm in}(\bk')\bigr)}
\end{align*}
with $\omega_{\rm in,out}$ some real functions. In the case where $F$ is a one-to-one map, corresponding for instance the end points of trajectories which do not cross each others
like for an inviscid flow, then the phases can be transported along trajectories and strict equivariance is satisfied with respect to the whole group of transformations, including phase shifts.
The phase mixing example considered in Section~(\ref{sec:FGF}) actually corresponds to a special case of such equivariant dynamical process,
in which $\gamma=\beta$, $f(k'/k,\alpha-\alpha')=\delta(k-k')\delta(\alpha-\alpha')kk'$ and $\omega_{\rm out}(\bk) -\omega_{\rm in}(\bk') = -\nu(\bk)t$.

\subsection{Projection on the feature space}\label{sec:Proj_fspace}
Since we are doing a linear regression, the learning task amounts to finding a good projection of the underlying data-generating functional map onto an operator
space defined by our set of pre-defined features, in a similar way as previous works~\cite{furtlehner2023free}. 
Here we want to specify the nature of this projection.
Let us consider an arbitrary functional $F$ corresponding to equation~(\ref{eq:map}) and a set of $M$
features $\{f_p,p=1,\ldots M\}$. We look for a decomposition of this kernel as
\begin{align*}
F(\bk,\bk') = F^\parallel(\bk,\bk')+F^\perp(\bk,\bk')
\end{align*}
such that $F^\parallel$ corresponds to the part that can be recovered with the features at hand, while $F^{\perp}$ is the orthogonal part in a sense to be specified.
The set of features defines a space $\EE_f$ of operators onto which we expect a decomposition of the form
\begin{align}\label{eq:Fdecomp}
F^\parallel(\bk,\bk') = \sum_{p=1}^M w_p f_p(\bk,\bk'),
\end{align}
with coefficients $w_p\in\R$.
These are obtained by orthonormal projection onto $\EE_f$, w.r.t. an inner product defined on a larger operator space containing both $F$ and $\EE_f$. For this it is required 
that two operators $F_1$ and $F_2$ are orthogonal whenever the corresponding $\psi_1[\varphi]$ and $\psi_2[\varphi]$
obtained from~(\ref{eq:map}) are orthogonal in average, \ie that:
\begin{align*}
\sum_{\bk\in\Omega} \vert\bk\vert^\gamma\E_{\varphi\sim FGF(\beta)} \left[\psi_1[\varphi](\bk)\psi_2^*[\varphi](\bk)\right] =0
\end{align*}
where $*$ denotes complex conjugation.
Here we implicitly use the inner product associated to the energy~(\ref{def:Energy}) of the $FGF(\gamma)$ distribution. 
This then corresponds to have
\begin{align*}
\sum_{\bk,\bk'\in\Omega} \frac{k^\gamma}{k'^\beta}F_1(\bk,\bk')F_2^*(\bk,\bk')  = 0.
\end{align*}
Consequently we define an inner product between operators adapted to input fields $\varphi\sim FGF(\beta)$ and output fields 
$\psi\sim FGF(\gamma)$ as
\begin{align}\label{def:inner_product}
\langle F_1,F_2\rangle_{\beta,\gamma} \egaldef \frac{1}{L^2}\sum_{\bk,\bk'\in\Omega} \frac{k^\gamma}{k'^\beta}F_1(\bk,\bk')F_2^*(\bk,\bk'),
\end{align}
where the $L^2$ normalization stands here for later convenience.
Equipped with this definition, the weights of the projection~(\ref{eq:Fdecomp})
are given by
\begin{align}\label{eq:wp}
w_p = \sum_{q=1}^M[C^{-1}]_{pq}\langle F,f_q\rangle_{\beta,\gamma},\qquad p=1,\ldots M.
\end{align}
with $C$ the covariance of the features, those coefficients are given by 
\begin{align}\label{eq:population_matrix}
C_{pq} = \langle f_p,f_q\rangle_{\beta,\gamma},\qquad p,q=1,\ldots M.
\end{align}
This corresponds to perform the regression with an infinite amount of data. In practice of course we have a finite number of data points and the regression is the result of a learning process which is 
described in the next section.

\subsection{Linearized learning dynamics}\label{sec:linearized_dyn}
Let us now state the learning dynamics in full generality. 
From now on we take into account the fact that we are on a finite lattice $\Omega$, 
so that all integrals are replaced by summation of $\bk$ over $\Omega$. 
The data consists in a set of $N$ pairs of fields $\{(\varphi^{(s)},\psi^{(s)}),s=1,\ldots N\}$ related by
\begin{align}\label{eq:data_map}
\psi(\bk) = \sum_{\bk'\in\Omega} K(\bk,\bk')\varphi(\bk')+\sigma(\bk) \epsilon(\bk)
\end{align}
with $\varphi\sim {\rm FGF}(\beta)$ and $\psi\sim {\rm FGF}(\gamma)$ the joint distribution being denoted $\sim {\rm FGF}(\beta,\gamma)$,
$\Omega\subset{\mathbb Z}^2$, with $\vert \Omega\vert = L^2$ and
where $K$ a kernel of the form~(\ref{eq:scale_form}) and $\epsilon(\bk) = {\mathcal N}(0,1)$ is iid,
with both $K$ and $\sigma(\bk)$ satisfying the necessary constraints to ensure that $\psi\sim {\rm FGF}(\gamma)$. In particular we have $\sigma(\bk) \propto 1/k^{\gamma/2}$.  
The model $\psi_\theta[\varphi]$ to be learned  is defined by
\begin{align}\label{eq:model}
\psi_\theta[\varphi](\bk) =& \sum_{\bk'\in\Omega} F_\theta(\bk,\bk')\varphi(\bk')\\
\text{with} \quad F_\theta(\bk,\bk') =& \sum_p^M f_p(\bk,\bk') \theta_p
\end{align}
where $\mathbf{\theta}$ is the parameter vector, assumed to be of dimension $M$.
In full generality,  $F_\theta(\bk,\bk')$ could in principle be a more complex architecture, with more parameters.
Here we are interested by the dynamics of the linear regime, where $f_p(\bk,\bk')$'s are independent from $\theta$.
We minimize the $L_2$ loss w.r.t. $\theta$:
\begin{align}\label{def:lossL2}
\LL(\theta) = \frac{1}{2} \E_{(\varphi,\psi)\sim \widehat{\rm FGF}(\beta,\gamma)}\left[\sum_{\bk\in\Omega} \bigl\vert \psi_\theta[\varphi](\bk) - \psi(\bk)\bigr\vert^2\right],
\end{align}
where $\widehat{\rm FGF}(\beta,\gamma)$ represents the empirical joint distribution of $\varphi$ and $\psi$ related by~(\ref{eq:data_map}).

Let us turn now to learning dynamics. For reasons which will become  more apparent when considering the scaling limits,
we introduce a learning rate scaling like $\eta/L^2$ with $\eta = \mathcal{O}(1)$ non-negative. The gradient dynamics then reads
\begin{align*}
\dot\theta = -\frac{\eta}{L^2}\frac{\partial \LL(\theta)}{\partial \theta} 
  &= -\frac{\eta}{L^2}\E_{(\varphi,\psi)\sim \widehat{\rm FGF}(\beta,\gamma)}\Bigl[\\[0.2cm]
  &\sum_{(\bk,\bk')\in\Omega^2} \mathbf{f}(\bk,\bk')\varphi(\bk')
\bigl( \psi_\theta[\varphi](\bk) - \psi(\bk)\bigr)\Bigr],
\end{align*}
with $\mathbf{f}(\bk,\bk')$ the vector of features  $f_p(\bk,\bk')$.

The linear dynamics is of the form
\begin{align*}
\eta^{-1} \dot\theta = - \widehat C\theta +\widehat Z
\end{align*}
with $\widehat C$  an $M\times M$ empirical matrix and $\widehat Z$ an $M$-dimensional empirical vector given by
\begin{align}
\widehat C_{pq} &= \frac{1}{L^2}\sum_{\bk,\bk',\bk''\in\Omega^3} f_p(\bk,\bk')f_q(\bk,\bk'')\nonumber \\[0.2cm]
&\qquad\qquad\qquad\times\E_{\varphi\sim {\rm FGF}(\beta)}\bigl[\varphi(\bk')\varphi(\bk'')\bigr] \label{eq:C_pq}\\[0.2cm]
\widehat Z_p &= \frac{1}{L^2}\sum_{\bk,\bk'\in\Omega^2} f_p(\bk,\bk') \E_{\varphi,\psi\sim \widehat{\rm FGF}(\beta,\gamma)}\bigl[\psi(\bk)\varphi(\bk') \bigr] \label{eq:Z_p}
\end{align}
leading to
\begin{align*}
\theta(t) = \widehat G(t)\widehat Z,
\end{align*}
with
\begin{align}\label{eq:time_propagator}
\widehat G(t) \egaldef {\widehat C}^{-1}\bigl(1-e^{-\eta\widehat C t}\bigr)
\end{align}
the propagator in time of the learning process. The difference with the usual linear regression setting, discussed in~\cite{advani2020high} is that we have now $\vert\Omega\vert = L^2$
regressions in parallel indexed by $\bk$ instead of a single scalar one, and the feature matrix is averaged over all these distinct regressions
by summing over $\bk$ in~(\ref{eq:C_pq},\ref{eq:Z_p}). Also as discussed in~\cite{advani2020high}, we see in~(\ref{eq:time_propagator}) that early stopping acts as a regularization, since at time $t$,
only modes with eigenvalues larger that $1/(\eta t)$ are learned. In our case this can lead to 
a strong spectral bias issue, when the modes of $\widehat{C}$ decay as a power law in $k$. Just notice
that the population matrix $C$ of $\widehat{C}$ corresponding to taking the limit $N\to\infty$ at fixed $M$ is actually given by~(\ref{eq:population_matrix}). This indicates
that in order to efficiently address the spectral bias issue, it may be sufficient to use a preconditioner that would orthonormalize the features w.r.t. the
inner product introduced above in~(\ref{def:inner_product}), leading to $C=\I$. 
A second issue is the extrapolation one. As previously stated we have $L^2$ regressions problems indexed by $\bk$ and for this we need to share information between
those regression problems. From the previous discussion on equivariance, the obvious way to proceed is to look for features in the form of smooth function of $\bk$
that diagonalize the kernel $K$, \ie functions  
having the functional form~(\ref{eq:scale_form}).
Hence our use of the Fourier-Mellin basis in the building of our FM Network to address the spectral flow task.

\subsection{Asymptotic limits of learning trajectories}\label{sec:Asymp_dynamics}
We are now interested to obtain learning trajectories in some meaningful asymptotic limits. From the previous discussion, it appears that 
the meaningful scaling parameter is given by 
\begin{align}\label{def:rho}
\rho = \frac{N N_s}{P}    
\end{align}
where $N$ is the number of independent samples, $N_s$ the number of independent scalar observable per sample and $P$ the number of parameters of the model. In the context of FGF regression,
the Fourier components are independent so in principle we should have $N_s =L^2$. Actually we only have $N_s = s^2L^2$ for the spectral flow experiment with $s<1$
since only the output modes $\psi(\bk)$ of the output with $\max(\vert k_x\vert,\vert k_y\vert) < s/2$ contain information. In the context of extrapolation scenario 
considered for the FGF experiments, some of the components are hidden during the train and the effective number of observations is further reduced as will
be detailed in Section~\ref{sec:Extrapolation_asymp}. In absence of inductive bias for the 
inference model, the feature space is supposed to span an operator space of dimension $L^4$, which means that $P = {\mathcal O}\bigl(L^4\bigr)$
in this case. 
If instead some prior knowledge is given, like scale and rotation (equ-)~invariance,
the operator could be considered in the diagonal form of some specific representation, like Fourier or Fourier-Mellin in our study cases, 
and then the number of parameters is reduced to $P = \mathcal{O}(L^2)$. This allows us to consider a meaningful asymptotic regime where 
$N = {\mathcal O}(1)$, enabling the analysis of global quantities such as training and test errors.
In such  case  we recover a dynamical behavior corresponding to the population limit when $N$ becomes large.  When considering instead local quantities like 
mode dependent train and test error, these should become deterministic when $N,P\to\infty$ in the proportional scaling regime $N/P$ fixed.
This will become more obvious in the following.

Let us look into this in more details. Call $\psi_{\theta_t}[\varphi]$ the model learned at learning time $t$.
First, the evolution of the training and test errors as functions of time  can be given in a general form in terms of the time propagator~(\ref{eq:time_propagator}). This leads to equations~(\ref{eq:E_train_t},\ref{eq:E_test_t}) given in Appendix~\ref{app:dyna_Errors}. 
when considering the  mean squared error (MSE) as the default metric, consistent with the loss function~(\ref{def:lossL2}).
These equations actually simplify significantly when considering instead the metric based
on the FGF energy~(\ref{def:Energy}), leading to the following definition  for 
the train and test errors as a function of $t$
\begin{align*}
E_{\rm train}(t) &= \E_{(\varphi,\psi)\sim \widehat{\rm FGF}(\beta,\gamma)}\left[\sum_{\bk\in\Omega} \vert\bk\vert^\gamma\bigl\vert \psi_{\theta_t}[\varphi](\bk) - \psi(\bk)\bigr\vert^2\right],\\[0.2cm]    
E_{\rm test}(t) &=  \E_{(\varphi,\psi)\sim {\rm FGF}(\beta,\gamma)}\left[\sum_{\bk\in\Omega} \vert\bk\vert^\gamma\bigl\vert \psi_{\theta_t}[\varphi](\bk) - \psi(\bk)\bigr\vert^2\right].   
\end{align*}
while modifying the loss~(\ref{def:lossL2}) accordingly. The corresponding gradient descent has also  the advantage to compensate for the spectral bias which originates from the power-law behavior of the spectrum and to simplify the learning equations as explained in Appendix~\ref{app:dyna_Errors},
so we will concentrate only on this formulation which corresponds to equations~(\ref{eq:E_train2},\ref{eq:E_test2}). 
While these equations can be dealt with in principle in full generality, once we are given the spectrum of the population matrix, 
for sake of simplicity (in particular to deal with the first term of~(\ref{eq:E_test2})) we focus on the case where $C=\I$. In that case, using random matrix theory we eventually arrive at closed form expressions for the train and test errors as function of time (see Appendix~\ref{app:RMT}). The way to proceed is quite standard (see e.g.~\cite{furtlehner2023free,cataniatheoretical}  and refs. herein respectively in the context of scalar regression and energy based models). For this we need first to introduce the mean  spectral  density $\bar\nu$ of $\widehat C$, which for $C=\I$ is the ordinary Marchenko-Pastur distribution:
\begin{align*}
\bar\nu(x) = (1-\rho)\ind{\rho<1}\delta(x)+\frac{\rho}{2\pi x}\sqrt{(x-x_-)(x_+-x)},
\end{align*}
with 
\begin{align*}
x_\pm = \Bigl(1\pm\frac{1}{\sqrt{\rho}}\Bigr)^2.
\end{align*}
Then we need to introduce two quantities depending on the decomposition of
\begin{align*}
K = K^\parallel+K^\perp
\end{align*}
involved in~(\ref{eq:model}), corresponding respectively to the component $K^\parallel$ on the feature space and its orthogonal space.
Using the inner product~(\ref{def:inner_product}) defined above for operators we call
$\kappa$ the variance of the part of the signal captured by our feature space, 
\begin{align*}
\kappa \egaldef \langle K^\parallel,K^\parallel\rangle_{\beta,\gamma}
\end{align*}
 and $\sigma_{\rm eff}^2$ the effective variance of the noise
which emerges from the fact that the regression is misspecified. By definition $K^\perp$ cannot be recovered from the features and adds  an additional  contribution to the original noise entering in the definition of the data model~(\ref{eq:data_map})
\begin{align}\label{eq:effective_noise}
\sigma_{\rm eff}^2 \egaldef \sum_{k\in\Omega} k^\gamma\sigma^2(k) + \langle K^\perp,K^\perp\rangle_{\beta,\gamma}.
\end{align}
This noise variance can actually be decomposed component-wise onto the modes as
\begin{align*}
\sigma_{\rm eff}^2(k) = k^\gamma\sigma^2(k) +  \frac{1}{L^2}\sum_{\bk'\in\Omega}\frac{k^\gamma}{{k'^\beta}} K^\perp(\bk,\bk')^2.
\end{align*}
Assuming without loss of generality $\langle K,K\rangle_{\beta,\gamma} = 1$, we have $\langle K^\perp,K^\perp\rangle_{\beta,\gamma} = 1-\kappa$.

These two quantities fully determine the learning dynamics at large scale
for the train, and test errors. Thanks to the preconditioning corresponding to the input and output field scaling behavior,
the $k$ dependency in equations~(\ref{eq:E_train_t},\ref{eq:E_test_t}) disappears and we are eventually left with the following simplified expression 
for the train and test error in the asymptotic limit (see Appendix~\ref{app:RMT}):
\begin{align}
E_{\rm train}(t) &= \kappa\int_0^\infty \bar\nu(dy) y\bigl[1-yj(y,t)\bigr]^2 \nonumber\\[0.2cm]
&+\sigma_{\rm eff}^2\Bigl[1-\frac{1}{\rho}+\frac{1}{\rho}\int\bar\nu(dy)\bigl[1-yj(y,t)\bigr]^2\Bigr] 
\label{eq:E_train_asymp} \\[0.2cm]
E_{\rm test}(t) &= \kappa\int_0^\infty \bar\nu(dy)\bigl[1-yj(y,t)\bigr]^2 \nonumber\\[0.2cm]
&+\sigma_{\rm eff}^2\Bigl[1+\frac{1}{\rho}\int dy \bar \nu(dy) y j(y,t)^2\Bigr]\label{eq:E_test_asymp}
\end{align}
where 
\begin{align*}
j(x,t) = \frac{1-e^{-xt}}{x}
\end{align*}
with learning rate $\eta=1$ by convention,
is the function controlling the dynamics of $\theta$ in equation~(\ref{eq:time_propagator}).
As shown in the Appendix~\ref{app:Train-Test}, this simplified case actually corresponds to make use 
of a preconditioner  adapted to the spectrum of the output.

\subsection{The case of FGF spectral flow in the extrapolation regime}\label{sec:Extrapolation_asymp}
Let us consider now what happens in the extrapolation setting of the spectral flow experiment considered in Section~\ref{sec:spectral_flow_exp}.
The range of scales on the square lattice is $k_{x,y}\in [k_{\rm min},k_{\rm max}]$ with $k_{\rm min} = \frac{1}{L}$ and $k_{\rm max} = 0.5$.
When considering the rescaling, which maps the component $\bk$ to $s\bk$ with $s<1$, all the components $\psi(\bk)$ of the output, 
situated in the domain $\max(\vert k_x\vert,\vert k_y\vert)> sk_{\rm max}$ are randomly generated, which corresponds to inserting the amount of 
noise~(\ref{eq:effective_noise})
\begin{align}\label{eq:sigma_s}
\sigma_{\rm eff}^2 = 1-s^2\qquad\qquad (s<1)
\end{align}
into the output field, when this one is assumed to be normalized to one.  In addition, in the extrapolation regime, let us call $k_{\rm occ}$ the occultation scale,
which for the super-resolution experiment corresponds to hide the components $\bk$ in the domain $\max(\vert k_x\vert,\vert k_y\vert)> k_{\rm occ}$
in the fields, while the large scale extrapolation experiments corresponds to hide components 
in the domain $\max(\vert k_x\vert,\vert k_y\vert) < k_{\rm occ}$. The set of meaningful observation (in the training set) of the output field 
in the super-resolution case then corresponds to $\max(\vert k_x\vert,\vert k_y\vert) \in ]k_{\rm occ},sk_{\rm max}]$, others components either corresponding to noise or being hidden.
Instead in the large scale extrapolation the meaningful observations correspond to $\max(\vert k_x\vert,\vert k_y\vert) > sk_{\rm occ}$ which justifies the definition~(\ref{def:f_extrapol})
of the extrapolation factor given in Section~\ref{sec:FGF}. To see the effect of extrapolation on the performance of the prediction  
we introduce also the ratio 
\[
r \egaldef \frac{k_{\rm occ}}{k_{\rm max}}
\]
which is more convenient.
As a function of $r$ and $s$ the equations~(\ref{eq:E_train_asymp},\ref{eq:E_test_asymp})
remain unchanged except for the coefficient $\sigma_{\rm eff}(s)$ given by~(\ref{eq:sigma_s}) and the samples/parameters ratio~(\ref{def:rho})
of the global feature matrix
\[
\rho(r,s) = \rho_0\times
\begin{cases}
\DD    (s^2- r^2) \qquad\text{(large scale extrapolation)} \\[0.4cm]
\DD    (sr)^2 \qquad\qquad\text{(super-resolution)}
\end{cases}
\]
where $\rho_0$ is a default ratio. As we see, for the large scale extrapolation there is no information to process when $r\ge s$
with a linear decay close to this point while for super-resolution the information vanishes quadratically with $r$.

\begin{figure}[t]
    \includegraphics[width=\columnwidth]{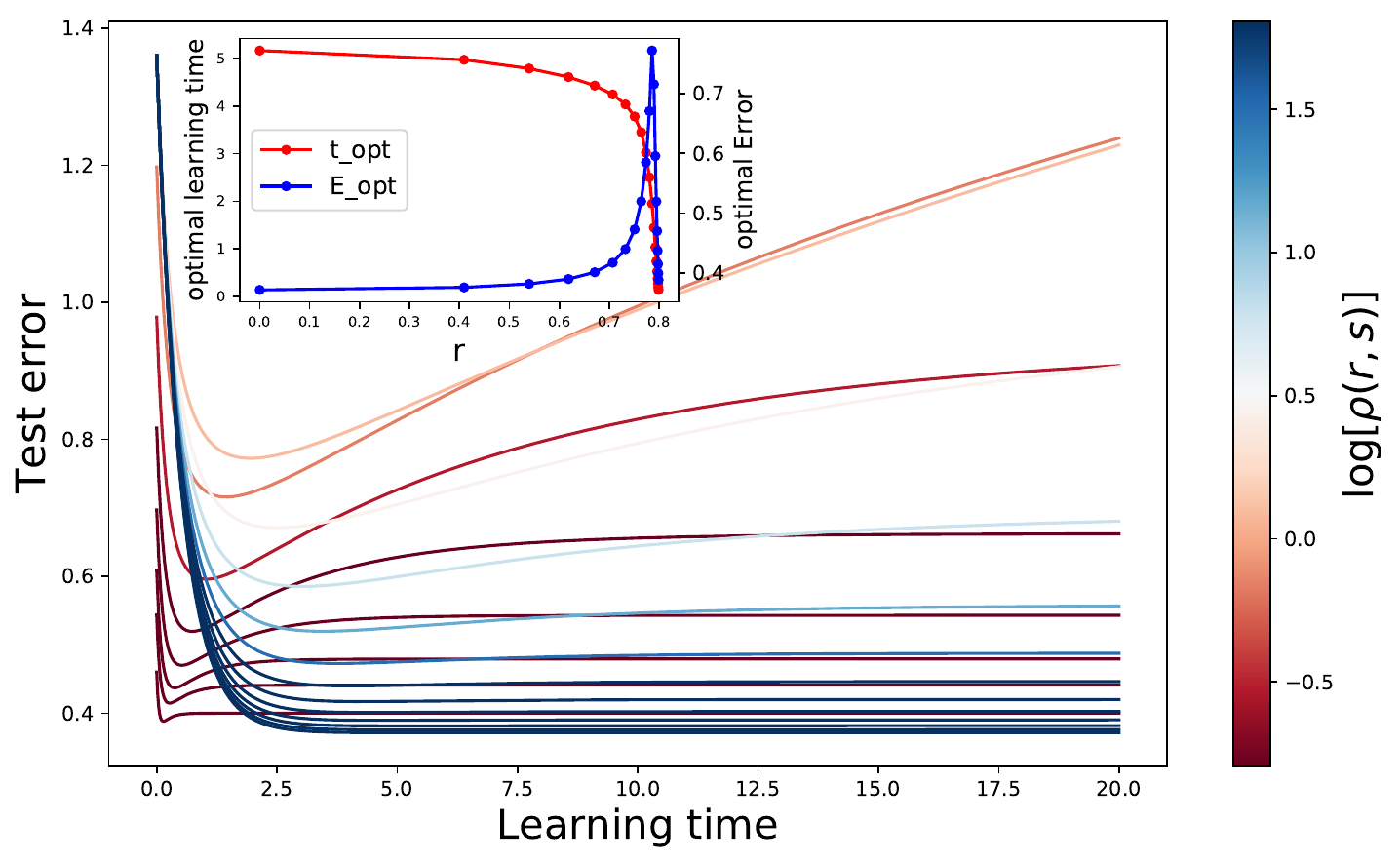}
    \caption{Behavior of the test error measured when varying the level of extrapolation $r$ for $s=0.8$ and $\kappa=1$ in the large scale extrapolation setting.
      The inset shows how the optimal stopping point and the corresponding test error behave with $r$. The spike of $E_{\rm opt}$ corresponds to the value of $r$ corresponding $\rho(r,s)=1$
      \ie the interpolation point separating the under $(\rho >1)$ from the over-parameterized regime.\label{fig:extrapol_theo} }
\end{figure}
The expected behavior of the test error in absence of any spectral bias of the features, \ie corresponding to $C=\I$ is displayed in Figure~\ref{fig:extrapol_theo}
for the large scale extrapolation setting (the super-resolution case being similar).
Note that this curves are generic and basically assume that there is some information, \emph{not local in scale}, which couples the various regressions of $\psi(\bk)$ to be done at different scales. 
Remarkably this tells us that in principle for sufficiently large input and output dimension ($L\gg 1$), inference in the extreme extrapolation regime ($r$ close to $s$
in the large scale extrapolation task or close to zero in the super-resolution task) should be achievable with only a few training instances ($N$ of the order of one).

In practice, in our experiments we do not find us exactly in this idealized situation. For the phase mixing problem, the phase shift function $g_\theta(\mathbf k)$  
to be learned in~(\ref{eq:learned_fn}) is not smooth enough to be decomposable into a few basic (Fourier modes in our case) delocalized functions of $\bk$. 
Concerning the FGF spectral flow process,  which in principle should be closer to the ideal case, it is still displaying spectral biases issues resulting from
finite size effect on the square lattice to the log-polar lattice interpolation. Our current implementation of the Fourier-Mellin transform is computationally prohibitive,
primarily due to the Dirichlet kernel and its pseudo-inverse, which represent fully connected layers. Developing a sparse implementation of these layers would enable us
to scale the model sufficiently to observe the behaviors
described on Figure~\ref{fig:extrapol_theo}.


\section{\label{sec:conclusions}Conclusion and Perspectives}

In this work we have investigated how scale invariance can be leveraged to perform extrapolation across scales on two study cases of scale invariant processes: linear FGF processes and the ASM.
At odds with most of the existing literature on this subject, we consider a setting in which scale invariance is realized at a probabilistic level, meaning that we look
for equivariance properties~(\ref{def:stat_equi}) of the architecture at the level of the probability distributions of the input and output fields. 

The two examples we consider represent two extreme cases in terms of difficulty. On the one hand, for the FGF we consider the two possible types of linear elementary processes leaving
the FGF distribution invariant: a phase shift, and a spectral flow corresponding to rescaling and rotation of the input in momentum space. On the other hand the ASM is a non-linear
map with input and output fields (height map and avalanche cluster) pertaining to very different kinds
of scale-free fields. The idea of considering FGF processes is to identify basic constraints that the neural operator should obey.
For each FGF process we develop new architectures -- the Fourier embedding and Fourier-Mellin network respectively -- which are adapted to each of these elementary transformations.
This opens the possibility to design more complex architectures by combining these building blocks, in order to address more complex inference tasks.
Already we see that these elementary tasks confound sophisticated models like the U-Net or the Riesz network. They suffer either from very strong spectral bias in the phase mixing task,
either fail entirely for the spectral flow task, not to mention the corresponding extrapolation tasks. For the FGF, while the basic tasks look straightforward,
they are not in practice and our work provides a proof of concept demonstrating that the proposed elementary building blocks are effective for extrapolation.
These building blocks could be used as basic layers of more complex deep models, in combination with non-linear activations like \eg ReLU
which would preserve the scale symmetry, in some way generalizing the Riesz network. This is left for further investigations.

From the theoretical analysis of Section~\ref{sec:theory}, we reach two important points. First it is clearly seen that a pre-conditioner for the gradient descent is mandatory to avoid filtering the small scale information. The relevant preconditioner can be directly deduced from the knowledge of the statistical properties of the input and output fields,  which can be either considered to be known in advance, as we do for the FGF, either estimated from the data. Second, imposing equivariance with respect to scale lets us reduce the number of features by a factor $L^2$, as discussed in Section~\ref{sec:Asymp_dynamics}, and ensures the possibility of extrapolation, in principle with only a few examples as stated in Section~\ref{sec:Asymptotics}. 

For the ASM instead the question of extrapolation remains open. Naive scale invariance alone (either Riesz or our GNN) is clearly not the solution and we did not yet attempt to test scale equivariant models, as we neither expect these to be sufficient to work. Some additional sophistication obviously needs to be added.
Indeed, we believe that the model needs to learn an equivariant latent representation analogous to what the spatial RG might deliver manually,
in order to be able to decode  across scales the output field  in a top-down auto-regressive manner, somewhat similar in spirit to  
what has been done successfully in~\cite{marchand2023multiscale} in the context of generative models.
In this respect the ASM is particularly complex as the coarse graining operation is highly non-trivial,  as shown in~\cite{vespignani1995renormalization}. By contrast, in the FGF tasks considered here, the coarse graining step is directly given by the anomalous scaling of the field and does not involve any latent representation. Future investigations should concentrate on problems of intermediate difficulty between these two, like for instance percolation problems where the coarse graining needs to be learned but is less involved than for ASM. Additionally a pragmatic strategy should consider first to learn the coarse grained latent representation in a supervised manner as a starting point, by learning directly pre-defined spatial RG flows.

\begin{acknowledgments}
    Authors thank Misaki Ozawa and Sergio Chibbaro for insightful discussions and acknowledge financial support by the French ANR grant Scalp (ANR-24-CE23-1320).
\end{acknowledgments}

\section*{References}
\bibliography{apssamp.bib}

\newpage

\appendix


\section{Model details}\label{app:model_details}

In this appendix we give additional details on some of the various architectures used throughout this work.
In particular, we include the derivation of the Dirichlet kernel method employed by the FM network to perform the Fourier-Mellin transform.

\subsection{U-Net}\label{app:unet}

U-nets are a type of CNN architecture created for image segmentation, which is the task of highlighting or extracting an object of interest from an image.

\begin{figure*}
\includegraphics[width=1.8\columnwidth]{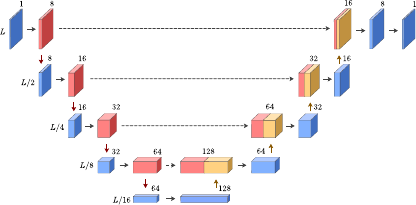}
\caption{\label{fig:unet}Diagram of the U-Net architecture used in all experiments. The dark faces of the volumes corresponds to the image or coarse-grained image sizes,
while the width represents the number of channels at each step. The right arrows correspond to convolutional layers, while the downwards and upwards ones are pooling and inverse pooling operations.
The dashed arrows indicate the skip connections, where the representation of the data at a given scale is reused later on during the upscaling process.}
\end{figure*}

These networks consist of a series of “downwards” layers followed by an equal number of “upwards” layers, forming in this way a hierarchical structure in which a representation at each scale is constructed from the previous one. This multiscale structure makes the U-Net a good baseline for the datasets treated in this work. They do not employ however any sort of scale invariance, as the weights learned at each scale are independent from each other. A diagram of our instantiation of this family of architectures is shown in Fig.~\ref{fig:unet}.

The downwards layers apply a convolution to the input, then they coarse-grain it via a pooling operation, reducing the spatial dimensions by half.
One of the key differences between the U-net and usual CNNs is that U-net records the features learned at each downwards step, to re-use that information later, in the upscaling steps.

The upwards layers perform an upscaling of the input, where a transposed convolution is applied as shown in figure \ref{fig:unet} in order to double the spatial size,
and this is convoluted with the output of the downwards layer of the same size, to aggregate the information corresponding to that scale.

\subsection{Riesz network}

\begin{figure}
    \includegraphics[width=\columnwidth]{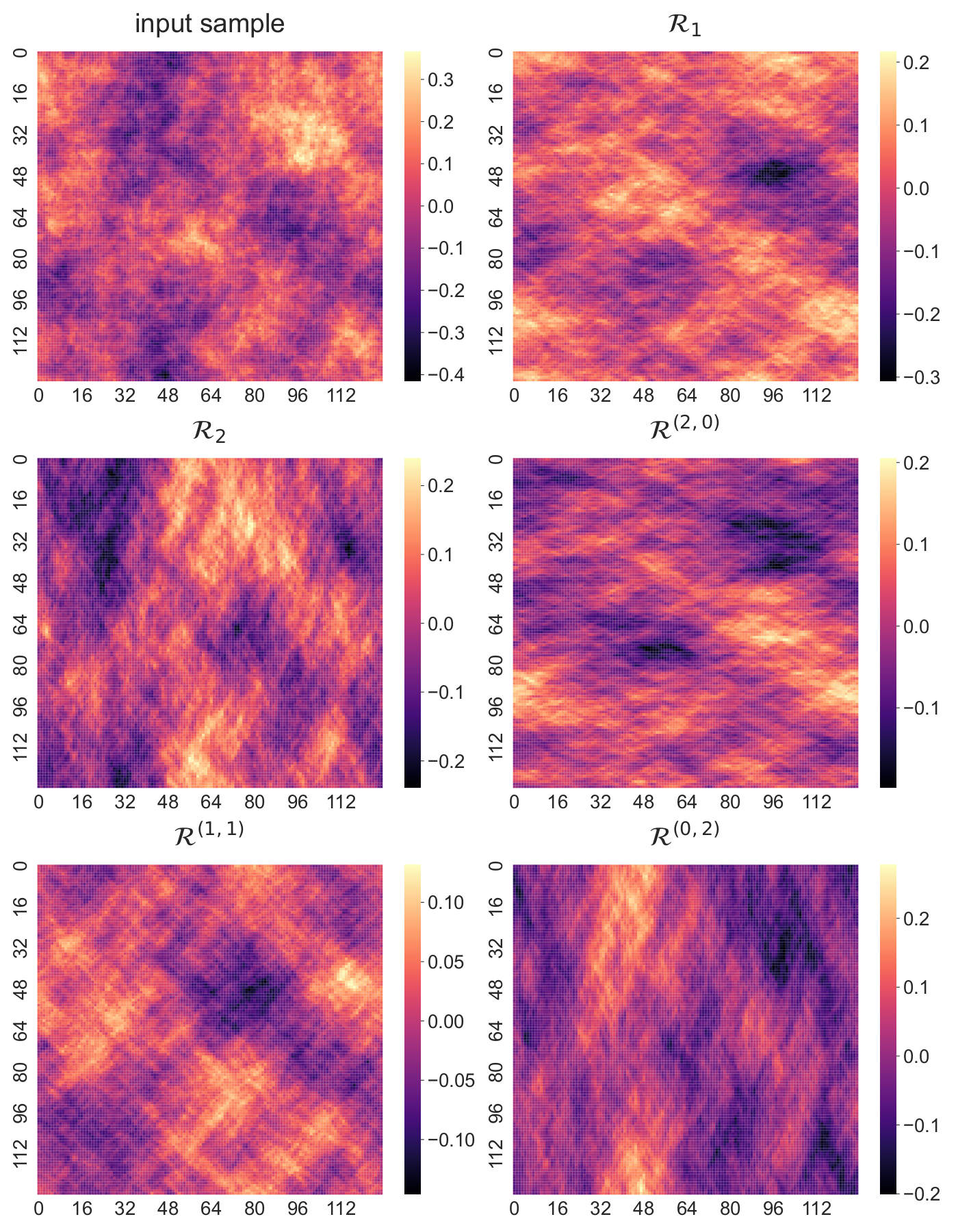}
    \caption{\label{fig:riesz_filters}
    Riesz transforms up to second order applied to a sample of the FGF.
    The orientation corresponding to each filter is clearly visible. 
    }
\end{figure}

The Riesz network is designed to be scale invariant due to its use of scale invariant filters. In Fig.~\ref{fig:riesz_filters} we show the five Riesz filters (up to order 2)
that we use in practice to build the network, applied to a sample of the FGF.

The shortcoming of this network is the limited expressivity, since the filters are fixed, and the network learns only their linear combinations. Two simple ways of increasing the expressivity are to add more layers and more channels at each layer, or to include higher order filters. However, both these options are computationally costly, due to the need to employ Fourier transforms at each step in order to apply the filters.

\subsection{FM network}
Here we derive the expression of the Dirichlet kernel, which we use to perform the interpolation of the FGF data from a cartesian lattice to logpolar coordinates.

Suppose we have a FGF sample on a square lattice of size $L$, denoted by $\{ \varphi(\mathbf r_i); \mathbf r_i\in \Omega_L \}$. We take the convention $\mathbf{r}_i=(x_j,y_j)$,
\[
x_j = -\frac{1}{2} -\frac{1}{2L} + \frac jL;\qquad 
y_j = -\frac{1}{2} -\frac{1}{2L} + \frac jL
\]
The discrete Fourier transform is given by
\[
\varphi(\mathbf k_j) = \frac1L \sum_{\mathbf x_i\in\Omega_L}
\varphi(\mathbf r_i) e^{i2\pi \mathbf k_j \cdot \mathbf r_i},
\]
with $\mathbf k_j\in \Omega_L$ the reciprocal square lattice. 

We consider the analytic extension of the discrete Fourier transform, given by
\[
\widetilde\varphi(\mathbf k) = \frac1L \sum_{\mathbf r_i\in\Omega_L}
\varphi(\mathbf r_i) e^{i2\pi \mathbf k \cdot \mathbf r_i}
.
\]
To perform the interpolation between the square and logpolar lattice, we evaluate this field at the points of the logpolar lattice of size $N$, which we will denote $\mathbf k'\in\Omega'_{N}$.
\[
\widetilde\varphi(\mathbf k') = \frac1L \sum_{\mathbf r_i\in\Omega_L}
\varphi(\mathbf r_i) e^{i2\pi \mathbf k_j' \cdot \mathbf r_i}
.
\]
We then write $\varphi(\mathbf r_i)$ in terms of its Fourier coefficients on the square lattice
\[
\widetilde\varphi(\mathbf k') = \frac{1}{L^2} \sum_{\mathbf r_i\in\Omega_L} \sum_{\mathbf k\in\Omega_L}
\widetilde\varphi(\mathbf k) e^{i2\pi (\mathbf k'-\mathbf k) \cdot \mathbf r_i}
\]
and define the Dirichlet kernel $D$ as
\begin{align}\label{eq:Dirichlet_interpolation}
\widetilde\varphi(\mathbf k') = \sum_{\mathbf k\in\Omega_L}D(\mathbf k'-\mathbf k)
\widetilde\varphi(\mathbf k),
\end{align}
so we have
\[
D(\mathbf k'-\mathbf k) = \frac{1}{L^2} \sum_{\mathbf r_i\in\Omega_L} e^{i2\pi (\mathbf k'-\mathbf k) \cdot \mathbf r_i}
\]
We can compute the geometric sum, using the convention for the values of $\mathbf r_i$
\[
D(\mathbf q) = D_\mathrm{1d}(q_x)D_\mathrm{1d}(q_y),
\]
with the 1d Dirichlet kernel $D_\mathrm{1d}$ defined as
\[
D_\mathrm{1d}(q) = \begin{cases}
\frac{1}{L} \frac{\sin (\pi L q)}{\sin(\pi q)} & q \ne 0 \\
1  & q = 0
\end{cases}
\]

The Dirichlet kernel interpolation also allows us to build the rescaling of the FGF samples. 
The frequencies of the rescaled field are defined by
\begin{align}
\varphi_s(\mathbf k) = s^{1+\beta/2} \varphi( s\mathbf k).
\label{eq:rescaledFourierEffective}
\end{align}
Concretely, we apply equation~\ref{eq:Dirichlet_interpolation} to interpolate between the original cartesian grid, and another cartesian grid rescaled by a factor $s$ and rotated by an angle $\alpha$. 
The high frequencies, corresponding to $s \mathbf k$ larger than the maximum $\mathbf k$ on the lattice $k_{\rm max}$, are then regenerated by drawing from the FGF distribution.

\section{Experiment details}\label{app:experiment_details}

In this section we detail the experimental setup for both FGF tasks and for the ASM experiments, including details on all the models used and some further results.

\subsection{FGF: phase mixing task}

\paragraph{U-Net.} 

We use a 4-layer U-Net with an initial convolutional layer that outputs 8 channels, and a channel distribution of $8\rightarrow16\rightarrow32\rightarrow64\rightarrow128$ for the downwards and upwards layers. All convolutions use $3\times 3$ complex-valued filters. A final convolution outputs one single channel from the 8 feature channels at the previous layer. The model is trained for 500 epochs with a learning rate of $10^{-3}$. 

\paragraph{Riesz network.} 

We use a 4-layer Riesz network with channel distribution of $1\rightarrow32\rightarrow40\rightarrow48\rightarrow1$. The model is trained for 50 epochs with a learning rate of $10^{-3}$. Note that this model is trained for a shorter time due to the higher computational cost. 

\begin{figure}[t]
    \includegraphics[width=.75\columnwidth]{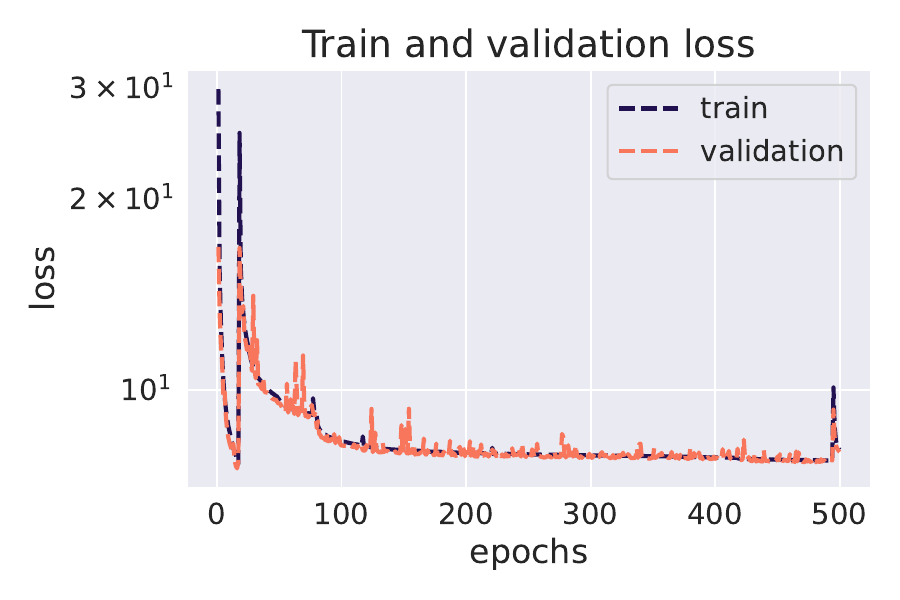}
    \includegraphics[width=.75\columnwidth]{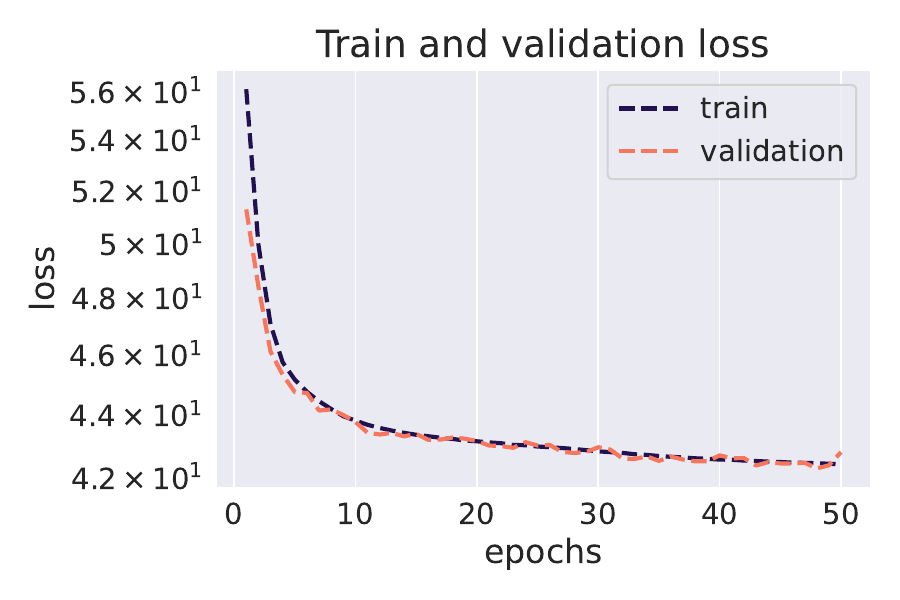}
    \includegraphics[width=.75\columnwidth]{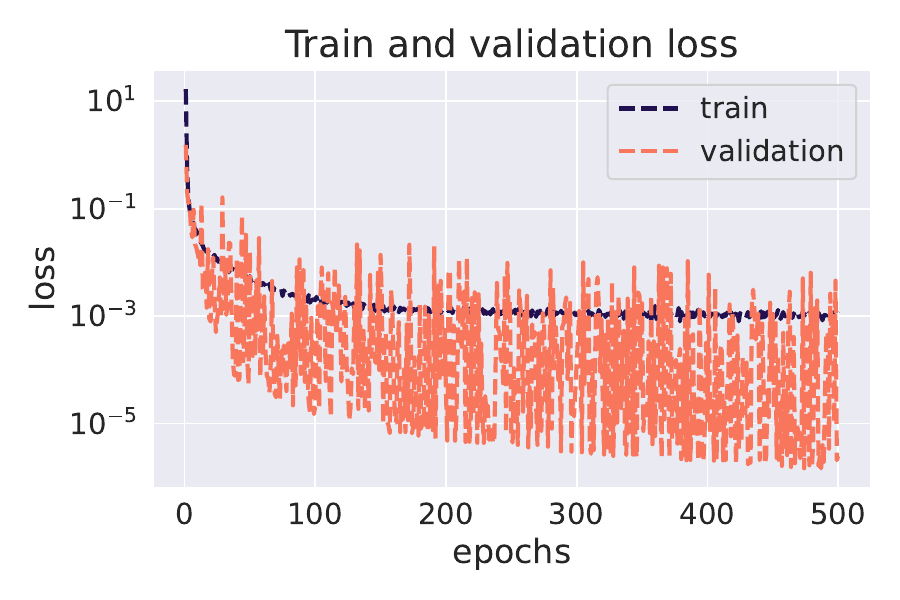}
    \caption{\label{fig:loss_phase_mixing} Train and validation loss for the U-Net, Riesz network, and Fourier embedding network respectively in the FGF phase mixing task, with no extrapolation.}
\end{figure}

\begin{figure}[t]
    \includegraphics[width=\columnwidth]{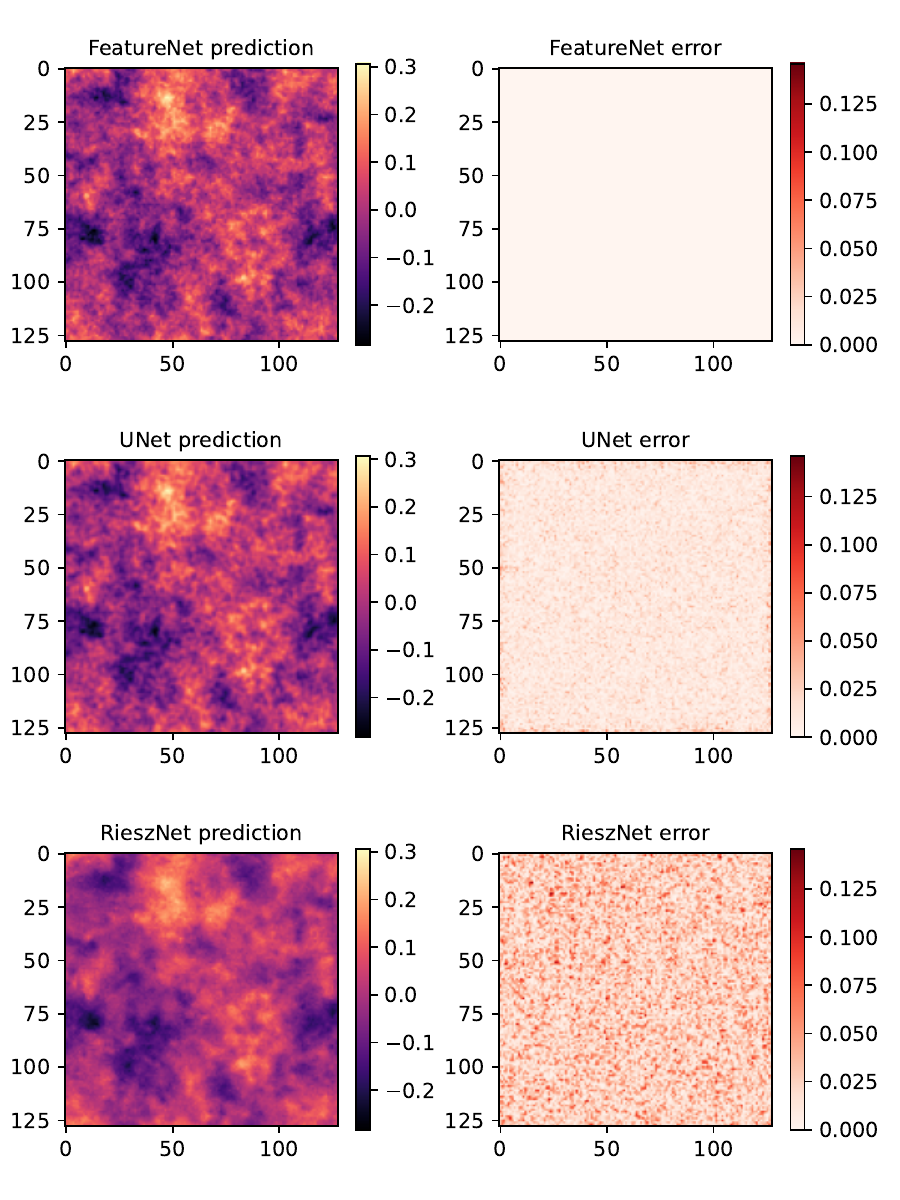}
    \caption{\label{fig:fgf_sample_outputs} Sample output of the Riesz network, U-net and Fourier embedding network on the FGF.}
\end{figure}

In Fig.~\ref{fig:loss_phase_mixing} we show the train and validation loss for each of the three models. 
Fig.~\ref{fig:fgf_sample_outputs} shows a sample output of the three models and the corresponding absolute error. 
While the output of the Fourier embedding matches the target, there is a visible loss of the high-resolution details for both the Riesz network and the U-net, which matches the quantitative result stated in the main text, that they fail to overcome the spectral bias problem. 

\subsection{FGF: spectral flow task}

\begin{figure}[t]
    \includegraphics[width=.75\columnwidth]{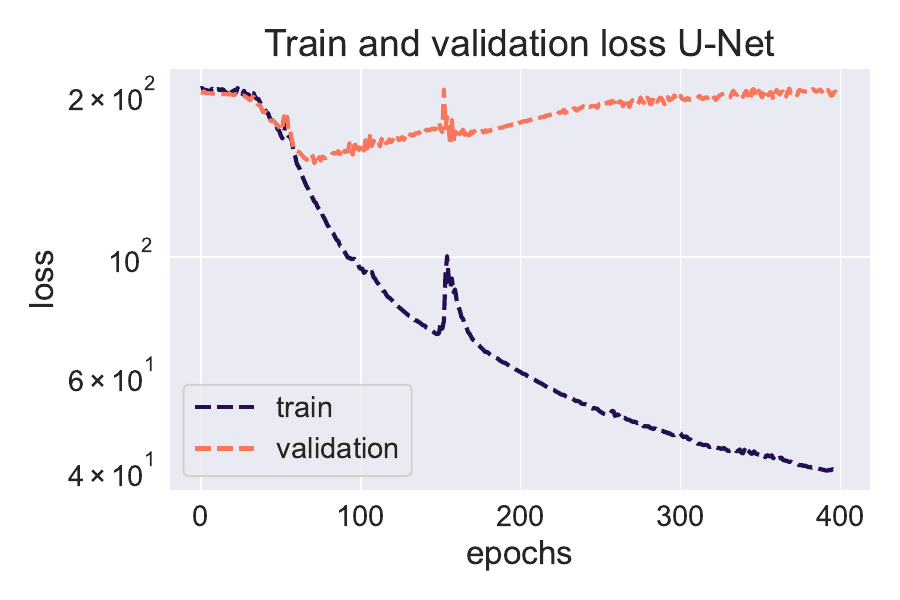}
    \includegraphics[width=.75\columnwidth]{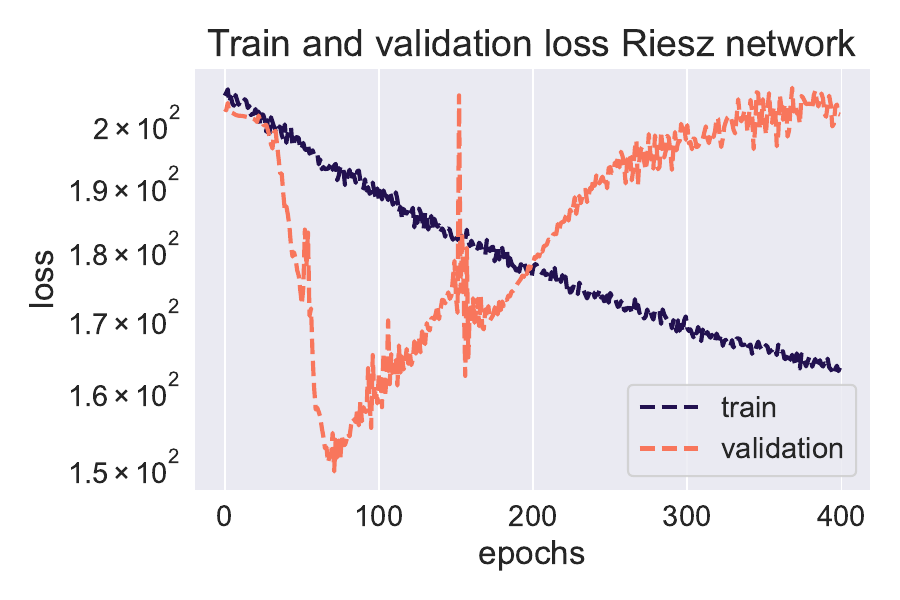}
    \includegraphics[width=.75\columnwidth]{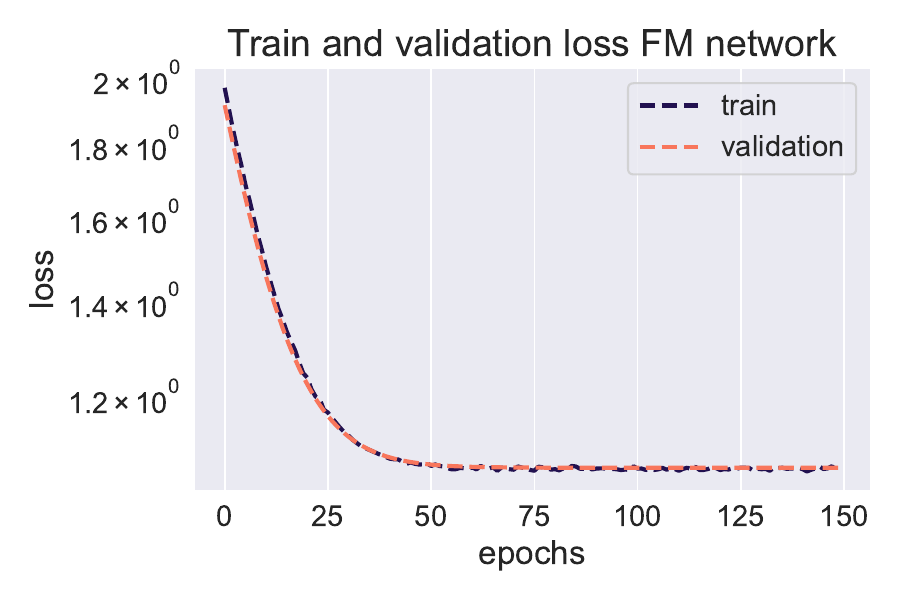}
    \caption{\label{fig:loss_rescaling} Train and validation loss for the U-Net, Riesz network, and FM network in the FGF spectral flow task, with no extrapolation. Note that the loss is not normalized, and the first two correspond to the MSE loss in direct space, while the third one is the MSE loss in Fourier-Mellin space.}
\end{figure}

For the spectral flow task, we trained a U-Net and a Riesz network with the same parameters as the one used for the phase mixing task. Note that although in this case the dataset consists of only 250 samples, this is still enough to learn, as each sample contains $64\times 64$ pixels. In Fig.~\ref{fig:loss_rescaling} we show the train and validation loss for the three models (U-Net, Riesz network and FM network). We observe severe overfitting for the first two models.
Even with our use of early stopping, the validation error is still very important.

\begin{figure}[h]
    \includegraphics[width=\columnwidth]{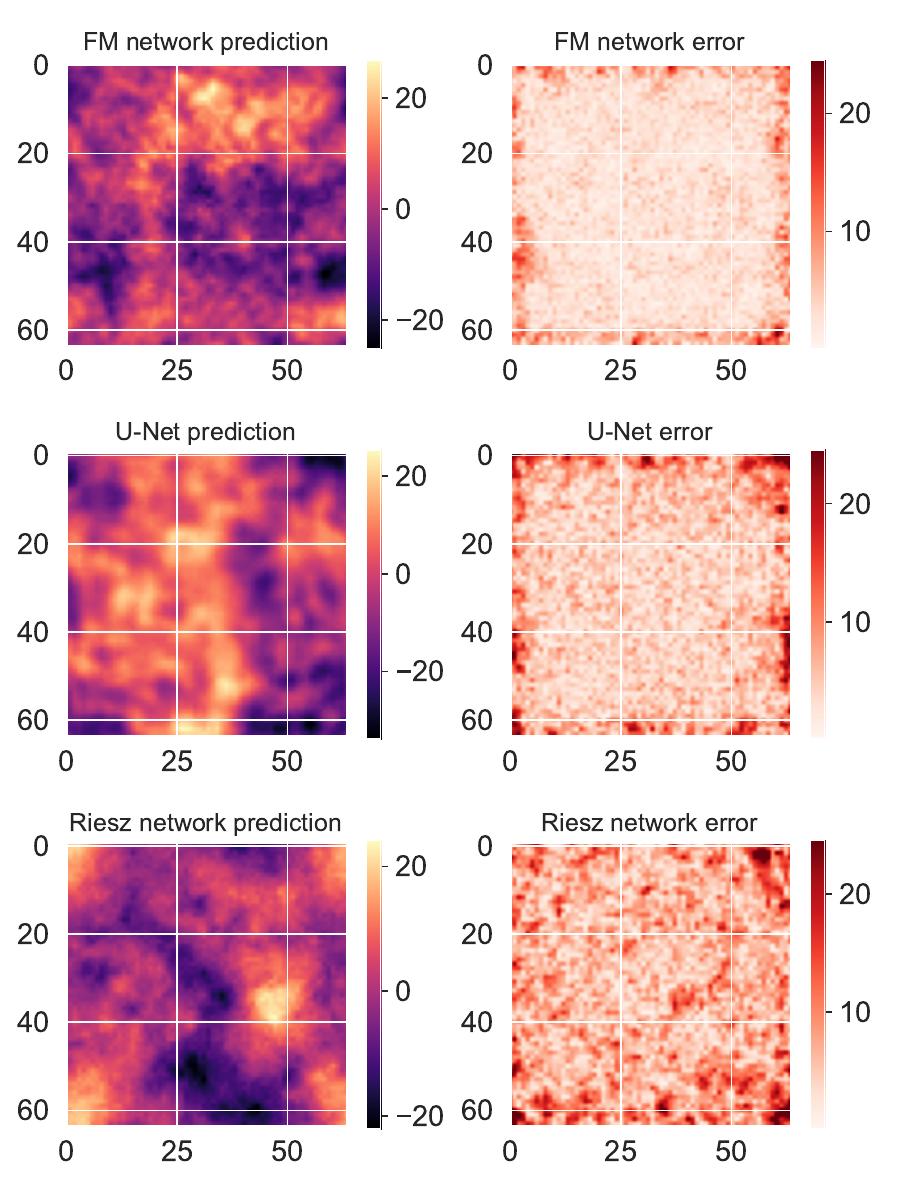}
    \caption{\label{fig:fgf_sample_outputs_rescaling} Sample output of the Riesz network, U-net and FM network on the FGF spectral flow task.
    The absolute error with respect to the ground truth is shown with equal color scale between the 3 models to allow for comparison.
    }
\end{figure}

Fig.~\ref{fig:fgf_sample_outputs_rescaling} shows, for the three models, a sample output along with its absolute error. 
The Fourier-Mellin (FM) has the best performance out of the three, with the difference being concentrated around the edges and at the smaller scales, while the U-Net and Riesz network only manage to capture some of large scale structure.

\begin{figure}[t]
    \includegraphics[width=\columnwidth]{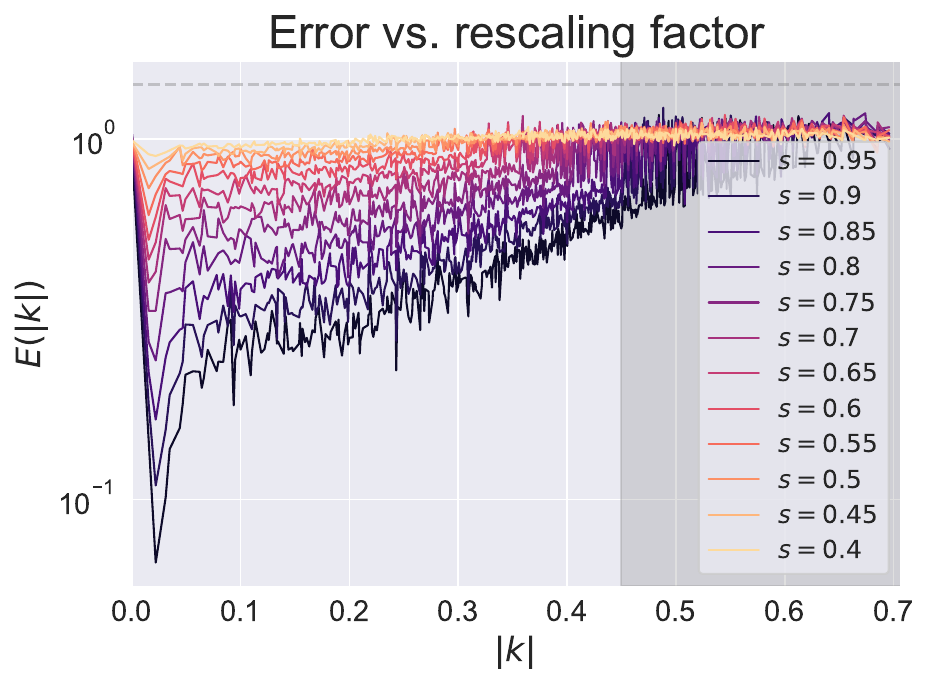}
    \caption{\label{fig:error_vs_rescaling} 
    Normalized Mean Squared Error $E(|\mathbf k|)$ (equation~(\ref{eq:normalizedPS}))
    for the FM network on the spectral flow FGF task, varying the rescaling factor between $s=0.4$ and $s=0.95$, and a rotation angle of $\pi/4$. In grey is the region corresponding to the noise regime for a scaling factor of $s=0.9$ (the one used in the experiments shown in the main text).
    This region extends to the left as $s$ decreases.
    The white line at $E(|\bk|)=1=10^0$ corresponds to the constant predictor $\widehat \psi=0$. 
    }
\end{figure}

Finally, in Fig.~\ref{fig:error_vs_rescaling}, for the FM network,  we consider the 
normalized Mean Squared Error $E(|\mathbf k|)$ (equation~(\ref{eq:normalizedPS}))
 in Fourier space  
as we vary the rescaling factor. As expected, we observe a decrease in performance as the rescaling becomes stronger (smaller $s$). 
Note that the noise region becomes larger as the rescaling factor becomes smaller, decreasing the effective amount of data available to the model in the learning process, and extending the region $k\in [sk_{\rm max}, k_{\rm max}]$ where it is impossible to perform better than random (since at these frequencies the ground truth target is generated and independent from the input).

\subsection{ASM}

For the avalanche prediction task, we used a U-Net and a Riesz network with the same architecture as for the FGF. Here we present some additional results of the extrapolation experiments. To evaluate the performance of the models, we look at the (per-pixel) balanced accuracy, and at the $\mathrm{IoU}$, defined as
\begin{equation}
    \mathrm{IoU} = \frac{\mathrm{TP}}{\mathrm{TP}+\mathrm{FN}+ \mathrm{FP}}.
\end{equation}
This metric is stricter than the balanced accuracy, which is relatively forgiving to predicting too large events, since False Positives are diluted into the large number of Negatives.

\begin{figure}[h]
    \includegraphics[width=\columnwidth]{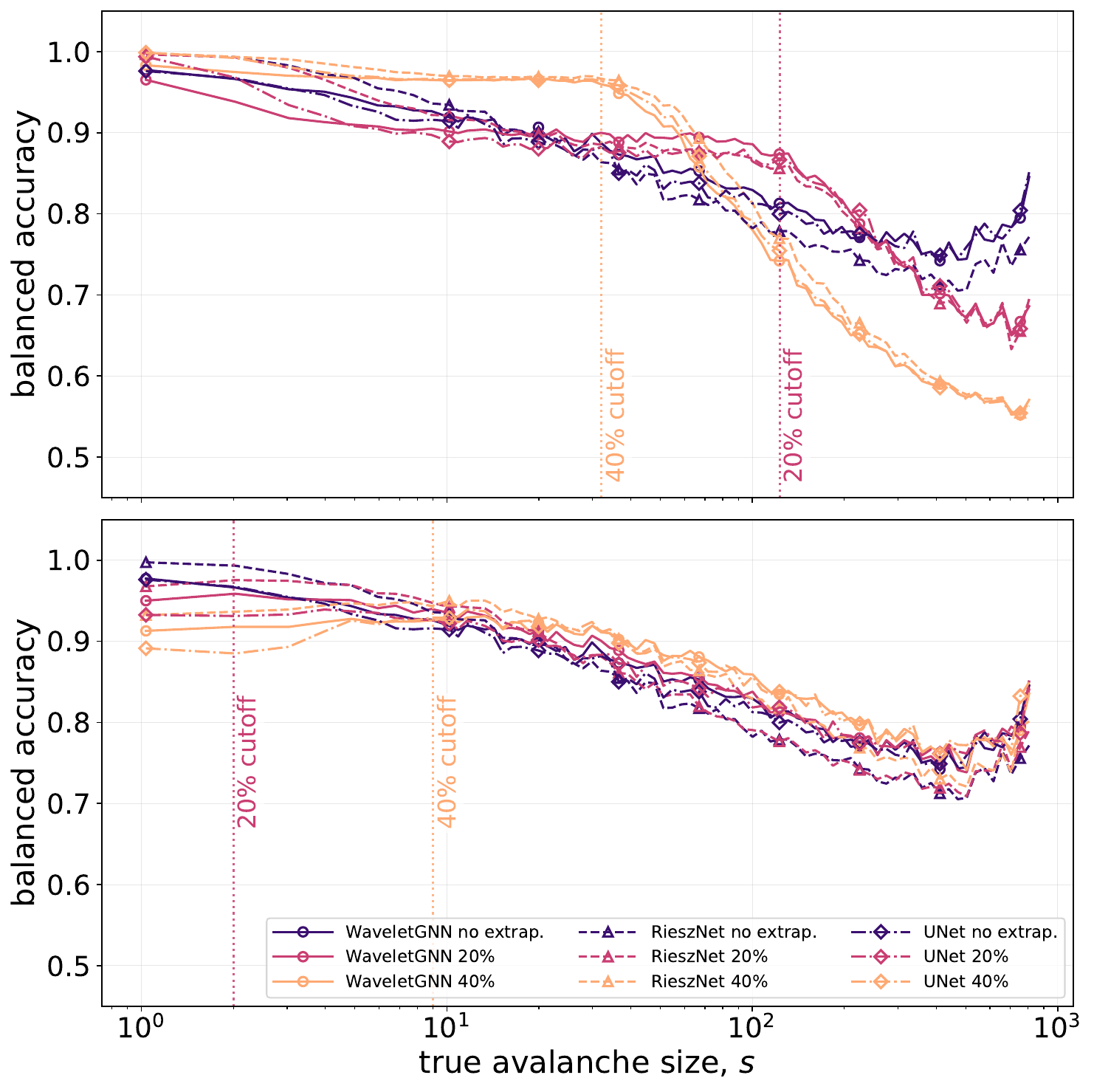}
    \caption{\label{fig:asm_balanced_accuracy} 
    Balanced accuracy as a function of avalanche size in the ASM large-scale extrapolation experiments \textbf{(top)}, and small-scale extrapolation experiments \textbf{(bottom)}. For each of the three models we show the balanced accuracy on a test set containing all avalanche sizes. We show three extrapolation regimes marked by the three different colors: the baseline model trained on the full dataset, and two extrapolation cases where we cut $20\%$ and $40\%$ of the largest/smallest avalanches.
    }
\end{figure}

In Fig~\ref{fig:asm_balanced_accuracy} we show the balanced accuracy as a function of avalanche size,   for all the extrapolation experiments. 
In the baseline case, results are very good, especially for the smaller sizes, while remaining above $0.7$ even for large avalanches. Note that the baseline for this metric is $0.5$, which would be the result for trivial uniform or random predictors.

In the large-scale extrapolation cases however,  the performance drops, for all models, in their extrapolation regime. The drop is clearly situated right after the largest available size for training.
In the small-scale extrapolation experiments, the drop is less visible, first because the absolute value remins high, and because the sheer span of events filtered out is smaller (even at $40\%$, events of size $s\geq10$ are available to train on).

\begin{figure}[h]
    \includegraphics[width=\columnwidth]{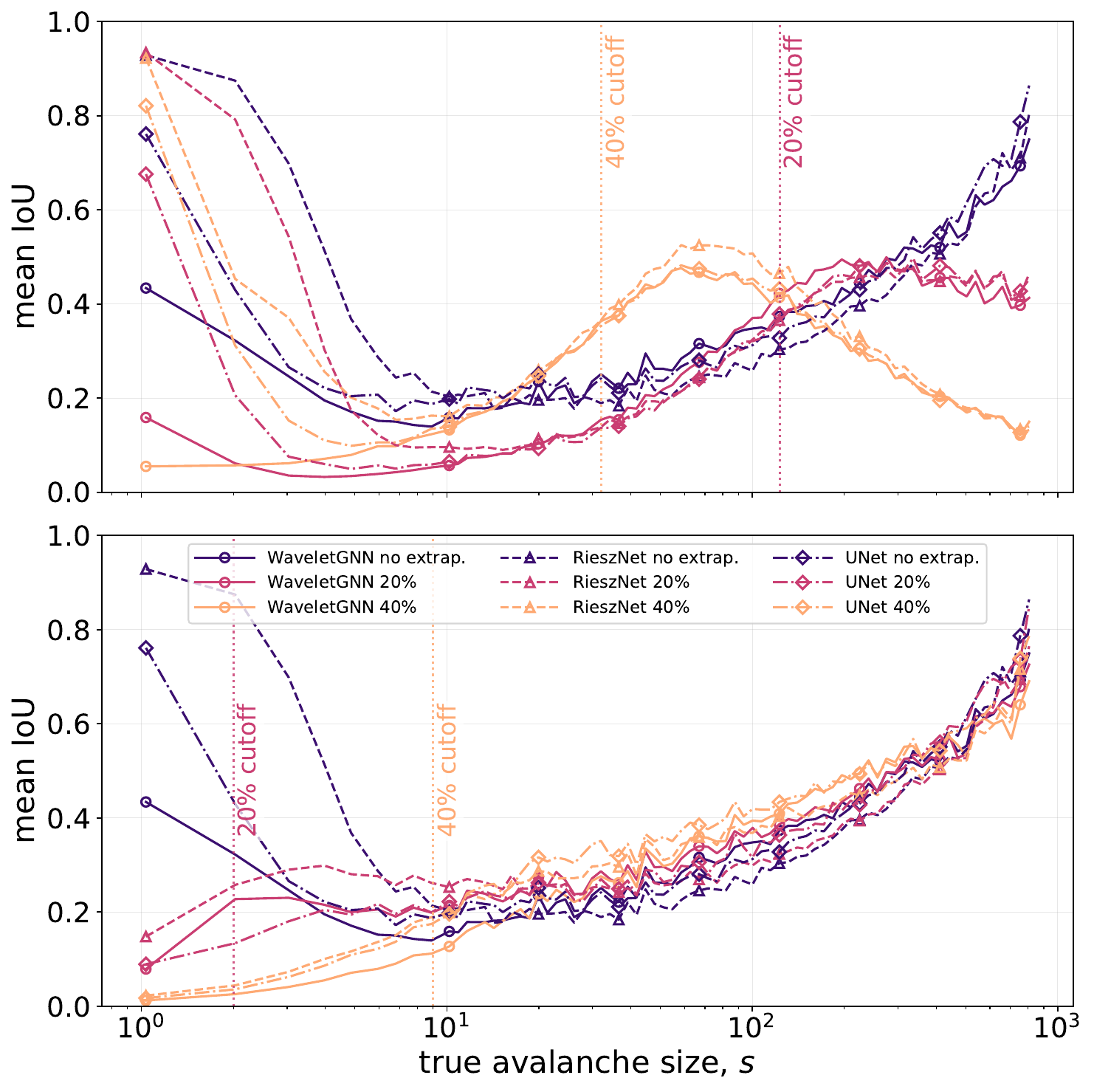}
    \caption{\label{fig:asm_iou}
    IoU as a function of avalanche size in the ASM large-scale extrapolation experiments \textbf{(top)}, and small-scale extrapolation experiments \textbf{(bottom)}. For each of the three models we show the IoU on a test set containing all avalanche sizes. We show three extrapolation regimes marked by the three different colors: the baseline model trained on the full dataset, and two extrapolation cases where we cut $20\%$ and $40\%$ of the largest/smallest avalanches.
    }
\end{figure}

In Fig~\ref{fig:asm_iou} we show similar results for the $\mathrm{IoU}$. 
Even in the baseline case, the tendency to predict overly large avalanches causes a steep drop in the $\mathrm{IoU}$, going down to around $0.2$. 
This effect was not visible in the balanced accuracy, which is a less strict metric.
These low performances can also be related to our choice of a threshold (that defines the active sites), that we chose so as to maximize the Balanced Accuracy, not the $\mathrm{IoU}$.

\section{Scale equivariance} \label{app:scale_equivariance}

\subsection{General properties}
\subsubsection{Reminders}
In the context of Neural Networks, the notion of (deterministic) scale equivariance (applied to images) has been introduced in~\cite{worrall2019deep, sosnovik2019scale}.
Here we aim to address the case of statistically self-similar fields and self-similar processes, and thus introduce the appropriate definitions, using less formal notations.

For simplicity, we focus solely on dilations centered at the origin, excluding translations and rotations.
Let us denote $T_s\in G$ an element of the group~\footnote{If we exclude the singular scaling that reduces space to a point, homoteties centered on the origin form a group.} of scale transformations $G$ with $s>0$ the scaling factor associated to $T_s$. 
For points or vectors $\bx\in \R^d$, the element $T_s$ acts on $\bx$ as follows: 
\begin{align}
    T_s \cdot \bx = s\bx 
\end{align}
For some field $\varphi: X\to \R$ with $X\subset \R^d$, we have a straightforward representation of the group action given by
\begin{align}
    T_s \cdot (\varphi(\bx)) &\egaldef \varphi(s\bx)= \varphi(T_s\cdot  \bx)
    \\
\text{\ie} \quad    T_s \cdot (\varphi(.)) &=  \varphi(T_s\cdot .)
\end{align}
Taking the Fourier transform of $\varphi(s\bx)$ we obtain the known property of the time scaling (in 1D), or more generally in $d$ dimensions:
\begin{align}
      {\left(T_s\cdot\widetilde\varphi\right)} (\bk) 
    \egaldef \widetilde \varphi_s(\bk) 
    = \frac{1}{s^d} \widetilde\varphi\left(\frac{\bk}{s} \right)
\label{eq:fourierscaling}
\end{align}

Now, in addition to these generic facts, we may assume our fields and/or processes to respect some \textit{statistical}  equivariance conditions.
We recall that the field $\varphi$ is (statistically) self-similar with Hurst exponent $H\in \R$ iff:
\begin{align}
\forall \bx, \quad T_s\cdot (\varphi(\bx)) \egaldef \varphi(s \bx) \egallaw s^H \varphi(\bx) 
\label{eq:appC5}
\end{align}
where   the equality is in distribution.
This expression is equivalent to that of the covariance of the increments between points very close together:
\begin{align*}
    \mathbb{E}_\varphi\left[ \vert\varphi(\bx+\mathbf{\delta x})  - \varphi(\bx)\vert^2 \right] \sim \vert\mathbf{\delta x}\vert^{2H}, \quad \text{ for } \vert\mathbf{\delta x}\vert\sim 0
\end{align*}
In Fourier, this scaling relation corresponds to the high-frequency limit of the energy spectral density (often simply called power spectrum in physics):
\begin{align*}
    \mathbb{E}_\varphi\left[ \vert\widetilde \varphi(\bk)\vert^2 \right] \propto \vert\bk\vert^{-(2H+d)}
\end{align*}
Note that for stationary stochastic processes, in general $\varphi$ is not square integrable and there is a nuance between energy density and power density
(evaluated over a finite window). Here in practice we will always work on a finite grid with finite resolution, so the difference amounts to a constant prefactor, that we do not care about.

Concretely, for the FGF with Hurst exponent $H=(\beta-d)/2$, in Fourier and in $d=2$ dimensions, the components $\varphi(\bk)$ are Gaussians with standard deviation $k^{-\frac{2H+d}{2}} = k^{-\beta/2}$,
hence, taking the Fourier transform on both sides of  \eqref{eq:appC5} and applying \eqref{eq:fourierscaling} on the left, we have:
\begin{align}
   \widetilde\varphi_s  =  \frac{1}{s^d} \widetilde\varphi\left(\frac{\bk}{s} \right) 
    &\egallaw s^H \widetilde\varphi(\bk) 
    \label{eq:scalingOfFourierField}
\end{align}
which is equivalent to \eqref{eq:eq8scaling_FGF2}. It's worth noting that the scaling of the Fourier transform of the rescaled field is thus:
$\widetilde\varphi_s(\bk) \egallaw s^H \widetilde\varphi(\bk)$.

In this appendix, we started from the direct space definition of the rescaling to derive what it corresponds to in Fourier space:
$\widetilde\varphi_s(\bk) \egallaw s^{\beta/2-d/2} \widetilde\varphi(\bk)$ (reminder: $H=\frac{\beta-d}{2}$), to study the self-similarity properties of the field $\varphi$. 
When defining the spectral flow however (flow designed to correspond to rescaling, with the proper prefactor to preserve distribution), for convenience we worked directly in Fourier space.
Our definition (\ref{def:Tscale_eq13}) implies (\ref{eq:eq16_spectralflow}): $\varphi(\bk,t)=s^{\beta/2+d/2}\varphi(s\bk,0)$, or with a change of convention $s\to s^{-1}$,
the equivalent equation (\ref{def:varphiseq52}).
The rescaling process defined by our spectral flow dynamics, $\varphi_s(\bk)\egaldef \varphi(\bk,t) = s^{\beta/2+d/2}\varphi(s\bk,0)$, thus corresponds to a  modified
direct space rescaling $\varphi_s(\bx) = s^{\beta/2-d/2}\varphi(\bx/s) = s^H \varphi(\bx/s)$.
Note that our process still preserves distribution, also in direct space: using (\ref{eq:scaling_FGF1}) we get $\varphi_s(\bx) \egallaw \varphi(\bx)$. 
To summarize, the way the field rescales (depending on $H$) and our definition of spectral flow (designed to preserve distribution for that field) should not be confused.

The Boltzmann distribution $P(\varphi)$ ought to be conserved under re-scaling, as we assume $\varphi$ to be self-similar. In particular, the stochastic integral $\int d^2 \bk |\bk|^\beta |\widetilde \varphi(\bk)|^2$ should be invariant under the change of variable $\bk'=\bk/s$, which is non trivial for a stochastic integral (since the scale at which we observe the field changes its scaling, an effect that competes with the Jacobian of the change of variable).
One can indeed check that this property holds for the integral   $I(\varphi) = \int d^2 \bk |\bk|^\alpha |\widetilde \varphi(\bk)|^2$ iff $\alpha=\beta$, using equation \ref{eq:eq8scaling_FGF2}.

\subsection{Scale equivariance for linear operators}\label{app:scale_equivariance_linear}
For linear operators, imposing scale equivariance in addition to rotation invariance leads
to a simple functional form which could  be used as a basic building block of deep architectures, 
when combined with ReLU activation functions for instance. Let us recall here a standard argument used to obtain the form of scaling function. Considering the kernel $F$ defining the functional map~(\ref{eq:map}), assuming rotational invariance imposes the form
\begin{align*}
F(\bk,\bk') = F(k,k',\alpha-\alpha')
\end{align*}
Rewriting~(\ref{eq:map}) for $k\to k/s $ and making the change of variable $k'\to k'/s$ in the integral we get
\begin{align}\label{eq:scale1}
\psi(\bk/s) = s^{-2}\int d^2\bk' F_\theta(k/s,k'/s,\alpha-\alpha') \varphi(\bk'/s)
\end{align}
By definitions~(\ref{def:varphiseq52},\ref{def:ys}), we have 
$\varphi_s\sim {\rm FGF}(\beta)$ and $\psi_s\sim {\rm FGF}(\gamma)$ and these 
should therefore be related by the exact same relation~(\ref{eq:map}) if this one is equivariant, so we have
\begin{align*}
\psi_s(\bk) &= \int d^2\bk F(k,k',\alpha-\alpha')\varphi_s(\bk')\\[0.2cm]
&= s^{(\beta-\gamma)/2-2} \int d^2\bk' F(k/s,k'/s,\alpha-\alpha')\varphi_s(\bk')
\end{align*}
using~(\ref{eq:scale1}) in the second line. 
Since this is supposed to be valid for any  field $\varphi\sim {\rm FGF}(\beta)$
the model should verify
\begin{align}\label{eq:scale_relation}
F(k,k',\alpha-\alpha') =  s^{(\beta-\gamma)/2-2} F(k/s,k'/s,\alpha - \alpha')
\end{align}
for any $k,k'$ and $s$.
Hence choosing $s =  k$  leads to the general form given in~(\ref{eq:scale_form}) after symmetrizing
by convention in $k$ and $k'$.

\subsection{Fourier-Mellin basis}\label{app:Fourier_Mellin}
The decomposition of the operator $F=D_{\rm F}^{-\gamma/2} D_{\rm FM}D_{\rm F}^{\beta/2}$
involves a the strictly scale invariant operator $D_{\rm FM}$ which becomes diagonal in the ``Fourier-Mellin'' representation. 
This representation corresponds to use the Fourier-Mellin basis~(\ref{def:Fourier-Mellin}).
When $F$ reduces to $D_{\rm FM}$ 
we have the decomposition
\begin{align*}
F(\bk,\bk') = \sum_{\mu\in\Z}\int d\lambda D(\lambda,\mu)\phi_{\lambda,\mu}(\bk)\phi_{\lambda,\mu}(\bk') 
\end{align*}
with 
\begin{align*}
D(\lambda,\mu) = \frac{1}{\sqrt{4\pi\log(L)}}\int_{-\log(L)}^{\log(L)} du\int_0^{2\pi} f(e^u,\alpha)e^{i(\lambda u+\mu\alpha)}
\end{align*}
valid in the limit where $\log(L)\to\infty$. For finite $L$ some numerical problems can be encountered 
when using this basis in practice, as orthonormality of the Fourier-Mellin basis is strictly ensured with the definition~(\ref{def:Fourier-Mellin}),
but in practice, after discretization, the states corresponds to momenta  distributed on the logpolar lattice. As a result they are not exactly orthogonal w.r.t.
the scalar product defined on the square lattice domain,
which is the reason why we have to resort to a log polar lattice interpolated from the square with a Dirichlet kernel.

\section{Asymptotic behavior of the linear model with RMT}\label{app:RMT}
\subsection{Train and test error as function of time}\label{app:Train-Test}
In order to write the train and test error along the learning process it is convenient first  to introduce the $M\times L^4$ feature matrix $J$ defined by its matrix
elements indexed by $p$ and implicitly by $(\bk,\bk')$ (which can take $L^4$ possible values):
\begin{align*}
J_p(\bk,\bk') \egaldef \frac{1}{k'^{\beta/2}}f_p(\bk,\bk').
\end{align*}
Given that, the population matrix~(\ref{eq:population_matrix}) and the coefficients~(\ref{eq:wp}) rewrite
\begin{align*}
C = JJ^\top\qquad\text{and}\qquad  w= {J^\top}^\dagger \widetilde K
\end{align*}
where $\widetilde K$ is the $L^4$ dimensional vector defined as
\begin{align*}
\widetilde K(\bk,\bk') \egaldef \frac{1}{k'^{\beta/2}} K^\parallel(\bk,\bk'). 
\end{align*}
In terms of all these quantities the train and test error take the following form as a function of learning time:
\begin{align}
E_{\rm train}(t) &= \Tr\Bigl[\widehat C\bigl(\I-\widehat G(t)\widehat C\bigr)^2{J^\top}^\dagger \widetilde K\widetilde K^\top J^\dagger\Bigr] \nonumber\\[0.2cm]
&+ \sum_k\sigma_{\rm eff}(k)^2 \Bigl[1-\frac{2}{N}\Tr\Bigl(\widehat G(t)\widehat C(k)\Bigr)\nonumber
\\[0.2cm]
&+\frac{1}{N}\Tr\Bigl(\widehat G(t)\widehat C\widehat G(t)\widehat C(k)\Bigr)\Bigr]\label{eq:E_train_t}
\\[0.2cm]
E_{\rm test}(t) &=  \Tr\Bigl[\Bigl(\I-\widehat C\widehat G(t)\Bigr) C 
\Bigl(\I-\widehat C\widehat G(t)\Bigr) {J^\top}^\dagger \widetilde K\widetilde K^\top J^\dagger\Bigr]\nonumber\\[0.2cm]
&+ \sum_k\sigma_{\rm eff}(k)^2 \Bigl[1+\frac{1}{N}\Tr\Bigl(\widehat G(t)  \widehat C(k) \widehat G(t) C\Bigr)\Bigr]\label{eq:E_test_t}
\end{align}    
with 
\begin{align*}
\widehat C_{pq}(k) = \sum_{\bk',\bk''\in\Omega^2} f_p(\bk,\bk')f_q(\bk,\bk'') \E_{\varphi\sim \widehat{\rm FGF}(\beta)}\bigl[\varphi(\bk')\varphi(\bk'')\bigr] 
\end{align*}
(by rotation invariance this actually depends only on the modulus of $\bk$)
and where $\sigma_{\rm eff}^2$ is the effective variance of the noise
which emerges from the fact that the regression is misspecified. As mentioned in the plain text this noise comes from the contribution of $K^\perp$ in~(\ref{eq:data_map}).
 
\subsection{Dynamics of train and test errors}\label{app:dyna_Errors}
Train and test errors given by~(\ref{eq:E_train_t},\ref{eq:E_test_t}) can be analyzed asymptotically in the so-called proportional regime, namely when $N$, $N_s$ and $P$ go to infinity at fixed $\rho$ in~\ref{def:rho}. To this end some preparatory work needs to be done.  
First, the longitudinal contribution represented by the bias term in~(\ref{eq:E_train_t},\ref{eq:E_test_t}), may be rewritten in a suitable form allowing in the asymptotic limit.
Since the features are not learned, $\widetilde K$ is 
not correlated with $\widehat C$ neither aligned with $C$. As a result one of the contribution to the bias term reads \eg in the train error:
\begin{align*}
  \Tr\Bigl[\widehat G(t) \widehat C \widehat G(t) {J^\top}^\dagger \widetilde K\widetilde K^\top J^\dagger\Bigr] 
  &\approx \frac{\kappa}{M}\Tr\Bigl[\widehat G(t) \widehat C \widehat G(t) 
  {J^\top}^\dagger J^\dagger\Bigr]\\[0.2cm] 
  &= \frac{\kappa}{M}\Tr\Bigl[\widehat G(t) \widehat C \widehat G(t) C^{-1}\Bigr]\\[0.2cm]  
\end{align*}
where $\approx$ stands for ${\mathcal O}\Bigl(\frac{1}{\sqrt{M}}\Bigr)$ errors.
As can be noticed, a $k$ dependent term shows up in the variance term of both 
the train and test errors~(\ref{eq:E_train_t},\ref{eq:E_test_t}). This term renders the asymptotic
analysis much more involved as will be detailed in the following. But this actually  simplifies when considering  a  different setting 
of the learning where a preconditioner is used to flatten the contributions of all scales
\begin{align*}
\LL(\theta) = \frac{1}{2} \E_{(\varphi,\psi)\sim \widehat{\rm FGF}(\beta,\gamma)}\left[\sum_{\bk\in\Omega} k^\gamma\bigl\vert \psi_\theta[\varphi](\bk) - \psi(\bk)\bigr\vert^2\right],
\end{align*}
and assume that the rescaled noise $k^\gamma\sigma_{\rm eff}(k)^2 = \bar\sigma_{\rm eff}$
is constant. Doing so requires to change accordingly the definitions of $\widehat C$ and $\widehat Z$
in~(\ref{eq:C_pq},\ref{eq:Z_p}) by an extra factor $k^\gamma$ inside the sum over $\bk$. 
In turn the population matrix is now defined as 
\begin{align*}
C_{pq} = \langle f_p, f_q \rangle_{\beta,\gamma}
\end{align*}
corresponding to a change in the definition~(\ref{def:inner_product}) of the scalar product between kernels to 
\begin{align*}
\langle F_1, F_2\rangle_{\beta,\gamma} \egaldef \int d^2\bk d^2\bk' \frac{k^\gamma}{k'^\beta} F_1(\bk,\bk') F_2(\bk,\bk')
\end{align*}
and accordingly in the definitions of $\widetilde K$ and $J$ above
\begin{align*}
J_p(\bk,\bk') &\egaldef \frac{k^{\gamma/2}}{ k'^{\beta/2}}f_p(\bk,\bk'), \\[0.2cm]
\widetilde K(\bk,\bk') &\egaldef \frac{k^{\gamma/2}}{ k'^{\beta/2}} K^\parallel(\bk,\bk'),
\end{align*}
and finally the new definition for $\kappa$:
\begin{align*}
\kappa \egaldef \langle K^\parallel,K^\parallel\rangle_{\beta,\gamma}.
\end{align*}
With all that, the train and test errors now takes the form
\begin{align}
E_{\rm train}(t) &\approx \kappa\Tr\Bigl[\widehat C\bigl(\I-\widehat G(t)\widehat C\bigr)^2
C^{-1}\Bigr] \nonumber\\[0.2cm]
&+ \sigma_{\rm eff}^2 \Bigl[1-\frac{1}{\rho}+\frac{1}{N}\Tr\Bigl(\I-\widehat G(t)\widehat C\Bigr)^2\Bigr]\label{eq:E_train2}\\[0.2cm]
E_{\rm test}(t) &\approx  \kappa\Tr\Bigl[\Bigl(\I-\widehat C\widehat G(t)\Bigr) C 
\Bigl(\I-\widehat C\widehat G(t)\Bigr) C^{-1}\Bigr]\nonumber\\[0.2cm]
&+ \sigma_{\rm eff}^2 \Bigl[1+\frac{1}{N}\Tr\Bigl(\widehat G(t)  \widehat C \widehat G(t) C\Bigr)\Bigr]\label{eq:E_test2}
\end{align}    
which can be straightforwardly analyzed asymptotically, at least when $C=\I$.
Let us call
\begin{align*}
j(x,t) = \frac{1-e^{-xt}}{x}
\end{align*}
the function governing the dynamics of the vector parameter $\theta$.
Using the Cauchy integral representation we may now express the train and test errors 
in terms of the resolvent
\begin{align*}
\widehat G(z) = \frac{1}{z\I-\widehat C}.
\end{align*}
We have
\begin{align}
  E_{\rm train}(t) &\approx \frac{\kappa}{2i\pi}\oint_{\C} dz z[1- zj(z,t)]^2 \Tr\bigl[\widehat G(z) C^{-1}\bigr]\nonumber \\[0.2cm]
  &+ \sigma_{\rm eff}^2 \Bigl(1-\rho^{-1}\nonumber\\[0.2cm]&
  +\frac{1}{2i\pi N}\oint_{\C} dz [1-zj(z,t)]^2\Tr\bigl[\widehat G(z)\bigr]\Bigr)
  \label{eq:E_trainz} \\[0.2cm]
  E_{\rm test}(t)   &\approx \frac{\kappa}{2i\pi}\oint_{\C} dz_1 [1-z_1j(z_1,t)]\oint_{\C} dz_2 [1-z_2j(z_2,t)]\nonumber\\[0.2cm] 
  &\times\Tr\bigl[\widehat G(z_1)C\widehat G(z_2)C^{-1}\bigr] \nonumber \\[0.2cm]
  &+\sigma_{\rm eff}^2 \Bigl(1+\frac{1}{2i\pi N}\oint_{\C} dz zj(z,t)^2\Tr\bigl[\widehat G(z)C\bigr]\Bigr)\label{eq:E_testz}
\end{align}
and the contour $\C$ is an  anti-clockwise contour around the real positive axis.
In this form we can obtain asymptotic expressions thanks to basic random matrix theory (RMT) formulaes~\cite{MP_law,ledoit2011eigenvectors} at least for the train error
and for the test error when $C = \I$. For $C \ne \I$ the bias term of the test error requires more advances techniques to be taken care of, and we postpone this discussion for the moment and concentrate on the case $C=\I$.  

\subsection{Asymptotic limits of learning trajectories}\label{sec:Asymptotics}
Let us first derive the RMT equations in this context. This is a little bit non-standard because
the covariant matrix takes the form of a Wishart matrix with non-iid samples. Such case has been 
addressed in~\cite{couillet2011deterministic} for instance.
We have
\begin{align*}
\widehat C = \frac{1}{N}\sum_{s=1,\bk\in\Omega}^N \x^{(s)}(\bk)\x^{(s)\top}(\bk),
\end{align*}
where 
\begin{align}\label{eq:xpsk}
x_p^{(s)}(\bk) = \sum_{\bk'\in\Omega} f_p(\bk,\bk')\varphi^{(s)}(\bk'),
\ p=1,\ldots M, s=1,\ldots N.    
\end{align}
It is a sum of $NL^2$ rank one matrices formed with the 
vectors $\x^{(s)}(\bk)$ indexed by $s$ and $\bk$. These are not iid because
of the $\bk$ dependency.
The essence of the method resides in finding if there exists a deterministic equivalent~\cite{hachem2007deterministic} $G(z)$  of the resolvent 
$\widehat G(z)$. This would mean that  for any sequence of $\R^M\times \R^M$ matrices $Q_M$ 
of Frobenius norm equal to one, $\Tr\Bigl[(G(z)-\widehat G(z))Q_M\Bigr]$ goes to zero typically like $\sqrt{\log(N)/N}$ (see e.g.~\cite{chouard2023sample}).
If it exists, $G(z)$ should coincide with the average of $\widehat G(z)$ when taking the proportional limit. 
As a $2$-points function, this obeys a Dyson equation 
\begin{align*}
G(z) = \frac{1}{z}+\frac{1}{z}\Sigma(z) G(z)
\end{align*}
where $\Sigma(z)$ corresponds to the self-energy which in the context of RMT reduces to a sum over planar diagrams. Let us re-derive this 
informally with the Feynmann diagrams expansion~\cite{burda2004signal,furtlehner2023free} when  
both the features and the data are correlated. This corresponds to a data matrix
of the form 
\begin{align}\label{eq:dependency}
X = A^{\frac{1}{2}}Z B^{\frac{1}{2}},     
\end{align}
where $X$ is $M\times N$ matrix, $A$ is $M^2$ while $B$ is $N^2$, representing respectively the feature and data linear dependency. $Z$ is a $M\times N$
random matrix with iid elements. We are interested by the limit spectrum  of
\begin{align*}
\widehat C \egaldef \frac{1}{N} XX^\top \qquad\text{and}\qquad \widehat C^\star \egaldef \frac{1}{N} X^\top X
\end{align*}
Note that the relation between $A$, $B$ and the population matrix is 
\begin{align*}
C = A\lim_{N\to\infty}\frac{1}{N}\Tr[B].
\end{align*}
Let us consider the  resolvent corresponding to the $(N+M)^2$ matrix
\begin{align*}
M(z) \egaldef 
\left(
\begin{matrix}
  \sqrt{z}\I_M  & -\frac{X}{\sqrt{N}}  \\[0.2cm]
  -\frac{X^\top}{\sqrt{N}} & \sqrt{z}\I_N 
\end{matrix}\right)
= \sqrt{z} \I_{N+M} - \frac{\widehat S}{\sqrt{N}}
\end{align*}
with $\widehat S$ denoting now the data matrix written in this two-component representation. 
Its inverse is given by
\begin{align}\label{eq:Mzinv}
M(z)^{-1} = \sqrt{z}
\left(
\begin{matrix}
  \widehat G(z) & -\frac{1}{\sqrt{z}} \widehat G(z)X \\[0.2cm]
  -\frac{1}{\sqrt{z}}\widehat G^\star(z)X^\top & \widehat G^\star(z)
\end{matrix}
\right)
\end{align}
with
\begin{align*}    
\widehat G(z) &= (z\I_M-\widehat C)^{-1}   \\[0.2cm]
\widehat G^\star(z) &= (z\I_N-\widehat C^\star)^{-1} 
\end{align*}
Expanding $M^{-1}$ in terms of the data matrix and averaging over $Z$, leads, thanks to the Wick theorem (see e.g.~\cite{burda2004signal})
to obtain the self energy $\Sigma(z)$ of the Dyson equation
\begin{align*}
{\mathcal G}(z) = {\mathcal G}_0(z) + {\mathcal G}_0(z) \Sigma(z){\mathcal G}(z) 
\end{align*}
concerning the resolvent ${\mathcal G}(z)$ of $M(z)$ in the asymptotic limit and where ${\mathcal G}_0(z) = \frac{\I_{N+M}}{\sqrt{z}}$.
Thanks to the planar diagrams approximation we get for the self energy
\begin{align*}
  \Sigma(z) &= \lim_{N,M\to\infty}\frac{1}{N}\E\bigl[ \widehat S\mathcal G(z) \widehat S\bigr] \\[0.2cm]
  &= \lim_{N,M\to\infty}
\left(
\begin{matrix}
\frac{1}{N}\Tr[G^\star(z)B]A & 0 \\[0.2cm]
0 & \frac{1}{N}\Tr[G(z)A]B
\end{matrix}
\right)
\end{align*}
where $G(z)$ and $G^\star(z)$ are the deterministic equivalents respectively for 
the resolvent for $\widehat C$ and $\widehat C^\star$.
Since we know from~(\ref{eq:Mzinv})  that 
\begin{align*}
{\mathcal G}(z) =
\sqrt{z}\left(
\begin{matrix}
  G(z) & 0 \\[0.2cm]
  0 & {G^\star}(z)
\end{matrix}
\right)
\end{align*}
we end up with the RMT equations
\begin{align*}
\Gamma(z) &= \frac{1}{\rho}\int\nu(dx) \frac{x}{z - x\Gamma^\star(z)} \\[0.2cm]
\Gamma^\star(z) &= \int\nu^\star(dx) \frac{x}{z - x\Gamma(z)}
\end{align*}  
where
\begin{align*}
  \Gamma(z) &= \lim_{N,M\to\infty}\frac{1}{N}\Tr\bigl[\widehat G^\star(z)A\bigr] \\[0.2cm]
  \Gamma^\star(z) &= \lim_{N,M\to\infty}\frac{1}{N}\Tr\bigl[\widehat G(z)B\bigr]   
\end{align*}
and $\nu$ and $\nu^\star$ are the spectral density respectively associated to $A$ and $B$.
If we specify this now to our problem, $X$ is to be seen as a $M\times (N L^2)$
matrix with elements $x_p^{(s)}(\bk)$, $p=1,\ldots M$ and $(s,\bk)\in\{1,\ldots N\}\times\Omega$.
From~(\ref{eq:xpsk}) the dependency structure is in general more entangled than~(\ref{eq:dependency})
and leads to RMT equations where $\Gamma(z)$ and $\Gamma^\star(z)$ have matrix form of respective sizes
$M\times M$ and $L^2\times L^2$, which cannot be given a sense in the asymptotic limit. In order to 
recover finite dimensional mean-field equations we need to impose that dependency factorizes in the following way:
\begin{align*}
\E\bigl[x_p^{(s)}(\bk)x_q^{(s')}(\bk')] = A_{pq} B(\bk,\bk')\ind{s=s'}
\end{align*}
which corresponds to assume that the features obey the following relations
\begin{align*}
\sum_{\bk\in\Omega} \frac{f_p(\bk,\bk_1)f_q(\bk',\bk_1)}{k_1^\beta}
= A_{p,q}B(\bk,\bk').
\end{align*}
This condition is obeyed for instance when the feature are rank one operators based on a set of orthonormal functions w.r.t. to the inner product associated to the density $\mu(d\bk) = k^\beta$.
In that case we have $A=\I$.

Along the contour of the Cauchy integrals in~(\ref{eq:E_trainz},\ref{eq:E_testz}) we integrate over $z=y+i\epsilon$ with $\epsilon$ infinitesimal.
In the limit $\epsilon\to 0$, both $\Gamma$ and $\Gamma^\star$ may acquire a finite imaginary part
which we write as
\begin{align*}
    \lim_{\epsilon\to 0^\pm}\Gamma(z) &= \Gamma_r(y)\pm\Gamma_i(y) \\[0.2cm]
    \lim_{\epsilon\to 0^\pm}\Gamma^\star(z) &= \Gamma^\star_r(y)\pm\Gamma^\star_i(y).
\end{align*}
Letting $\bar\nu(dx)$ the asymptotic limit of the empirical spectrum in the proportional regime, 
its Stieltjes transform is given by the trace of the resolvent:
\begin{align*}
g(z) \egaldef \int \frac{\bar\nu(dx)}{z-x}.
\end{align*}
Then the  spectrum is given by the Stieltjes transform
\begin{align*}
g(y+i\epsilon) = g_r(y)+i\pi\frac{\epsilon}{\vert\epsilon\vert}\bar\nu(y),
\end{align*}
which thanks to the identity 
\begin{align*}
\Gamma(z)\Gamma^\star(z) = \frac{1}{\rho}[1+zg(z)]
\end{align*}
rewrites 
\begin{align}\label{eq:bnu}
\bar\nu(y) = (1-\rho)\ind{\rho<1}\delta(y)+ 
\rho\frac{\Gamma_i(y)\Gamma^\star_r(y)+\Gamma_r(y)\Gamma^\star_i(y)}{\pi y}.
\end{align}
In terms of these quantities we obtain equations~(\ref{eq:E_train_asymp},\ref{eq:E_test_asymp}) for the train and test errors up 
to some quite technically involved additional corrections when $C \ne \I$ to evaluate exactly the bias term of the test error,  \ie the 
first term of~(\ref{eq:E_test_t}).

\end{document}